\documentclass[preprint2]{aastex}

\usepackage[letterpaper]{geometry}
\usepackage{apjfonts}
\usepackage{graphicx}
\usepackage{epstopdf}
\shorttitle{Chemistry in cometary nuclei}
\shortauthors{R. T. Garrod}

\begin{document}

\title{Simulations of ice chemistry in cometary nuclei}

\author{Robin T. Garrod\altaffilmark{1}}
\email{rgarrod@virginia.edu}

\altaffiltext{1}{Departments of Astronomy and Chemistry, University of Virginia, Charlottesville, VA 22904}

\begin{abstract}

The first computational model of solid-phase chemistry in cometary nuclear ices is presented. An astrochemical kinetics model, {\em MAGICKAL}, is adapted to trace the chemical evolution in multiple layers of cometary ice, over a representative period of 5 Gyr. Physical conditions are chosen appropriate for ``cold storage'' of the cometary nucleus in the outer Solar System, prior to any active phase. The chemistry is simulated at a selection of static temperatures in the range 5 -- 60 K, while the ice is exposed to the interstellar radiation field, inducing a photochemistry in the outer ice layers that produces significant formation of complex organic molecules. A treatment for the chemistry resulting from cosmic-ray bombardment of the ices is also introduced into the model, along with a new formulation for low-temperature photochemistry. Production of simple and complex molecules to depth on the order of 10~m or more is achieved, with local fractional abundances comparable to observed values in many cases. The production of substantial amounts of O$_2$ (and H$_2$O$_2$) is found, suggesting that long-term processing by high-energy cosmic rays of cometary ices {\em in situ}, over a period on the order of 1 Gyr, may be sufficient to explain the large observed abundances of O$_2$, if the overall loss of material from the comet is limited to a depth on the order of 10 m. Entry into the inner solar system could produce a further enhancement in the molecular content of the nuclear ices that may be quantifiable using this modeling approach.

\end{abstract}


\section{Introduction}

Cometary nuclei are thought to be the most primitive and pristine objects in the solar system (Whipple 1950, Delsemme 1982), and are composed of organic ices and rocky material (i.e. dust grains) in approximately equal amounts (Mumma \& Charnley 2011). The composition of the ices of dozens of cometary nuclei has been inferred through millimeter-band and infrared observations of the gaseous comae around active comets, which form by the sublimation of volatile nuclear material. These measurements indicate that the nuclear ices are dominated by water, with CO and CO$_2$ comprising up to 20--30\% (with respect to water) in some cases, while trace amounts of other organics, such as methanol (CH$_3$OH) and methane (CH$_4$), are detected (Cordiner et al. 2016, Ehrenfreund \& Charnley 2002). However, so-called {\em complex organic molecules} (COMs, typically defined as organics of 6 or more atoms; van Dishoeck \& Herbst 2009), such as methyl formate (CH$_3$OCHO; Bockel{\'e}e-Morvan et al. 2000), glycolaldehyde (CH$_2$OHCHO; Biver et al. 2015) and ethylene glycol ((CH$_2$OH)$_2$; Crovisier et al. 2004, Remijan et al. 2008), have also been identified through remote observations, while the {\em Stardust} mission to comet Wild 2 returned samples containing the simplest amino acid, glycine (NH$_2$CH$_2$COOH; Elsila et al. 2009). The same molecule was more recently identified in comet 67P as part of the {\em Rosetta} mission (Altwegg et al. 2016).

Simulations of the gas-phase chemical kinetics occurring within cometary comae have shown that some molecules (e.g. HNC; Irvine et al. 1998) must be considered ``daughter'', rather than ``parent'', species, forming through gas-phase processes in the expanding coma. However, observations indicate that the more complex species originate in the nuclear ices themselves; high-spatial resolution ground-based observations of comet C/1995 O1 (Hale-Bopp) by Remijan et al. (2008) show that the emission from ethylene glycol molecules, first detected by Crovisier et al. (2004), must be located close to the comet's surface, rather than in an extended region that would imply a delayed, gas-phase formation process. 

The recent {\em Rosetta} mission to the Jupiter-family comet 67P/Churyumov--Gerasimenko has further enhanced our knowledge of the chemical composition of cometary nuclei, and has provided yet more evidence that many complex organic species must come directly from the ice. Reaching its target in 2014, the mission culminated in the touchdown of the {\em Philae} lander on the comet's surface. Data from the {\em COSAC} mass spectrometer was used to identify 16 molecules that were released from the surface as a result of the impact and heating associated with the landing (Goesmann et al. 2015), with abundances on the order of 0.1--1\% of water abundance. Four of these molecules -- propanal (C$_2$H$_5$CHO), acetone (CH$_3$COCH$_3$), acetamide (CH$_3$CONH$_2$) and methyl isocyanate (CH$_3$NCO) -- represented entirely new detections of complex organic molecules in a comet. The {\em Ptolemy} instrument, sampling the ambient gas-phase coma, detected yet more complex material, such as large CHO-group bearing compounds (Wright et al. 2015). Using {\em Rosetta}'s {\em ROSINA} spectrometer,  Altwegg et al. (2016) detected glycine in the coma of 67P near perihelion.

One of the most striking features of the chemical compositions of cometary ices is their similarity to interstellar abundances (e.g. Bockel{\'e}e-Morvan et al. 2000, Gibb et al. 2000) for the small number of species that can be detected through IR absorption spectroscopy, indicating that cometary material may have a direct origin in interstellar ices, or in ices of similar composition formed during the protoplanetary disk stage. The difficulty in ascertaining the chemical composition of protoplanetary disks makes the determination of the relative contributions difficult (Mumma \& Charnley 2011), although models of protoplanetary disk chemistry by Cleeves et al. (2014) indicate that the deuterium fractionation ratios observed in terrestrial water are a direct inheritance from interstellar chemistry, and not from later processing in the disk. 

Similarities between cometary and interstellar ices extend beyond the simpler ice constituents like H$_2$O, CO and CO$_2$, to more complex species such as methyl formate and ethylene glycol, which are also detected in abundance in the gas phase of high-mass star-forming cores known as ``hot cores'' (e.g. Blake et al. 1987, Hollis et al. 2002, Brouillet et al. 2015). The recently-detected cometary molecule methyl isocyanate (CH$_3$NCO; Goesmann et al. 2015) has also been detected in the chemically-rich high-mass star-forming region Sgr B2N (Halfen et al. 2015; Cernicharo et al. 2016) and toward the Class 0 protostar IRAS 16293-2422 (Ligterink et al. 2017). Experimental studies suggest that few such complex molecules can be formed in sufficient quantity in the gas-phase through ion-molecule reactions (Horn et al. 2004, Luca et al. 2002, Geppert et al. 2006), although neutral-neutral processes have now been proposed in some cases (e.g. Vasyunin \& Herbst 2013). However, coupled gas-grain chemical simulations show that grain-surface and solid-state ice processes are capable of reproducing observed values in interstellar clouds and hot cores (e.g. Garrod \& Herbst 2006, Garrod et al. 2008, Garrod 2013). For example, methyl formate (CH$_3$OCHO) may be formed through the UV photolysis of methanol in interstellar ices, leading to radical--radical reactions within and upon the ice mantles at temperatures at which one or other of the key functional-group radicals (i.e., HCO and CH$_3$O) become mobile. The dissociation of methanol and several other major ice constituents appears to be crucial to the formation of various complex molecules in star-forming sources, in combination with the elevated dust/ice temperatures ($\sim$20--40 K or higher) produced by protostellar heating that allow increased diffusional mobility.

Various laboratory experiments have also considered the formation of complex molecules in organic ices (e.g. Greenberg 1982, Bennett \& Kaiser 2007, {\"O}berg et al. 2009a, Abou Mrad et al. 2016). Other experiments have explored the formation of glycine and more complex amino acids by UV or charged-particle irradiation of ice mixtures appropriate to interstellar and cometary compositions (Bernstein et al. 2002, Holtom et al. 2005, Lee et al. 2009); a number of different compositions have been found capable of producing such molecules, although the reaction kinetics are poorly understood. However, it appears that the Strecker synthesis mechanism, which requires liquid water, is not necessary for glycine formation in ices (Elsila et al. 2007). Chemical-kinetics models of hot-core chemistry (Garrod 2013) indicate that several possible reaction routes may instead be capable of producing glycine by radical addition during the star-formation process, although no unambiguous detections of glycine in the interstellar medium have yet been made (Snyder et al. 2005).

While some diversity is observed between the chemical abundances observed in different comets, the abundances in cometary ices of the possible pre-cursors (e.g. CH$_3$OH, H$_2$CO, CH$_4$, NH$_3$) of complex organic molecules are somewhat less than those detected in many interstellar and protostellar sources; indeed, the abundance of the important molecule methanol (CH$_3$OH) is not found to exceed 7\% with respect to water ice in comets (Mumma \& Charnley 2011), in contrast to the $\sim$30\% abundance detected toward some low-mass protostars (Boogert et al. 2008). It is plausible that this statistic may, of itself, be suggestive of processing of pre-cometary and/or cometary ices toward greater chemical complexity. Indeed, the observed cometary abundances of complex organics ($\sim$$10^{-3}$ -- $10^{-2}$ with respect to water, in some cases) can be significantly greater than those observed toward interstellar sources, which typically do not exceed 10$^{-4}$ with respect to water, indicating that a further period of ice processing beyond the interstellar/star-formation stage could be at work.

The cometary reservoir known as the Oort cloud is thought to have been formed by the clearing and ejection of debris from the giant-planet region around 3.5--4.5 Gyr ago, with the gravitational effect of passing stars and other material helping to stabilize the orbits of the ejected planetesimals (e.g. Stern 2003; Dones et al. 2015). The Oort cloud is estimated to comprise on the order of $10^{11}$ comets, with an uncertainty around a factor of two (Brasser \& Morbidelli 2013). Definitions vary as to the inner and outer radii of the Oort cloud; simulations by Duncan et al. (1987) assumed 3000 and 200,000 au, respectively, while Brasser \& Morbidelli (2013) adopted an inner value of 1000 au. Models commonly show radial distributions that peak in the range 10$^4$--10$^5$ au (e.g. Dones et al. 2004). Work by Brasser \& Morbidelli (2013) suggests that the overall population of the Oort cloud reaches its peak within around 0.5--1 Gyr, gradually declining by around half over the following 3--3.5 Gyr. Gravitational interactions with passing stars can alter the orbits of Oort cloud comets, reducing their perihelion distances (Dones et al. 2015) and making possible a closer approach to the Sun.

Thus, comets may be stored in the Oort cloud for billions of years before entering the inner solar system and becoming active; this dormant status in the Oort cloud is sometimes referred to as ``cold storage''. However, during this period, comets may be irradiated by the interstellar UV field, and impacted by galactic cosmic rays, with both being capable of causing chemical changes (Meech \& Svore{\v n} 2004; Guilbert-Lepoutre et al. 2015). Thermal processing is also suggested to occur as a result of close encounters with luminous O and supergiant stars, or by nearby supernova events (Meech \& Svore{\v n} 2004; Stern \& Shull 1988). This could act to raise significantly the temperatures of comets in cold storage, especially those in the Oort cloud, from ambient temperatures of 5--6 K up to as high as 60 K (Stern 2003), albeit for a relatively brief period in a comet's overall lifetime. As well as inducing sublimation, and activating UV-driven chemistry in the outer ice layers, this thermal processing may activate latent radical-radical chemistry between otherwise immobile photo-fragments. Supernovae could also have heated comets to similar temperatures, over much shorter timescales (Stern \& Shull 1988; Stern 2003). 

Thus, the outer layers of most comets originating in the Oort cloud are likely to have been heavily processed during the cold-storage phase. Many of the same processes may have occurred also to Kuiper belt objects, although these are thought to be ``younger'' objects, due to their much higher degree of surface re-processing caused by collisions. Upon later solar approach, active-phase comets are further heated and irradiated by the Sun, causing more drastic chemical and physical changes.

The similarities between the type of processing experienced by interstellar ices (during the star-formation process) and cometary ices (during the cold-storage and/or active phases), i.e. heating, and UV and cosmic-ray processing, mean that the computational techniques used to study the time-dependent chemistry of the former should also be applicable to the latter. However, while there exist {\em physical} models of cometary ices, until now no {\em chemical} simulations of cometary ice evolution have been carried out, and the solid-phase chemical processing that occurs between the formation of the comet and its ultimate destruction remains computationally unexplored. 

In this paper is presented the first such set of models, concentrating initially on the processing that occurs during the ``cold storage'' phase of Oort cloud objects, with a view to investigating the potential for production of both simple and complex organic molecules in the cometary ices, as well as their survival during this period. The model is based on the interstellar gas-grain chemical kinetics model {\em MAGICKAL} (Garrod 2013), which includes solid-phase and surface chemistry for a range of complex molecules including the simplest amino acid, glycine. The new model expands the physical treatment of the ice from a relatively simple gas/surface/mantle model to a multi-layered, stratified ice model. The model simulates UV irradiation and chemistry at various depths, taking account of the diminution of UV field strength with depth (due primarily to absorption by dust within the ice), thus moving the simulations from the microscopic scales of interstellar dust grains to the macroscopic scales appropriate to comets. A simple cosmic ray-induced radiolysis treatment is also introduced.

The specific methods used are explained in Section 2, with results of the basic models presented in Section 3. Section 4 presents results for models that include cosmic-ray processing. Discussion and Conclusions follow in Sections 5 and 6.

\section{Method}

The computational model used here to simulate cometary nuclear ice chemistry is based on the interstellar gas-grain chemical kinetics model {\em MAGICKAL} ({\em Model for Astrophysical Gas and Ice Chemical Kinetics And Layering}). This model uses a rate-equation approach, solving a set of coupled ordinary differential equations to determine the time-dependence of all chemical abundances. In addition to simulating ice-surface chemistry, the model also traces diffusion-mediated reactions within the bulk ice itself, and dissociation caused by UV radiation (as well as cosmic-ray impacts -- see Section 4). Unlike in the interstellar application, gas-phase chemistry is switched off in the comet model, and while surface species are allowed to desorb (both thermally and non-thermally) into the gas phase, no re-accretion of gas-phase material back onto the cometary surface is allowed in the simulations presented here; the surrounding gas would be of extremely low number density (less than 0.05 cm$^{-3}$; Scherer \& Fahr 1998), making accretion onto the comet surface essentially negligible during the cold-storage phase.

The model comet is stratified into 25 separate ``layers'', with thicknesses that increase with depth; in this work, the term ``layer'' is used exclusively to indicate one of those 25 model layers of varying thickness, while the term ``monolayer'' is used to indicate a unit of thickness of precisely one molecule. The outer layer in the model is one monolayer thick, the second is 3 monolayers, and each subsequent layer is three times thicker than the last, such that the deepest layer has a thickness of $\sim$$2.8 \times 10^{11}$ monolayers. The majority of the ice is composed of water, thus it is assumed here that a monolayer of generic cometary ice has a thickness corresponding to the size of that molecule. Under the assumption that cometary water ice is well represented by an amorphous solid water (ASW) structure, a typical density of 0.9 g~cm$^{-3}$ (e.g. Brown et al. 1996) is adopted, which assumes that there is no empty space within the ice, and which provides a mean volume per water molecule of 33.22 \AA$^3$. Taking this as the basic (cubic) unit volume per particle in the ice, the thickness of a single monolayer is taken as 3.215 \AA. The deepest ice layer therefore comprises $\sim$91 m of ice, and the total thickness of the modeled comet ice is $\sim$136 m. These depth calculations will vary if ice porosity is taken into account; see Secs. 2.1, 4, and 5.2.

Each of the 25 distinct layers is treated as a separate phase with its own set of kinetic rates and chemical abundances; neighboring layers are also coupled through thermal diffusion of chemical species from one layer to another. Losses caused by desorption into the gas phase from the surface layer are compensated by a transfer of material from the layer beneath, and similar transfers continue down between contiguous pairs of layers as necessary to ensure that the sum of all fractional abundances is conserved for each layer at all times. The total fractional abundance within each layer may also be affected by the photo-dissociation of a single species into two products, or by reactions between two species forming a single product; all such gains and losses are taken into account when calculating the total transfer rates between layers.

Beneath the deepest layer is a chemically-inert reservoir, with fractional abundances permanently fixed to the initial composition assumed for all ice layers. Any shortfall in the sum of abundances in the 25$^{\mathrm{th}}$ layer is made up by a transfer of pristine material from the reservoir. Likewise, any overabundance in the 25$^{\mathrm{th}}$ layer is absorbed by the reservoir, although the reservoir composition is assumed to be unchanged as a result. No thermal diffusion into or out of the reservoir is allowed.

For the purposes of the chemical modeling, the lateral size of each ice layer is undefined, and thus chemical abundances are calculated simply in terms of the number of monolayers of each species in any given layer. It should be noted that, due to the large size of a comet (as compared to an interstellar dust grain), all chemical species in the modeled ice layers may be assumed to have an absolute abundance significantly greater than one particle per layer, meaning that stochastic chemical effects sometimes observed in simulations of interstellar dust grains, which are related to having an average population less than unity, may be ignored (see e.g. Garrod 2008).

The ice surface/mantle chemical network used in the model is based on that presented by Garrod (2013), which specifically included radical-radical reactions relating to the formation of glycine. To aid in the rapid completion of the models, some species were removed from the interstellar network, leaving only those comprised of 5 or fewer carbon atoms. The network nevertheless retains all the main molecules typically observed in chemically-complex star-forming regions. Again, the gas-phase portion of the network is technically present, but its reaction/photo-dissociation rates are set to zero.

The network includes 201 solid-phase/surface species, with 1039 reactions and photo-dissociation processes. The chemical network is the same within each layer, but the calculated rates are specific to the conditions in each. Therefore, to distinguish the chemistry within each individual layer,  all 25 layers have their own set of 201 independent chemical abundances, totalling 5025. Including the 474 largely redundant gas-phase species, the model's differential equation solver is set up to solve for a total of 5499 chemical species. This is approximately five times as many species as were solved in total in the interstellar model of Garrod (2013).

\subsection{Diffusion, desorption and chemistry}

As mentioned above, molecules are allowed to diffuse thermally between the different layers, through a process of bulk diffusion. Garrod (2013) formulated rates for this process in the case of a single bulk-ice phase topped by one monolayer of surface ice. Diffusion from the bulk ice into the surface layer for species $i$ was dependent on a swapping rate coefficient, $k_{\mathrm{swap}}(i)$, for bulk-ice diffusion of a single molecule from one site to another adjacent site, determined by a diffusion barrier $E_{\mathrm{swap}}(i)$, such that

\begin{equation}
k_{\mathrm{swap}}(i) = \nu_{0}(i) \exp [-E_{\mathrm{swap}}(i)/T]
\end{equation}

\noindent where $\nu_{0}(i)$ is the characteristic vibrational frequency of a harmonic oscillator, and $T$ is the ice temperature. Surface and bulk-ice diffusion barriers, as well as desorption energies, are given by Garrod (2013) for various species used in this model. Note that this treatment assumes that the ice is both amorphous and chemically diverse, rather than having a pure water and/or crystalline structure, which could inhibit bulk diffusion.

In the present model, the rate of diffusion of species $i$ into the surface from the layer beneath (of whatever thickness) is given by

\begin{equation}
R_{\mathrm{swap,m}}(i)=N_{m}(i) \ \frac{N_{S}}{N_{M}} \ k_{\mathrm{swap}}(i) / 6
\end{equation}

\noindent where $N_{m}(i)$ is the mantle population of species $i$, and $N_{M}$ and $N_{S}$ are the total mantle and surface populations over all species (thus, in the comet model, $N_{S}/N_{M}=1/3$). The factor of 6 corresponds to the six competing directions available for diffusion assuming a simple cubic structure.

The (swapping) diffusion of a mantle molecule into the surface is in each case balanced by the diffusion of a surface molecule into the layer beneath; the choice of molecule to undergo this exchange is determined proportionately to the fractional abundances of each species in the surface layer. 

Thus the overall net rate of diffusion of molecules of all kinds into and out of each layer is zero, while the net rate for individual chemical species may not be. Following Garrod (2013), the complementary rates for the above-mentioned exchange of (in this case) material from the surface layer into the layer beneath are given by

\begin{equation}
R_{\mathrm{swap,s}}(i)= \frac{N_{s}(i)}{N_{S}} \cdot \sum_{all j} R_{\mathrm{swap,m}}(j)
\end{equation}\

A similar treatment is extended to all layers in the comet model; however, diffusion initiated in each sub-surface layer can lead to exchange with either the layer above or the layer below. Diffusion in each direction has its own version of Eq. (2) for each species in each layer, as well as complementary rates for the exchange partners. In general, the rates of inter-layer diffusion for all layers and chemical species will be mostly negligible at very low temperatures ($<$20~K). No diffusion into or out of the reservoir (via the lowest ice layer) is allowed.

Chemical reactions are allowed to occur within each ice layer, with the reactants meeting nominally through the same site-to-site diffusion process as outlined above and described by Eq. (1). Rates of reaction between arbitrary species $A$ and $B$ are therefore determined by the expression:

\begin{equation}
R_{\mathrm{AB}} = \kappa_{\mathrm{AB}} N_{m}(A) \ N_{m}(B) \ [k_{\mathrm{swap}}(A) + k_{\mathrm{swap}}(B)] / N_{\mathrm{M-D}}
\end{equation}

\noindent where $\kappa_{\mathrm{AB}}$ represents the reaction efficiency in cases where an activation energy barrier exists, corresponding to the competition between reaction of the two species and their diffusion away from each other.
Note that the denominator of Eq. (4) contains not $N_M$, but $N_{M-D}= N_{M}(1 - X_{\mathrm{dust}})$, the total ice abundance in the mantle layer in question with dust excluded. This substitution is made to account for the fact that diffusive species in the ice are unable to penetrate the space occupied by the dust. 

A similar expression to Eq. (4) pertains to reactions in the surface layer, although here a lower diffusion barrier is used, and all surface-diffusion sites are considered available, thus the denominator contains simply $N_S$. Following Garrod (2013), the surface and bulk diffusion barriers are set to fractions 0.35 and 0.7, respectively, of the desorption energy, $E_{\mathrm{des}}(i)$. Many of the radical-radical and neutral H-abstraction reactions included in the chemical network are listed by Garrod et al. (2008) and Garrod (2013).

The above treatment for bulk-ice diffusion and the related treatment for diffusion-mediated reactions do not explicitly consider the presence of porosity in the ice (nor in the grains suspended within them). No infra-red signatures of the OH dangling bond feature have so far been detected in the interstellar medium (Keane et al. 2001), suggesting that interstellar ices are not very porous; however, cometary ices are thought to be so, based on density determinations (e.g. Jorda et al. 2016; see also Secs. 4 and 5.2). This porosity may result from irradiation and the subsequent loss of volatiles (e.g. Johnson 1991). But it is unclear whether comets under cold storage conditions should initially have significant ice porosity -- nor, crucially, whether such pores would show significant connectivity throughout the overall ice structure on scales greater than, say, the characteristic size of a dust grain particle. The presence of a well-connected network of pores or larger cavities spanning different depth scales into the ice could allow a more rapid transport of material, via thermal diffusive hopping of individual molecules on those surfaces, which would be significantly faster than bulk processes. Also, even the presence of enclosed, unconnected pores would provide a surface upon which chemical reactions could occur more rapidly than within a bulk ice. In this initial cometary chemistry model, the mechanisms used in the interstellar treatments are retained. An explicit treatment of porosity and diffusion will be a focus for future work.

As well as thermal desorption from the surface, which is governed by an expression similar to Eq. (1), non-thermal desorption processes are also included for the surface layer (no desorption is allowed from deeper layers). Reactive desorption is included in the model according to the treatment of Garrod, Wakelam \& Herbst (2007). Single-product surface reactions are thus assumed to result in desorption of the product in a fraction of cases no greater than 1\% (i.e. $a_{\mathrm{RRK}}$=0.01).

UV photo-desorption is included for each surface species, $i$, with rate coefficients set by the expression:

\begin{equation}
k_{\mathrm{PD}}(i) = Y(i) \, \sigma \, F_{\mathrm{UV}} / 2
\end{equation}

\noindent where $\sigma$ is a generic geometric cross section equal to $10^{-15}$ cm$^2$, $F_{\mathrm{UV}}$ is the flux of UV photons, set to $10^8$ cm$^{-2}$ s$^{-1}$ to represent the standard interstellar radiation field (with the factor of 2 corresponding to exposure to photons from one side only), and $Y(i)$ is the photo-desorption yield per photon, as determined by {\"O}berg et al. (2009b). Where measured yields are unavailable (as in most cases), a value of $10^{-3}$ is assumed, in line with the order of magnitude of the measured values. In models where the UV field strength is altered (for the purposes of photo-dissociation), the value of $F_{\mathrm{UV}}$ above is also adjusted accordingly.

\subsection{Initial Abundances}

To approximate the composition of cometary ices at the beginning of the life of a comet, representative abundances are assumed based on the dominant, simple components of interstellar ices. Water is assumed to be the most abundant ice species, followed by CO and CO$_2$ (20\% with respect to water), methanol (5\%), and methane, formaldehyde and ammonia (1\% each). No other chemical species are initially assumed, although an inert dust component of unspecified composition is included in the model. Each of these abundances is assumed to be uniform throughout all layers of the comet at the beginning of the model. Since all chemical species in the initial ices are highly stable, the gradual evolution of chemistry within the ices in the model is driven by the dissociation of stable molecules into radicals or atoms that may react to form other species. 

Sulfur-bearing species are not included in the present models, as no sulfur-bearing pre-cursor is included in the initial abundances. Although H$_2$S has been detected in various comets with abundances in the range 0.1--1\% H$_2$O (see Mumma \& Charnley 2011), sulfur chemistry in dense and star-forming regions is still a matter of debate, due to the lack of a clear carrier for the majority of the sulfur budget. For reasons of simplicity, sulfur is therefore omitted from the model, which is nevertheless in keeping with the lack of detection of interstellar solid-phase H$_2$S. Future models will consider the inclusion of sulfur chemistry, by the adoption of more varied initial ice abundances including a range of sulfur-bearing species.

\subsection{Dust}

Unlike interstellar ices, which are simply formed on the surfaces of dust grains, the cometary ices are expected to contain dust-grain particles themselves, interspersed with the ice. The cometary ice model must therefore include a component corresponding to dust in each layer. As well as taking up physical space, the dust will also act to extinguish the external UV field that could otherwise induce photo-dissociation of molecules in the ice.

In the model presented here, dust is treated as a distinct, inert chemical species that is present in every ice layer. It may not diffuse within the ice and it is not allowed to undergo thermal desorption from the ice surface, but it may nevertheless be transferred between layers in the same way as other species, to compensate for gains and losses primarily caused by desorption of volatiles from the surface. In this way, it is possible for the dust to become more concentrated in the surface layers, as volatiles desorb while the dust is left behind.

The dust component is assigned a uniform initial abundance throughout the comet, which can be expressed as a fraction of each layer of the ice/dust mixture (i.e. a volume fraction). Because the majority of the cometary ice is composed of water, and thus depends on the availability of interstellar oxygen, the determination of the dust fraction in this model is based on an estimate of the mass ratio of dust-grain material to interstellar oxygen content. Assuming a canonical Milky Way gas-to-dust ratio (for atomic hydrogen) of 100:1, the volume ratio of dust to water ice is given by:

\begin{equation}
V_d / V_w = \frac{0.01}{X(\mathrm{O}) \,  m_w \,  F_w} \frac{\rho_w}{\rho_d}
\end{equation}

\noindent where $X$(O) is the elemental abundance of oxygen with respect to hydrogen nuclei, $m_w$ is the mass ratio of water to atomic hydrogen (=18), $F_w$ is the fraction of oxygen atoms locked up in water ice, and $\rho_d$ and $\rho_w$ are respectively the mass densities of dust-grain material and of water ice, the latter of which was assigned in Section 2 a value of 0.9 g/cm$^3$. Based on the initial ice abundances provided in Section 2.2, if the total oxygen budget is present in the form of H$_2$O, CO, CO$_2$, H$_2$CO or CH$_3$OH ice, then the amount locked up specifically in water is $F_w$=100/166. Following past interstellar chemistry models (e.g. Hasegawa et al. 1992), a dust density of $\rho_d$=3 g/cm$^3$ is assumed. The remaining quantity required to evaluate Eq. (6) is the overall interstellar oxygen abundance. Two values have commonly been assumed in past models: $X$(O)=$3.2 \times 10^{-4}$ (Garrod et al. 2008), and the so-called low-metal abundance value of $X$(O)=$1.76 \times 10^{-4}$ used by Graedel et al. (1982). These result in respective volume ratios of $V_{d}/V_{w}$ = 0.865 and 1.57, which are equivalent to ratios of dust to total ice content of 0.584 and 1.06 (based on the ice composition in Section 2.2), or ratios of dust to total-ice oxygen of 0.521 and 0.947. For the models presented here, an intermediate value of $V_{d}/V_{w}$ = 1.11 is assumed, which corresponds to a ratio of dust to total ice of 111/148 = 0.75. Based on the assumed initial abundances of all ice species, dust therefore comprises $\sim$42.9~\% by volume of each layer of cometary material (dust+ice) at the beginning of the model. As with other species, the dust fraction in the reservoir beneath the lowest ice layer is held constant over time. 

The precise initial value used here is not definitive, as it depends on the several assumptions above; the dust content could plausibly diverge by several tens of percent from the value assumed here. However, the absence or otherwise of porosity in the ices will not affect the above calculations.

Adopting a canonical interstellar dust-grain radius of 0.1 $\mu$m, the grains are around 300 times larger than the individual molecules that make up the remainder of each layer. In the model presented here, the transfer of material from layer to layer, as required to counteract any net losses or gains in each layer, is treated in the same way for grain material as for molecules. However, this could in practice lead to a layer becoming entirely composed of grain material, filling even every molecule-sized space. The rigid structure of the grains evidently precludes this, so a maximum fractional population for grains is imposed on each layer. This maximum is assumed to be equal to the well-known maximum space-filling factor for identically-sized spheres, i.e. $\pi / 3 \sqrt{2}$ $\simeq$ 74\% (e.g. Hales 2005). The dust component of each layer, which begins at a volume fractional abundance of $\sim$42.9\%, may therefore concentrate by a factor of up to $\sim$1.73 in this model before reaching the geometrical maximum. While the $\sim$74\% volume limit is invariant with the choice of grain size, the inclusion of a range of dust-grain sizes could allow a higher threshold (See also Sec. 2.4 regarding the size distribution). The solution of the close-packing problem for multiple sphere sizes is not trivial; however, computational determinations of upper bounds for binary sphere packing indicate maxima around 86\% (De Laat et al. 2014).

\subsection{UV field and photo-dissociation}

The model assumes that the simulated comet is exposed to a UV radiation field capable of inducing photo-dissociation within the ice, and which is similar in strength to the standard interstellar radiation field. Interstellar {\em gas-phase} photo-dissociation rates are used as a basis for the photo-dissociation of molecules in the ice layers (similarly to Garrod 2013, and earlier models). 

In the interstellar model, photo-dissociation rates are defined by two parameters, $a$ and $c$; these correspond, respectively, to the photo-dissociation rate coefficient (s$^{-1}$) produced by the unattenuated interstellar radiation field (of strength 1 Draine unit), and to a dimensionless modifier to the visual extinction ($A_{\mathrm{V}}$) that accounts for the wavelength dependence of absorption by dust (with typical values in the range 1--3), giving an expression:

\begin{equation}
k_{\mathrm{PD}}(j) = a(j) \, \exp[-c(j) \, A_{\mathrm{V}} ] \\
\end{equation}

\noindent where $k_{\mathrm{PD}}(j)$ is the rate coefficient of photo-dissociation process $j$ for some specific molecule.

In the present models, the surface layer of comet ice is exposed to the maximum field, producing photo-dissociation at the maximum rates. Photo-dissociation in the deeper layers is modified according to the amount of dust, as well as the amount of each individual molecule, in the layers above, both of which act to reduce the flux of dissociating photons. The attenuation of the UV field is determined independently for each photo-dissociation process. It is implicitly assumed that all photo-dissociation occurs through line absorption at specific freqencies; however, in general, the presence of dust mixed into the ice has a much larger effect on the attenuation of all photo-dissociation rates than do the individual absorption processes that are modeled implicitly here.

The attenuation of the radiation field associated with a specific photo-dissociation rate is calculated repetitively through each layer of ice, beginning in the outer surface layer and working inward. For layers that are thicker than 1 monolayer (i.e. all layers beneath the surface layer), the attenuation by each individual monolayer within that layer is considered. A mean photo-dissociation rate -- averaged over all depths within that layer -- is then calculated, which is applied (in the chemical calculations) equally to all material within that full layer. 

To obtain the total fractional attenuation due to absorption in a single monolayer ($f_{\mathrm{ML}}$) for a specific photo-dissociation process $j$, three fixed quantities are employed: the fractional attenuation per monolayer of dust for UV/visual-band photons, $f_{\mathrm{dust}}$; the wavelength-dependent modifier of the dust absorption for the specific photo-dissociation process, $c(j)$; and the fractional attenuation per monolayer due to absorption by the species being dissociated, $f_{\mathrm{PD}}(j)$. These are modified by the fractional abundance of dust (by volume) in that particular monolayer, $X_{\mathrm{dust}}$, and by the fractional abundance, $X(i)$, of the molecule being photodissociated, $i$, each abundance taking a value from 0 -- 1, and whose precise values are time-dependent variables calculated in the code. Thus:

\begin{equation}
f_{\mathrm{ML}}(j) = f_{\mathrm{PD}}(j) \, X(i) + f_{\mathrm{dust}} \, X_{\mathrm{dust}} \, c(j)
\end{equation}

\noindent where $f_{\mathrm{ML}}(j)$ is the total fractional loss in intensity due to transmission through an individual monolayer, and $(1 - f_{\mathrm{ML}}(j))$ is thus the fraction of the intensity transmitted by that monolayer. For an ice layer of total thickness $N$ in monolayers, the fraction of intensity transmitted after passage through the full layer is therefore $(1 - f_{\mathrm{ML}}(j))^N$. The mean rate of photo-dissociation applied to material within a particular layer is a factor $\Sigma_{n=1}^{N}(1 - f_{\mathrm{ML}}(j))^{n}/N$ multiplied by the total rate after attenuation by any layers above the layer in question. The value of $f_{\mathrm{PD}}(j)$ is assumed for all species to be 0.007, following the determination by Andersson \& van Dishoeck (2008) for the attenutation of the photo-dissociation rate of water, per water monolayer.

For a single, canonically-sized grain suspended in ice, a geometric cross-section of $\sim$$3.14 \times 10^{-10}$~cm$^2$ is presented. To calculate (as required by this model) the average attenuation of the UV/visual-band photon flux per monolayer of dust, $f_{\mathrm{dust}}$, an evaluation of the number of monolayers of grain material provided by the entire grain in the given cross-section is required. This is easily calculated as the height of the canonical grain in monolayers, $\simeq$622. However, a solid, spherical grain does not occupy all of the space in the cylinder traced out by its cross-section; the sphere of the dust grain occupies exactly 2/3 of this space, meaning that the average  fractional abundance of dust over each of the 622 monolayers of its height is $X_{\mathrm{dust}}$=2/3.

In this model, the attenuation per monolayer is assigned a value $f_{\mathrm{dust}} = 0.0055$, which results in the transmitted UV/visual-band intensity being reduced by a factor $(1-f_{\mathrm{dust}} \times 2/3)^{622} \simeq 0.1$ over a full grain-height (ignoring any contribution from $f_{\mathrm{PD}}(j)$), corresponding to an absorption efficiency of $\sim$0.9 over the whole grain. This latter value is approximately equal to the mean of the typical absorption coefficients, $Q_{\mathrm{abs}}$, for carbonaceous and silicaceous interstellar grains of radius 0.1~$\mu$m at a wavelength of 5500~\AA, which take values typically around 0.1 and 1.7, respectively (e.g. Kr{\"u}gel 2008). Note that the simple one-dimensional treatment presented here considers only absorption and not scattering, while the close proximity between dust grains in the comet ice will also influence the optical transmission. The assumed value of $f_{\mathrm{dust}} = 0.0055$, which is applied to the photo-dissociation rates via Eq. (8), is thus quite imprecise.

The adoption of a single grain size in the above calculations is itself a substantial simplification; choosing an appropriate representative size is therefore important. In the interests of consistency with interstellar treatments, the adoption of a grain radius of 0.1~$\mu$m, as above, makes sense for this initial chemical modeling study.

The dust in observed comets in fact displays a distribution of sizes, from submicron to macroscopic scales; as summarized by Agarwal et al. (2007), much of it is composed of aggregates of submicron-sized subunits. The subunits are understood to have mass densities of a few g/cm$^3$ (consistent with the value adopted in Sec. 2.3), while the agregates would be porous, with densities in the range 0.1--1 g/cm$^3$. Agarwal et al. analyzed the dust environment of comet 67P, based on a selection of remote observational data. Their models suggest that the total dust mass is concentrated in grains of size 10~$\mu$m or larger, with one model showing a distinct peak around this value. Recent work by Mannel et al. (2019) presents imaging and analysis of dust particles obtained directly from 67P, studied using Rosetta's MIDAS Atomic Force Microscope. This work reinforces the idea that the dust is composed of aggregates of small sub-units. Those authors find a mean size of $\sim$0.1 $\mu$m for the subunits. Experimental work by Price et al. (2010), which aimed to reproduce the impacts on detectors in the NASA Stardust spacecraft of small dust particles from comet 81P/Wild 2, also derived comparable sizes.

The fact that the larger grains observed in real comets appear to be low-density, loose and/or porous aggregates of smaller subunits of 0.1~$\mu$m therefore seems consistent with the simple UV-penetration model based around absorption by grains exclusively of that size. The adoption of a more comprehensive grain-size distribution in the chemical models, with implications for the UV treatment and the packing density of grains, is left for future studies.

The above treatment assumes all UV impinging on the comet is interstellar in origin, rather than solar. Although the relative contributions of interstellar and solar UV at a particular distance from the sun will vary with wavelength, a comparison at a representative wavelength is instructive. Curdt et al. (2001) presented solar spectra in the 670 -- 1609 {\AA} range (i.e. $\sim$18.5 -- 7.7 eV); the majority of the wavelength-integrated dissociation cross-sections for molecules of interest tend to fall within this range (see e.g. Heays et al. 2017). Interstellar UV is dominated by emission from stars hotter than the sun; thus, in general, solar UV will be most competitive with interstellar UV at the longer-wavelength end of the range. The Curdt et al. data provide a (quiet-sun) solar radiance of $\sim$0.6 W sr$^{-1}$ m$^{-2}$ \AA$^{-1}$ at 1600 \AA. At an assumed inner radius for the Oort cloud of 1000 au, this corresponds to a field strength of $\sim$3.3$\times$10$^3$ photons s$^{-1}$ cm$^{-2}$ \AA$^{-1}$. The interstellar UV field of Draine (1978) gives a field strength of $\sim$1.9$\times$10$^5$ photons s$^{-1}$ cm$^{-2}$ \AA$^{-1}$ at this wavelength, roughly 60 times greater than the solar value. Thus, to a first approximation, solar UV may be ignored, especially if the comet is significantly more distant than 1000 au. 

Interstellar UV photon flux drops sharply at energies above the ionization potential of atomic hydrogen (13.6 eV). It is therefore possible that there is some small contribution of solar UV to the photo-dissociation rates that is not otherwise captured in the interstellar rates. The UV output of the early sun was also likely higher than at present (e.g. Zahnle \& Walker 1982), perhaps sufficient to become competitive with interstellar values. A more detailed treatment of the combined interstellar and solar photo-dissociation rates are left for future work.

The penetration treatment used for cosmic ray-induced dissociation is presented in Section 4, along with the results of the models using that treatment.

\subsection{Photo-dissociation-induced reactions}

To complete the treatment of reactions within and upon the cometary ice, a further mechanism is included in this model beyond those typically considered in interstellar ice chemistry simulations. The calculated reaction rates described in Section 2.1 are purely diffusive, thus reaction depends on the mobility of the reactants and does not consider their origins, other than the implicit assumption that the reactants are not immediately in contact at the start of an individual diffusive reaction process. However, the production of reactive species as the result of photo-dissociation in the ice may lead to the spontaneous appearance of a reactive atom or radical in close proximity to another that is already present in the ice, leading immediately to reaction without any mediating (thermal) diffusion process. This possibility is especially important in the case of very low ice temperatures (i.e. 5--20~K), in which the thermal diffusion of atoms and/or radicals may be prohibitively slow. 

To incorporate into the model the immediate reaction of photo-dissociation fragments with other reactants present in the ice, the total rate of production for each chemical species, as a result of all photo-dissociation processes that can produce it, is calculated (which is done separately within each ice layer). This total production rate, $R_{\mathrm{prod}}(i)$, is then multiplied by the fraction of the ice composed of each other species with which the photo-fragment may react, providing a rate of immediate reaction between the two species. Thus, for some reaction $j$, occurring between chemical species A and B, the total rate (s$^{-1}$) at which it occurs as a direct and immediate result of the production of either reactant by photo-dissociation is given by:

\begin{equation}
R_{\mathrm{reac}}(j) = \kappa_{\mathrm{AB}} \ \left( R_{\mathrm{prod}}(A) \ \frac{ N_{m}(B) }{ N_{\mathrm{M-D}} } + R_{\mathrm{prod}}(B) \ \frac{ N_{m}(A) }{ N_{\mathrm{M-D}} }  \right)  \\
\end{equation}

\noindent where again, $R_{\mathrm{prod}}$ represents exclusively the production rate caused by photo-dissociation of pre-cursor species, with other quantities defined as in Sec. 2.1. A similar expression is used for the same process occurring in the upper surface layer.

As an example, a hydrogen atom may be produced by the photo-dissociation of a selection of molecules in the ice, including H$_2$O, CH$_3$OH and NH$_3$. The sum of these production rates is multiplied by the fractional abundance of, for example, CH$_3$, to give the rate for immediate production of CH$_4$. A further term corresponding to photo-produced CH$_3$ reacting with existing H would complete the prescription given in Eq. (9). Because the fractional abundance of CH$_3$ and all other species ranges between 0 and 1, the resultant reaction rate will in practice always be less than the overall production rate of photofragments. It should be noted that in this treatment, as with the calculation of bulk-ice reaction rates, the determination of fractional abundances within the layers beneath the surface ignores the presence of dust, because the dust and ice are distinct phases that cannot be fully mixed.

This new treatment is considered for all existing reactions in the network for which one or both reactants may be produced via photo-dissociation of other species. For reactions involving activation energy barriers, the usual barrier treatment is used, i.e. the fastest of thermal or quantum tunneling through the activation barrier, in competition with thermal diffusion of the reactants. Thus, the photo-products involved are implicitly assumed not to have excess thermal energy for overcoming activation barriers, nor to be in an electronically or vibrationally excited state that might otherwise alter the barrier.

\subsection{Physical Conditions}

This initial version of the cometary ice/dust model is intended to represent (stable) conditions in the Oort cloud. The main physical parameters are the choice of ice temperature and UV field strength. Seven temperature values are used in the models presented here: 5, 10, 20, 30, 40, 50, 60~K. While the two lowest values are more representative of the typical values in the Oort cloud, the higher temperatures are chosen to test how sensitive the solid-phase chemistry is to elevated temperatures, which might be achieved periodically through stochastic astrophysical events such as supernovae or the close passage of hot stars. It is also assumed here that the same temperature permeates the entire comet ice at every layer. Propagation of temperature structure within comets has been modeled in detail by other authors (see e.g. Prialnik et al. 2008, Guilbert-Lepoutre et al. 2016); the effects of temperature structure within the chemical models are left for later work.

Three UV field strengths are tested in the models, based around the standard interstellar Draine field, G$_0$: low (G = 0.1 G$_0$), medium (G = G$_0$) and high (G = 10 G$_0$).

\begin{deluxetable}{lrrrrrrrrrrrrrr}
\rotate
\tabletypesize{\small}
\tablecaption{\label{tab-rates} Rates and quantities of material lost from comet surface}
\tablewidth{0pt}
\tablehead{ \colhead{Model}  &  \multicolumn{2}{c}{ Total surface }                                                & \colhead{H$_2$O} & \colhead{CO}      & \colhead{CO$_2$}  & \colhead{H$_2$CO}   & \colhead{CH$_4$}   & \colhead{H}      & \colhead{H$_2$}   & \colhead{O}      & \colhead{OH}   & \colhead{O$_2$}  &  \colhead{ Total }   & \colhead{Dust}  \\
                     \colhead{ }             & \multicolumn{2}{c}{ loss rate }                                                        & \colhead{(\%)}         & \colhead{(\%)}     & \colhead{(\%)}          & \colhead{(\%)}              & \colhead{(\%)}           & \colhead{(\%)}  & \colhead{(\%)}       & \colhead{(\%)}  & \colhead{(\%)}  & \colhead{(\%)}       & \colhead{ lost }      & \colhead{depth}  \\
                     \colhead{ }             & \colhead{ (ML yr$^{-1}$) }  & \colhead{ (m$^{-2}$ s$^{-1}$) }  & \colhead{ }               &  \colhead{ }           & \colhead{ }                & \colhead{ }                    &   \colhead{ }                & \colhead{ }        &   \colhead{ }            & \colhead{ }         & \colhead{ }        & \colhead{ }             &  \colhead{ (m) }      & \colhead{ (m) }    }

\startdata
5~K, low UV          &    6.3(-5)     &    1.9(7)     &    55     &    19     &    2.9     &       0.16       &     1.4       &    --     &    --     &    3.8     &  11  &    0.78   &    1.0(-4)    & 1(-5) \\
5~K, mid UV          &    6.3(-4)     &    1.9(8)    &    55     &    19     &    2.8     &        0.16       &      1.4         &    --     &    --     &    3.8     &  11  &    0.79   &    1.0(-3)    & 1(-4) \\
5~K, high UV         &    6.3(-3)     &    1.9(9)     &    55     &    19     &    2.8     &       0.16       &      1.4          &    --     &    --     &    3.8    &  11  &    0.78 &    1.0(-2)     & 1(-3)    \smallskip  \\
     
10~K, low UV        &    6.9(-5)     &    2.1(7)     &    69     &    18     &    9.4     &       0.19      &     0.44            &    --     &    --     &    --     &  1.9(-2)  &    -- &    1.1(-4)     &  1(-5) \\
10~K, mid UV       &    6.9(-4)     &    2.1(8)     &    69     &    19     &    9.2     &        0.18      &     0.41           &    --     &    --     &    --     &  2.9(-2)  &    -- &    1.1(-3)     & 1(-4)  \\
10~K, high UV      &    6.8(-3)     &    2.1(9)     &    69     &    19     &    8.6     &        0.18      &    0.34            &    --     &    --     &    --     &  2.4(-1)  &    -- &    1.1(-2)     &  1(-3)    \smallskip \\
     
20~K, low UV        &    1.1(-2)     &    3.4(9)      &    --           &    0.76   &    -               &        --        &      --        &    63     &    36     &    --     &  --  &    --      &    1.9(-2)     & 1(-3)  \\
20~K, mid UV        &    2.6(-2)     &    7.9(9)      &    --           &    1.7    &    0.38         &        --       &     --          &    --       &    95     &    --     &  --  &    1.9   &    4.4(-2)     & 1(-2)  \\
20~K, high UV       &    6.1(-2)     &    1.9(10)    &    0.20     &    5.5    &    1.7           &        --       &      --         &    --       &    88      &    --     &  --  &    3.3   &    1.0(-1)     & 1(-2)    \smallskip \\
     
30~K, low UV        &    6.4(-3)     &    2.0(9)      &    0.59      &    --         &    0.25     &        --        &       --       &    50     &    20     &    10      & --   &    18 &    1.3(1)          & 1(0)  \\
30~K, mid UV        &    5.4(-2)     &    1.7(10)    &    0.53     &    --         &    0.41     &       --       &       --         &    50     &    23     &    3.3     &  --  &    22  &    1.3(1)        &  1(0) \\
30~K, high UV       &    5.0(-1)     &    1.5(11)    &    0.49     &    15.4   &    0.49      &        --         &      --       &    37     &    25     &    0.37   &  --  &    21  &    1.4(1)     &  1(0)   \smallskip \\
     
40~K, low UV        &    2.6(-2)     &    8.1(9)      &    --          &    --          &    --           &        1.8      &       --         &    63     &    26     &    8.4     &  --  &    0.16  &    1.3(1)     & 1(0)  \\
40~K, mid UV       &    2.3(-1)     &    7.0(10)    &    0.14    &    34     &    --           &        --         &     --        &    39     &    14     &    9.4         & --   &   3.0  &    1.3(1)            & 1(0)  \\
40~K, high UV      &    9.4(-1)     &    2.9(11)    &    0.48     &    16    &    0.12      &        --         &     0.82        &    47     &    13     &    12          &  --  &    11 &    1.4(1)              &  1(0)    \smallskip \\
     
50~K, low UV        &    2.0(-2)     &    6.2(9)      &    --     &    --               &    1.9        &          --        &      --      &    38     &    43     &    11     &  --  &    1.1  &    1.3(1)      & 1(0) \\
50~K, mid UV        &    1.2(-1)     &    3.6(10)    &    --     &    --              &    0.64      &         --        &      --       &    53     &    29     &    12     &  --  &    3.7  &    1.3(1)     & 1(0) \\
50~K, high UV       &    6.4(-1)     &    2.0(11)    &    --     &    12            &    0.66      &         --        &      1.2       &    49     &    16     &    12     &  --   &    8.1 &    1.4(1)      & 1(0)     \smallskip \\
     
60~K, low UV        &    2.2(0)     &    6.8(11)    &    --     &    14     &    84     &        0.69       &      0.35         &    0.40     &    0.47      &    --      &  --  &    --  &    1.7(1)     & 1(0)  \\
60~K, mid UV        &    2.5(0)     &    7.8(11)    &    --     &    12     &    74     &      5.2          &      0.61        &    2.4     &    4.8     &    0.67      &  --  &    --  &    1.7(1)     &  1(0) \\
60~K, high UV       &    5.2(0)     &    1.6(12)    &    --     &    13     &    36     &     14.3        &      0.60           &    12     &    18     &    3.3          &  --  &    0.12  &    2.2(1)     &  1(0)    \smallskip \\

  5~K, with CR     &    6.34(-4)     &    1.94(8)     &        55     &    19     &    2.8     &    0.15     &    1.4     &    --       &    --       &    3.8     &    11     &    0.79   & 1.0(-3) & 1(-4)  \\
10~K, with CR     &    6.93(-4)     &    2.12(8)     &        70     &    19     &    7.7     &    0.19     &    0.5     &    --       &    --        &    --       &    --      &    --        & 1.2(-2) & 1(-3)  \\
20~K, with CR     &    2.12(0)     &    6.50(11)    &        --       &    --      &    --        &    --         &    --       &    2.0     &    98      &    --       &    --      &    --        & 3.9(0) &  1(0)  \\

\enddata
\tablecomments{$A(B)=A \times 10^B$. Quoted values correspond to the end time of each model run, $t=5\times10^{9}$~yr. The second column under ``Total surface loss rate'' is the rate of loss in monolayers converted to particles per m$^2$ per second, applying the number of water molecules per m$^2$ implied in Sec. 2 for all species. Percentage loss rates for individual molecules are given for most species that achieve $>$0.1\% of the total rate; ``--'' indicates values less than this threshold. ``Total lost'' column indicates the amount of material lost from the surface during the entire model run, and is based on the assumed diameter of a water molecule (Sec. 2). ``Dust depth'' column indicates the approximate depth at which the dust reaches its maximum concentration.}
\end{deluxetable}

One model is run for each value of temperature and UV field strength, beginning from the same set of chemical initial conditions. Each model is run for a period of 5 Gyr, approximately the age of the Solar System. Further models are run for the 5, 10 and 20~K cases only, in which the action of cosmic-ray impacts is approximated, in tandem with the UV-related effects; see Section 4.

The CPU time required to run each model varies, dependent on the degree of interaction between different ice layers (which affects the stiffness of the differential equations); however, all are on the order of 1 week, using a single thread per model run.

\section{Results}

Figures \ref{fig-5K-early}--\ref{fig-40K-late} show the depth-dependent fractional abundance values for a selection of molecules of interest at an intermediate time ($t$=10$^6$ yr) and at the end of the model run ($t$=$5\times10^9$ yr), for each of temperatures 5, 10, 20, 30 \& 40~K. Similar plots for temperatures 50 and 60~K are included in the Appendix. Abundances are plotted as a fraction of the total layer composition. The depth of the ice is given in meters along the top axis and in the number of monolayers along the bottom. Data points for each layer are plotted at the mean depth of that layer. Although three separate UV field-strength models were run for each temperature, the behavior of the models may be described adequately by considering just the low- and high-UV cases. Thus, within each figure, the upper panels (a--c) show the results for the low-UV models, and the lower panels (d--f) correspond to high UV. The left-hand column of panels shows the abundances of the molecules that are initially present in the ice, as well as the dust-grain fraction, which is shown as an unlabeled, dotted black line. Each figure also shows the abundances of a selection of relatively simple product molecules (middle panels), and more complex molecules (right panels) of the sort typically detected in high-mass star-forming regions. Since there are many such molecules, the choice of which to plot is made based on several criteria, such as their detection in either comets and/or star-forming regions, their unique behavior in the present models, and their usefulness in demonstrating the overall behavior of the chemistry. For this reason, dimethyl ether (CH$_3$OCH$_3$), ethanol (C$_2$H$_5$OH), methyl formate (CH$_3$OCHO) and glycolaldehyde (CH$_2$(OH)CHO), for example, are plotted in the figures, each of which are now commonly detected in high-mass star-forming cores, although only the latter pair have yet been detected in a comet.

\subsection{5~K models}

\begin{figure*}
\begin{center}
\includegraphics[width=0.32\textwidth]{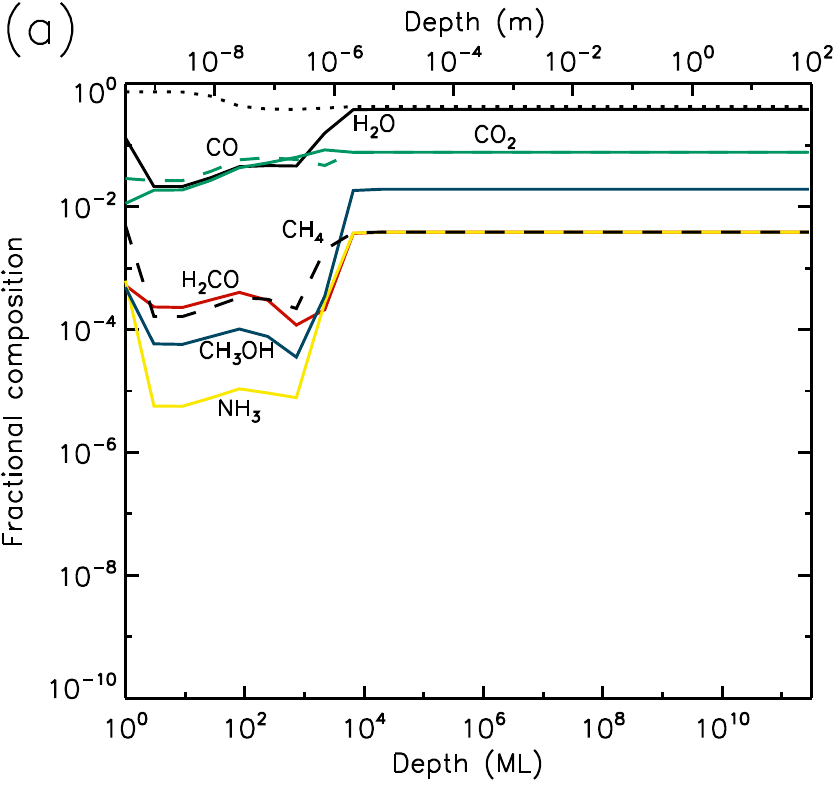}
\includegraphics[width=0.32\textwidth]{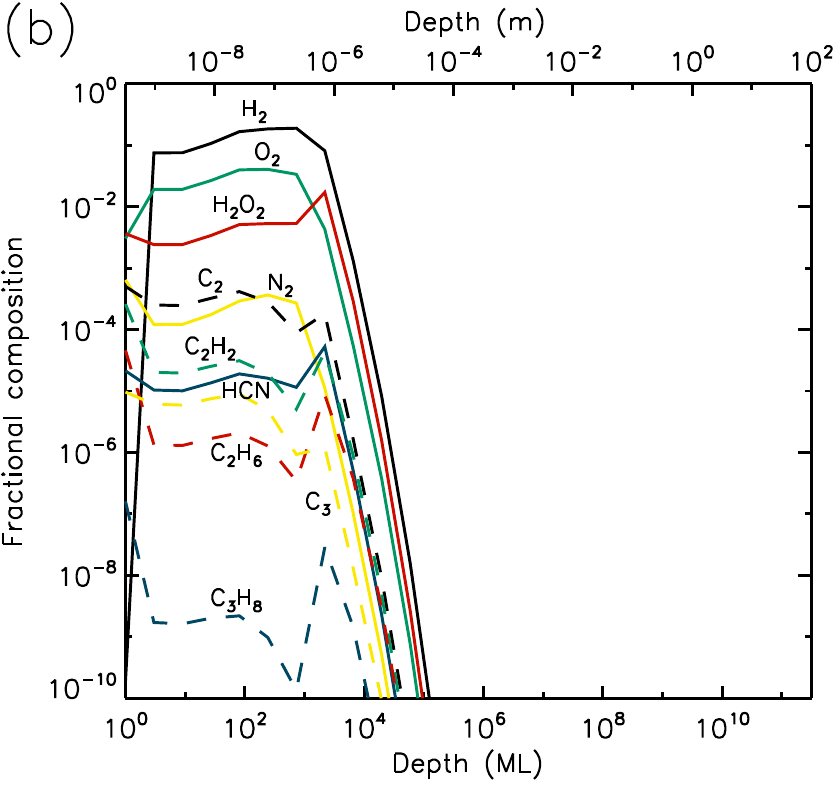}
\includegraphics[width=0.32\textwidth]{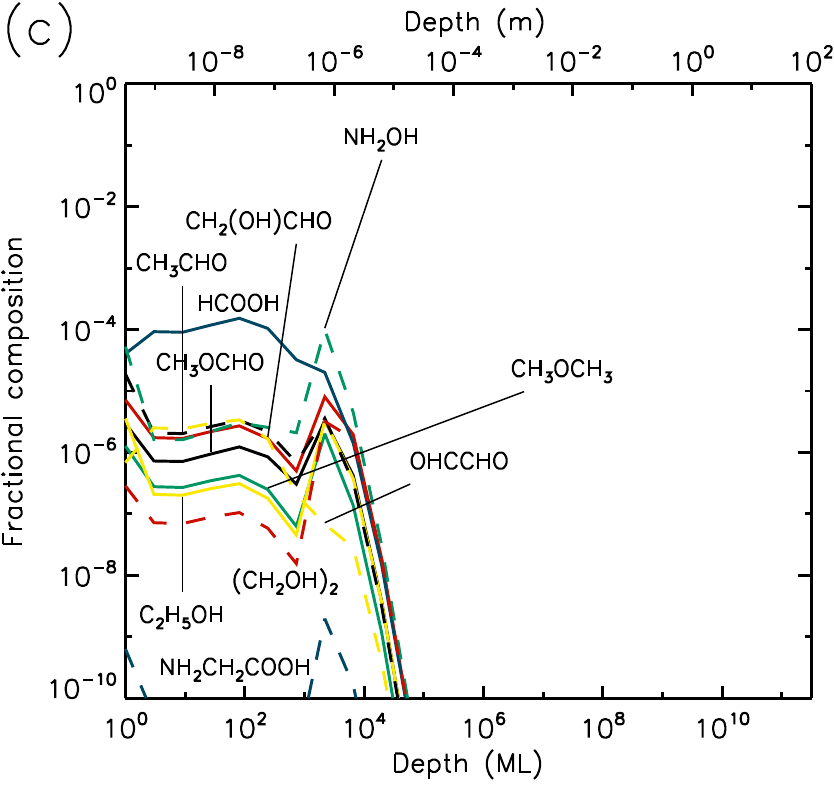}
\includegraphics[width=0.32\textwidth]{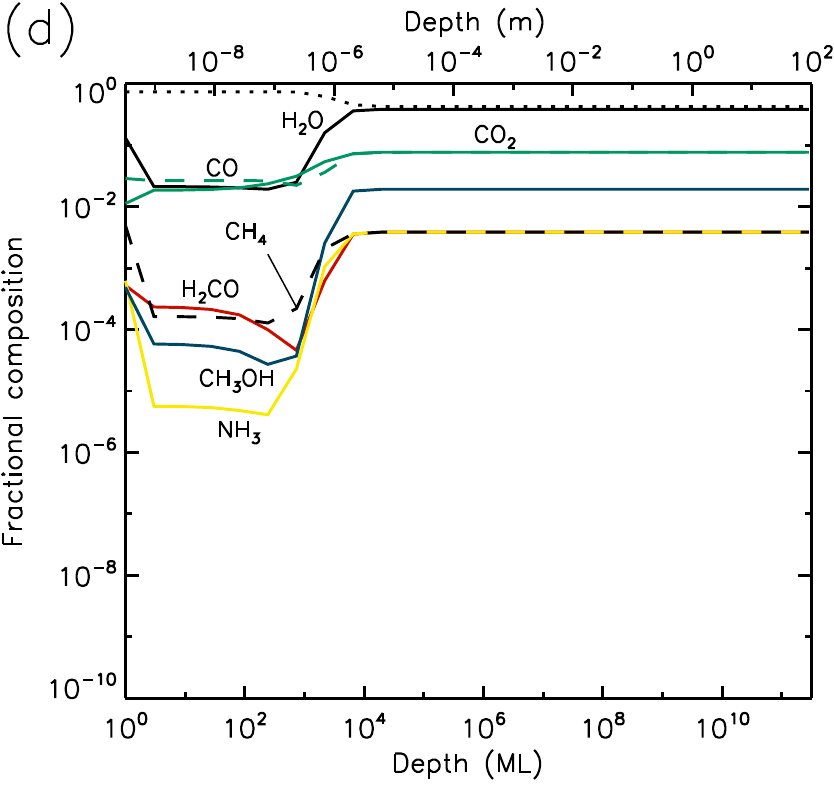}
\includegraphics[width=0.32\textwidth]{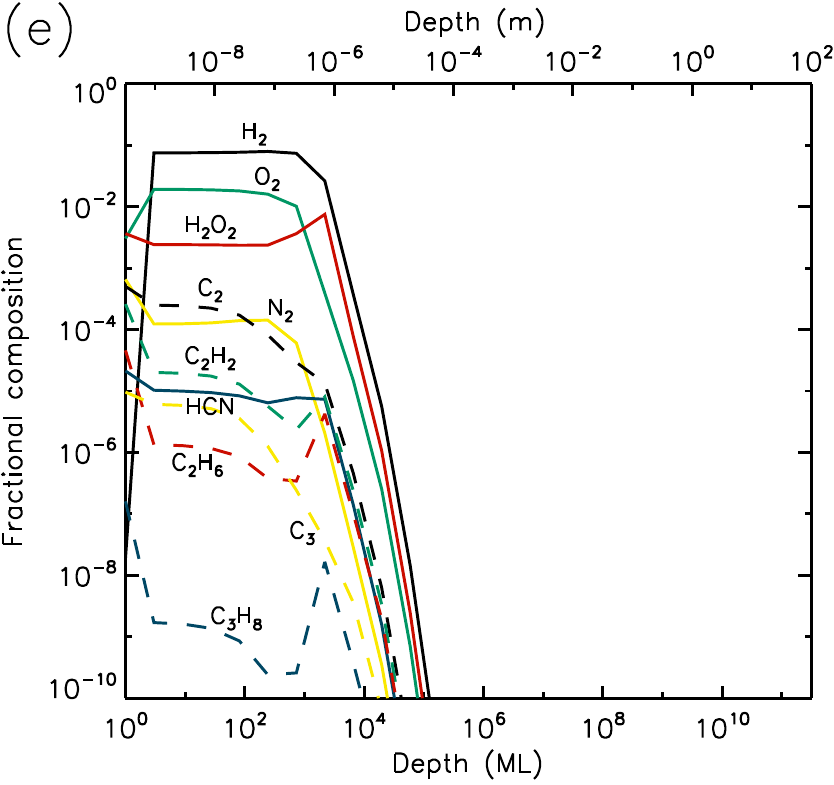}
\includegraphics[width=0.32\textwidth]{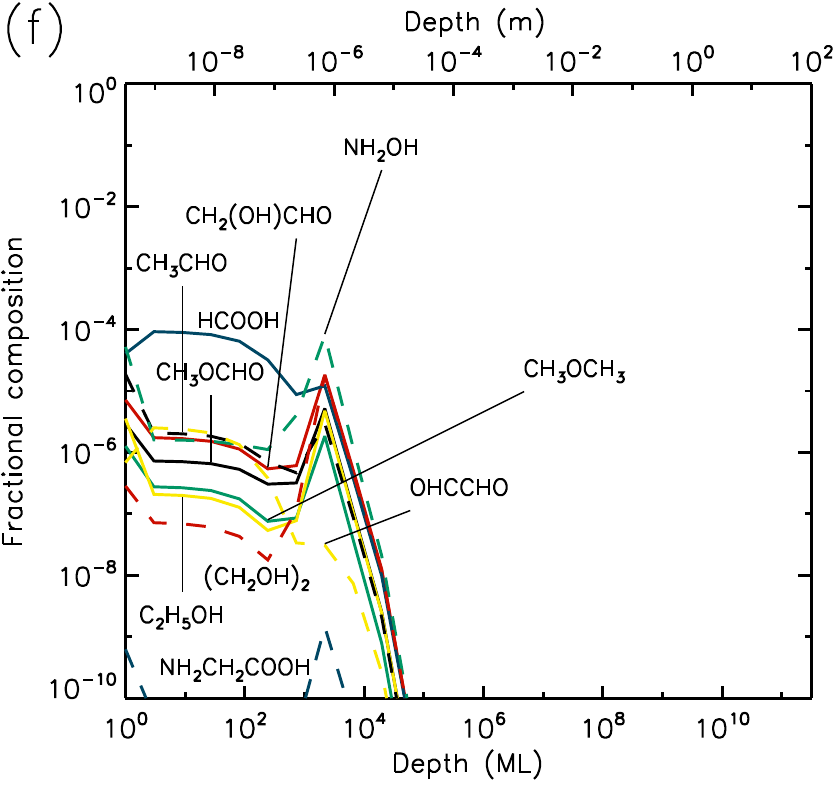}
\end{center}
\caption{\label{fig-5K-early} 5 K models at time $t=10^{6}$ yr. Abundances of the initial ice components are shown in the left panels, with dust shown as a dotted line. Results for two model setups are shown: low UV (upper panels) and high UV (lower panels).}
\end{figure*}

\begin{figure*}
\begin{center}
\includegraphics[width=0.32\textwidth]{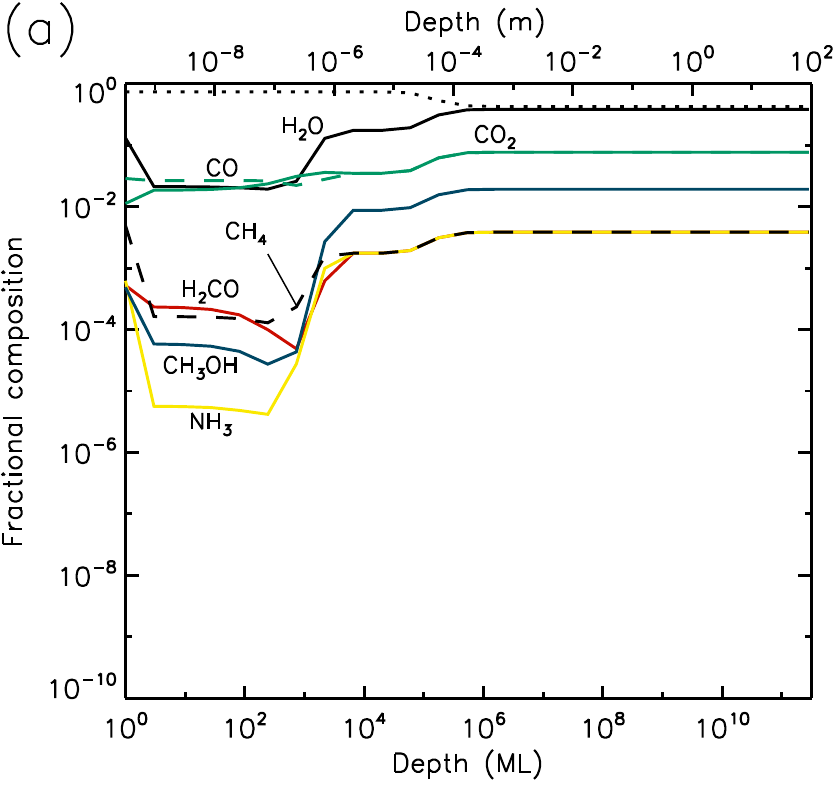}
\includegraphics[width=0.32\textwidth]{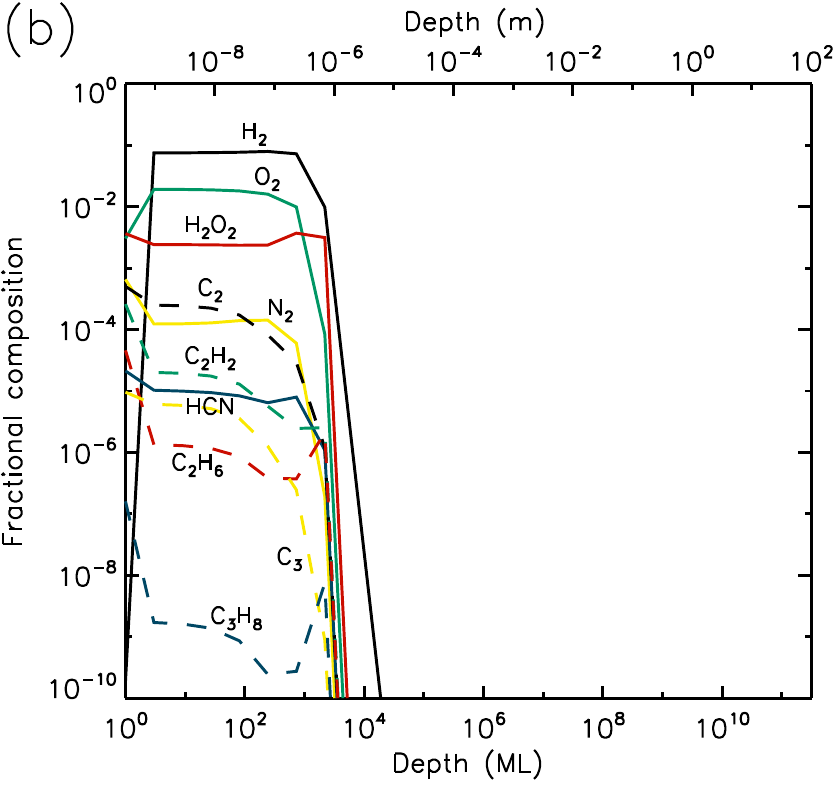}
\includegraphics[width=0.32\textwidth]{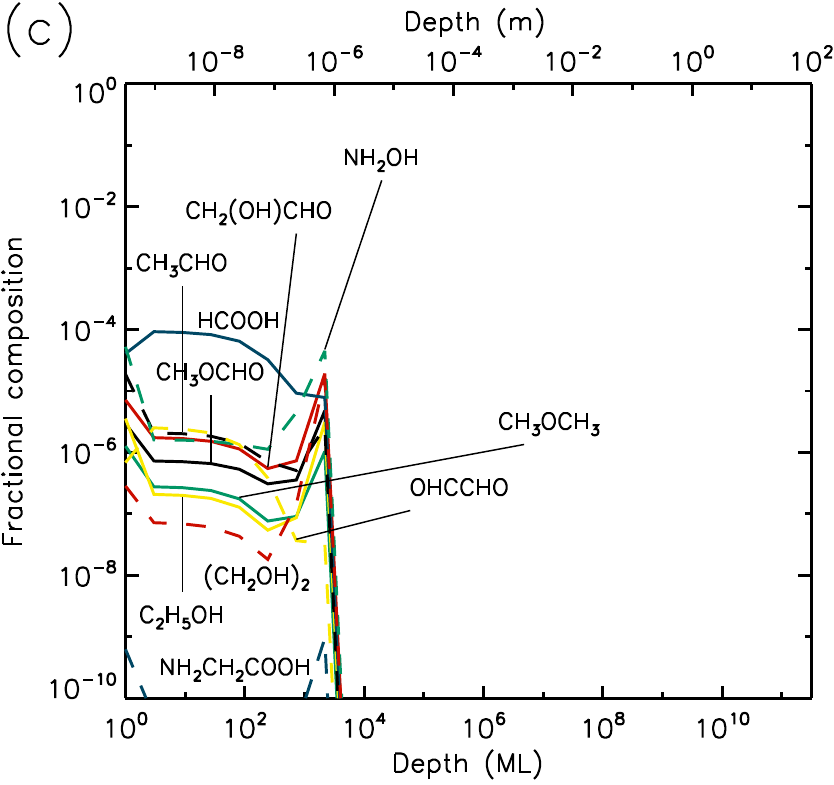}
\includegraphics[width=0.32\textwidth]{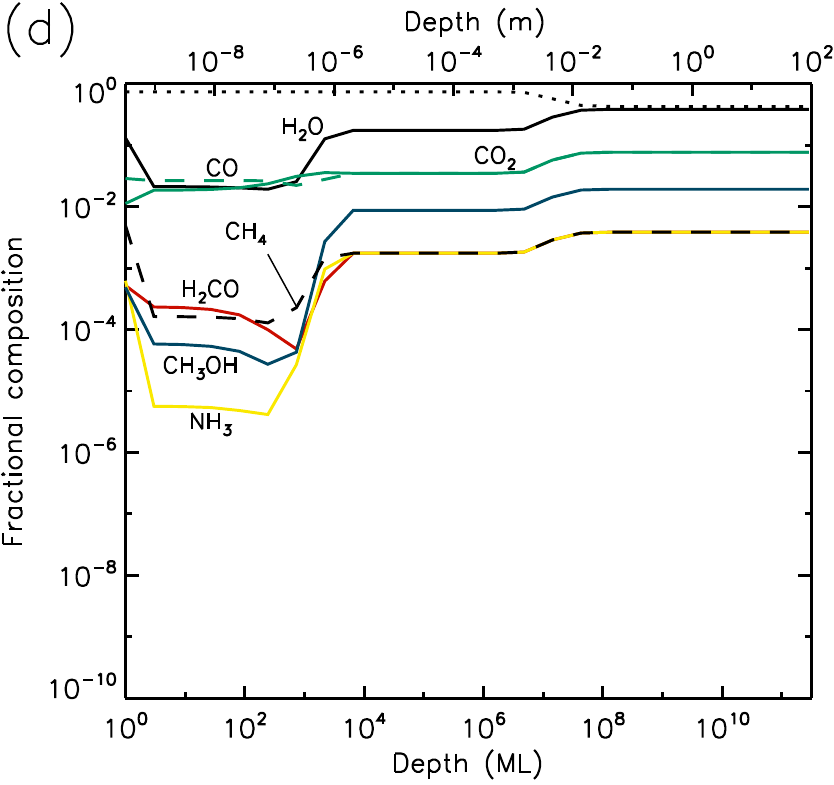}
\includegraphics[width=0.32\textwidth]{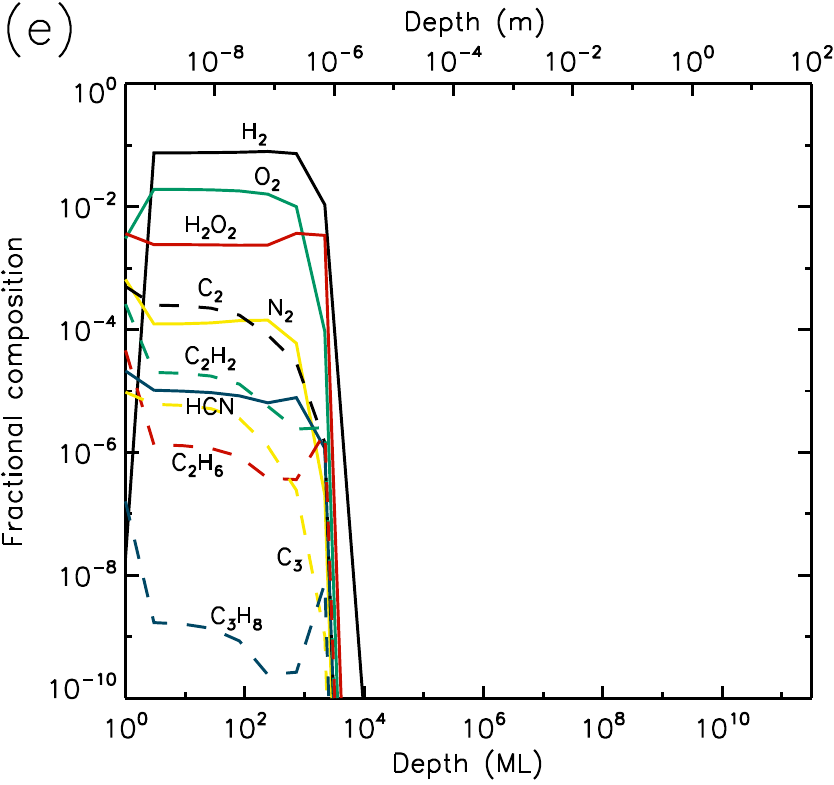}
\includegraphics[width=0.32\textwidth]{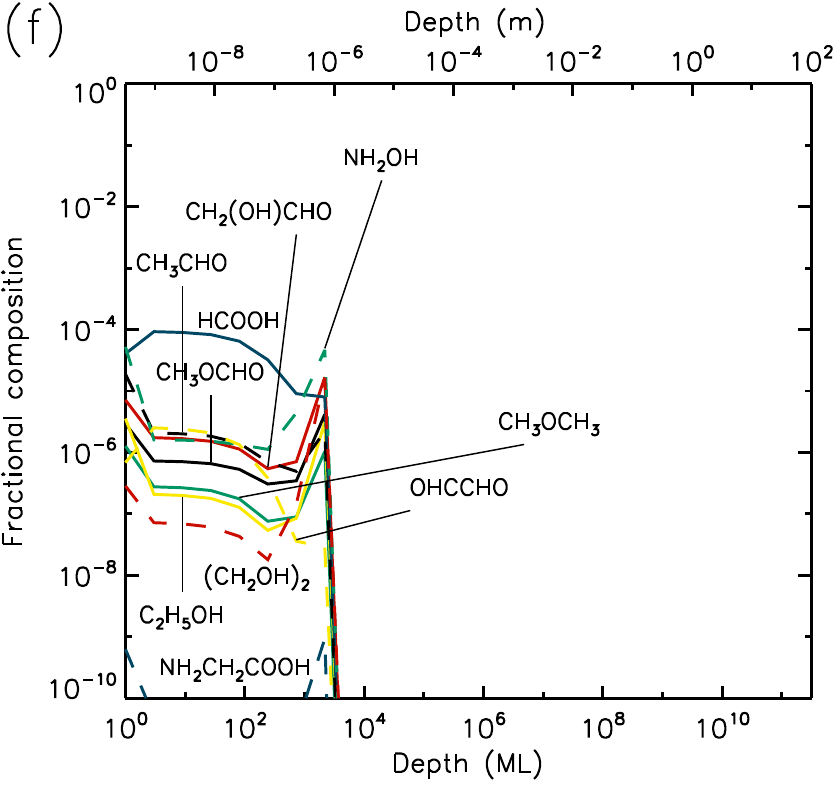}
\end{center}
\caption{\label{fig-5K-late} 5 K models at time $t=5 \times 10^{9}$ yr. Abundances of the initial ice components are shown in the left panels, with dust shown as a dotted line. Results for two model setups are shown: low UV (upper panels) and high UV (lower panels).}
\end{figure*}

\begin{figure*}
\begin{center}
\includegraphics[width=0.32\textwidth]{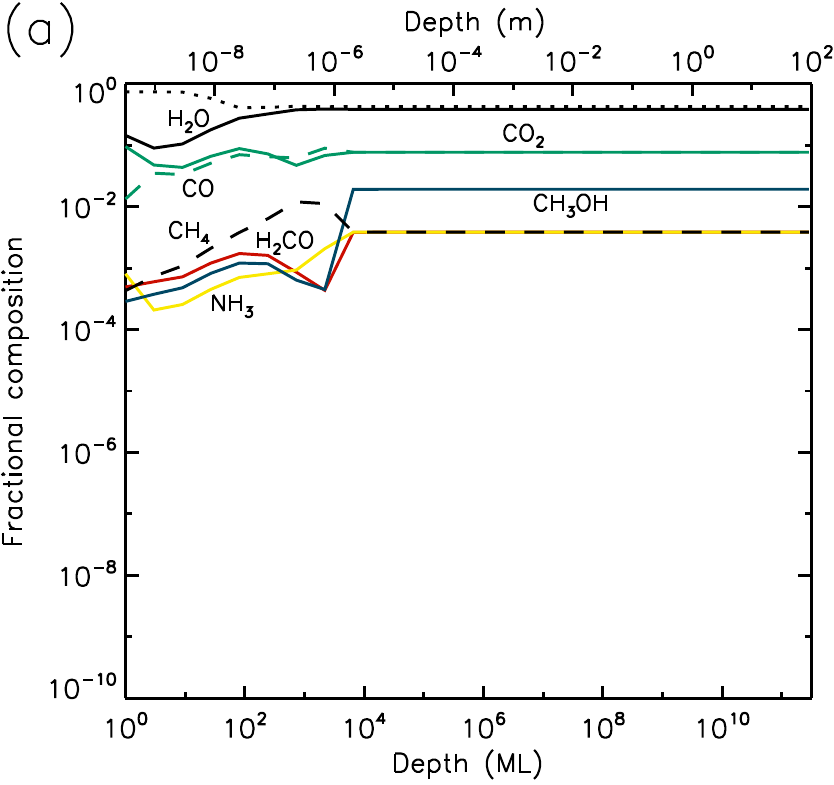}
\includegraphics[width=0.32\textwidth]{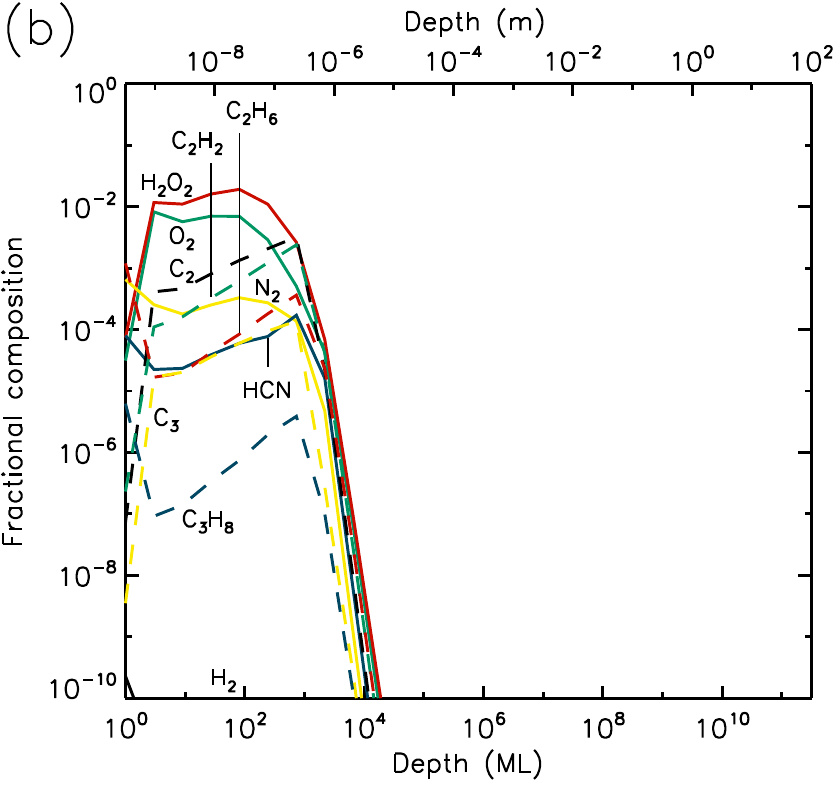}
\includegraphics[width=0.32\textwidth]{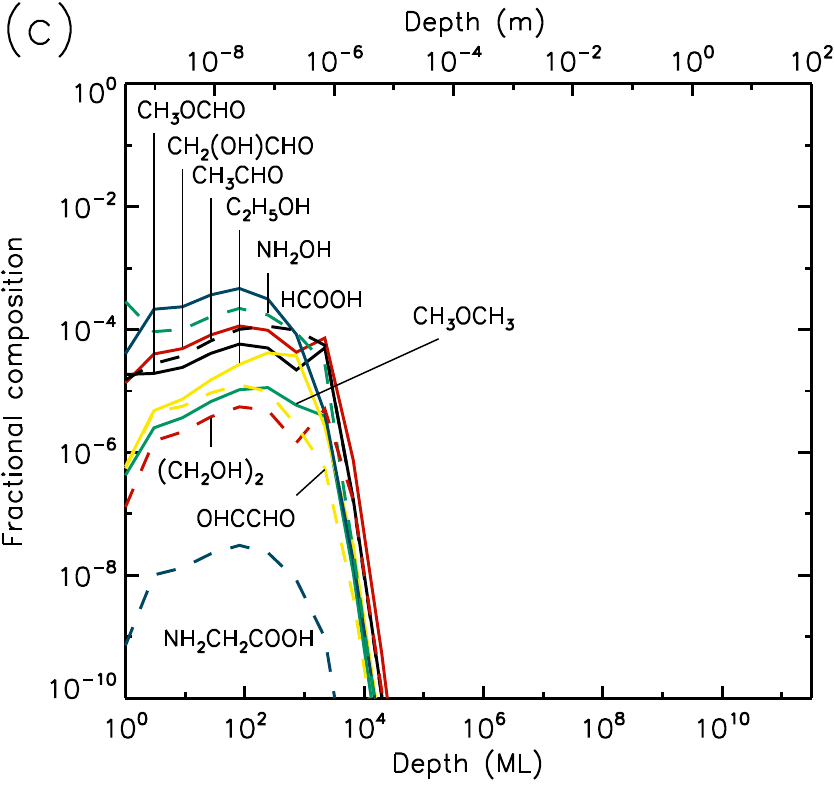}
\includegraphics[width=0.32\textwidth]{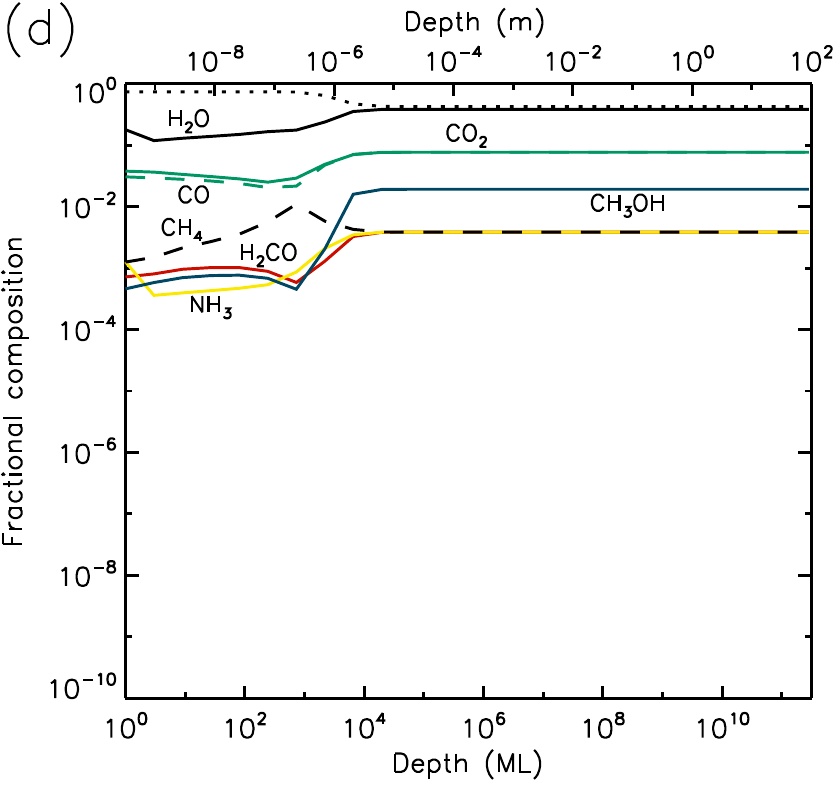}
\includegraphics[width=0.32\textwidth]{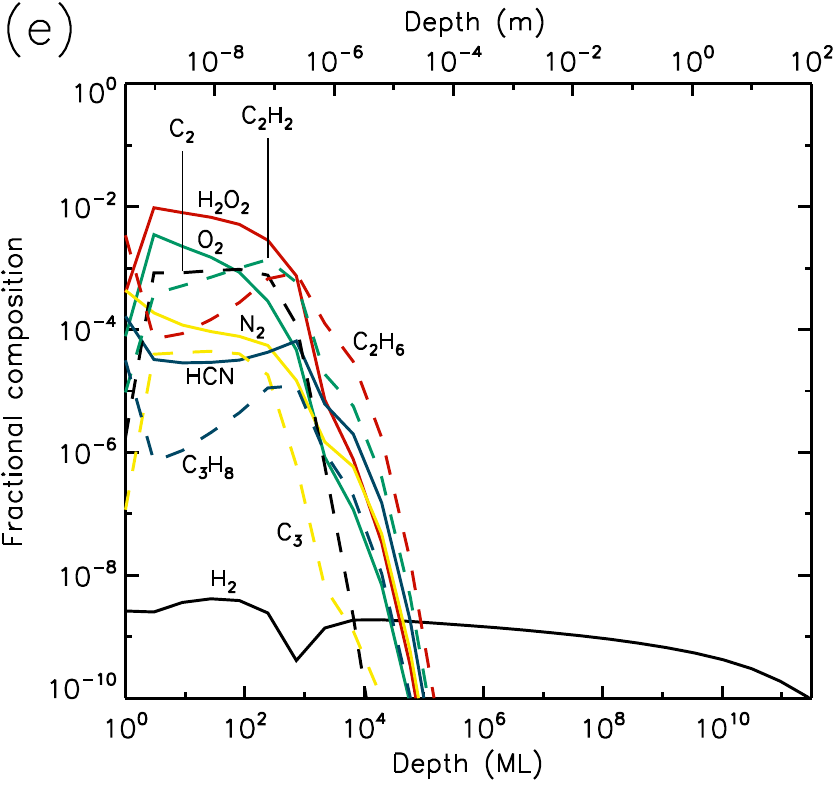}
\includegraphics[width=0.32\textwidth]{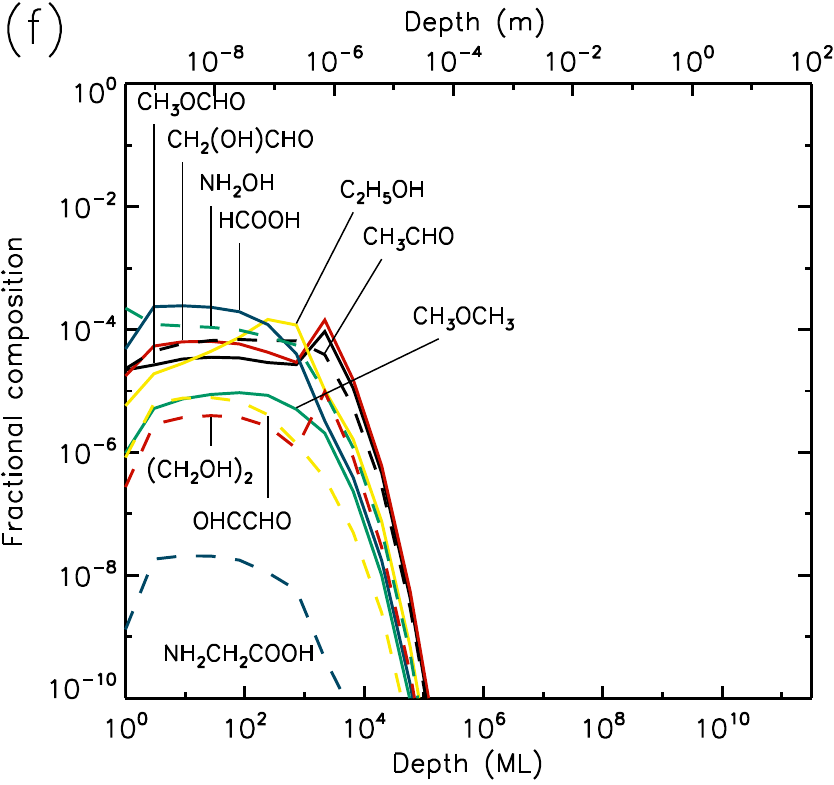}
\end{center}
\caption{\label{fig-10K-early} 10 K models at time $t=10^{6}$ yr. Abundances of the initial ice components are shown in the left panels, with dust shown as a dotted line. Results for two model setups are shown: low UV (upper panels) and high UV (lower panels).}
\end{figure*}

\begin{figure*}
\begin{center}
\includegraphics[width=0.32\textwidth]{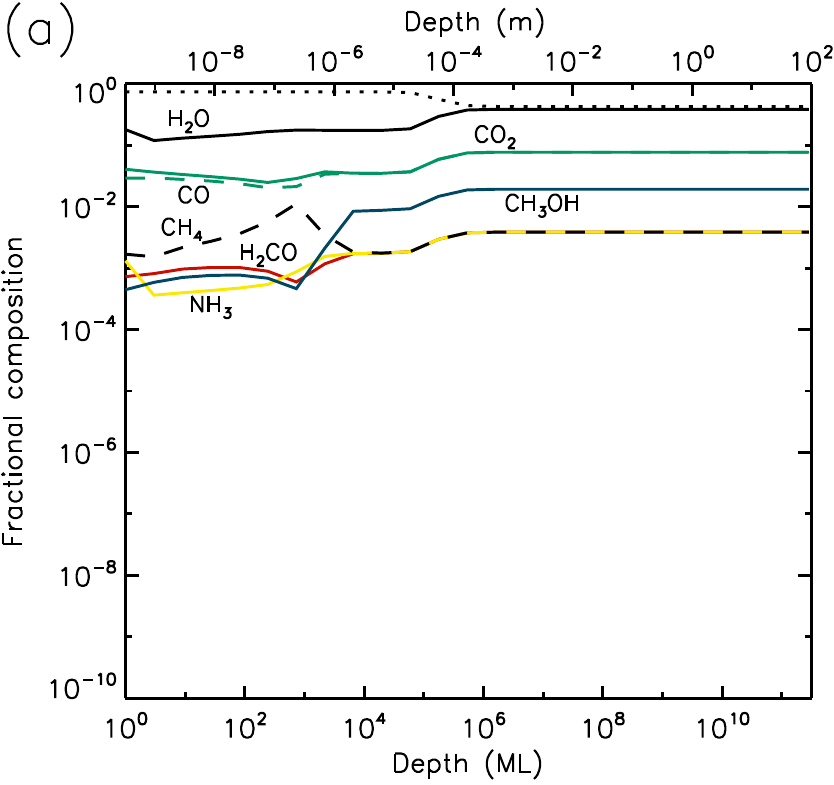}
\includegraphics[width=0.32\textwidth]{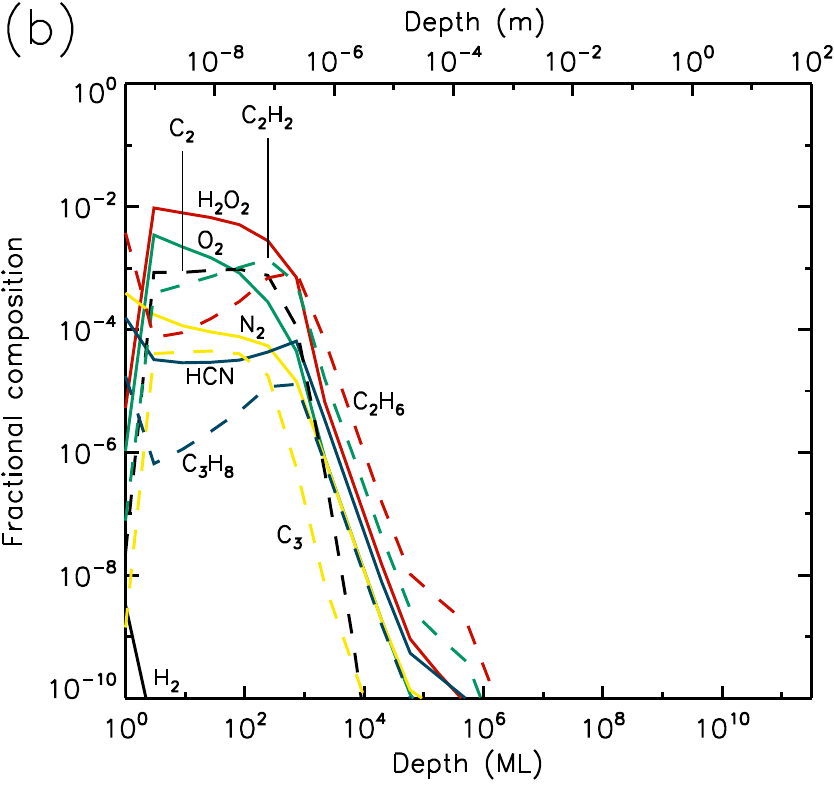}
\includegraphics[width=0.32\textwidth]{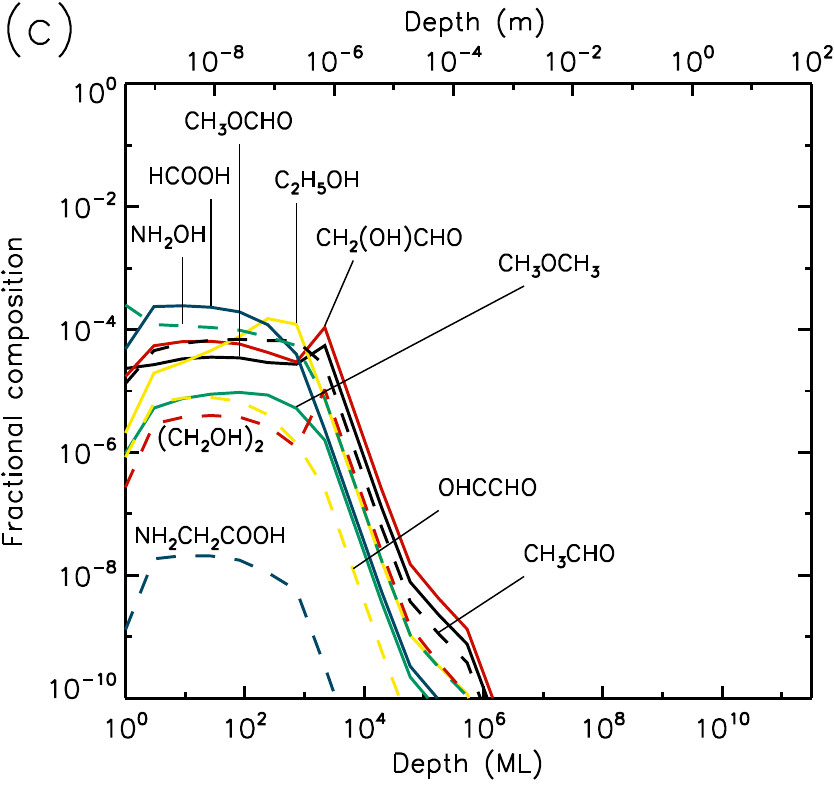}
\includegraphics[width=0.32\textwidth]{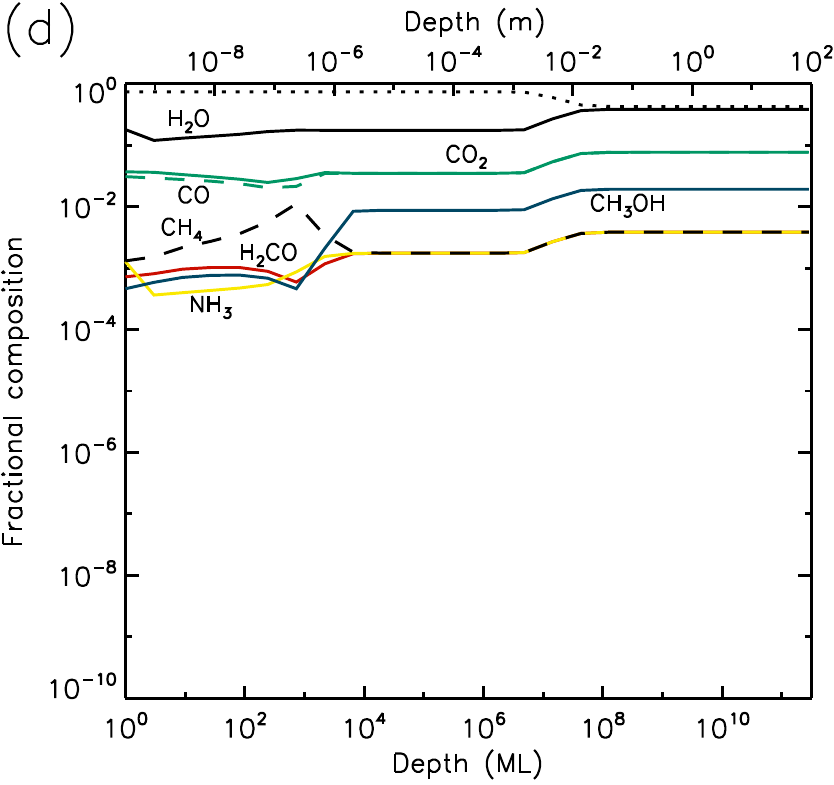}
\includegraphics[width=0.32\textwidth]{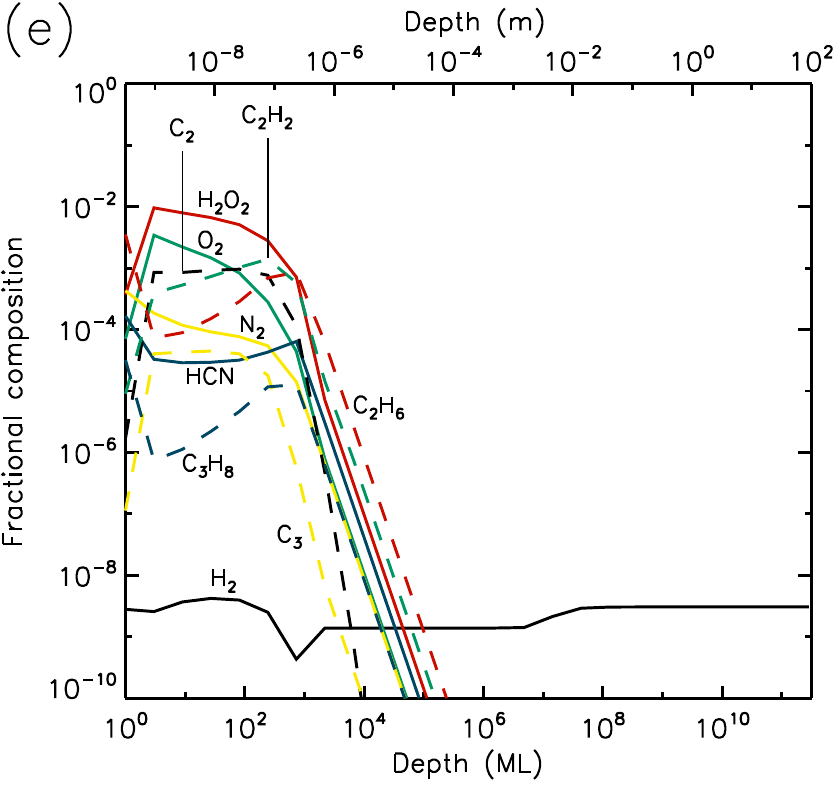}
\includegraphics[width=0.32\textwidth]{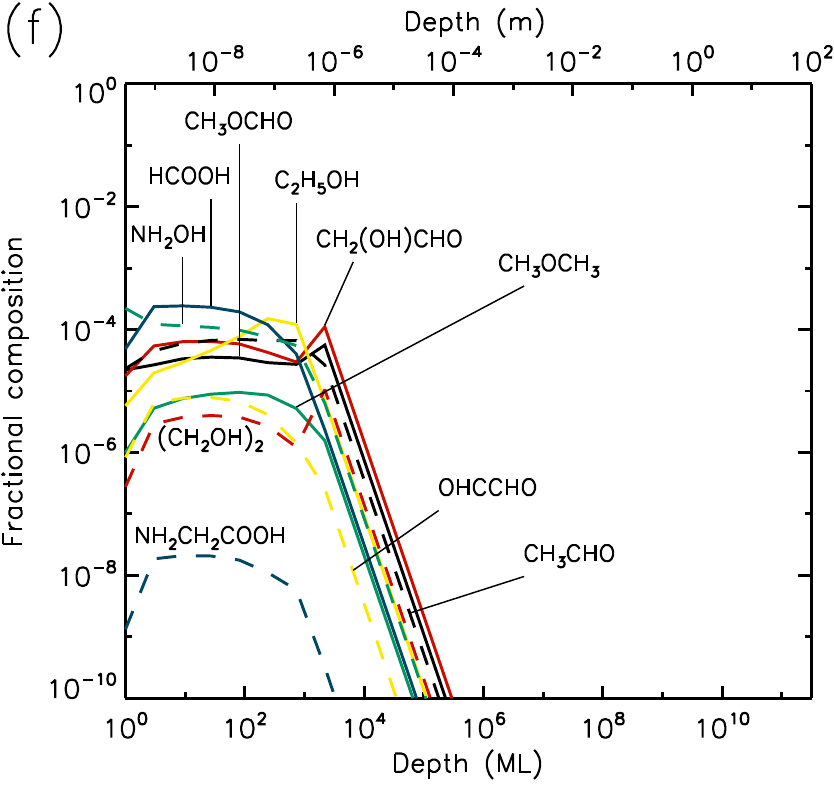}
\end{center}
\caption{\label{fig-10K-late} 10 K models at time $t=5 \times 10^{9}$ yr. Abundances of the initial ice components are shown in the left panels, with dust shown as a dotted line. Results for two model setups are shown: low UV (upper panels) and high UV (lower panels).}
\end{figure*}

\begin{figure*}
\begin{center}
\includegraphics[width=0.32\textwidth]{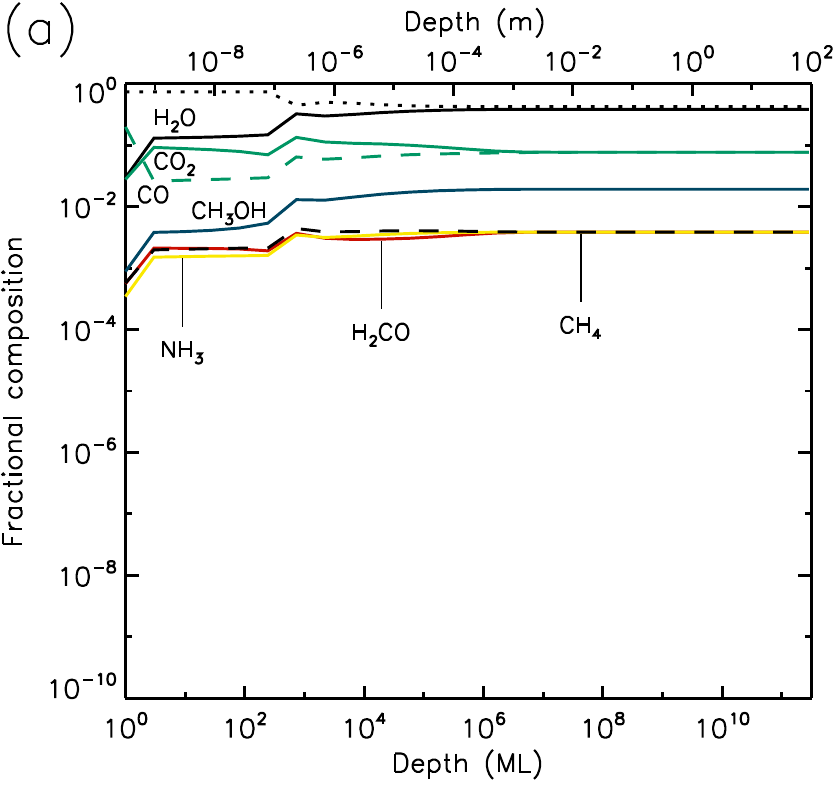}
\includegraphics[width=0.32\textwidth]{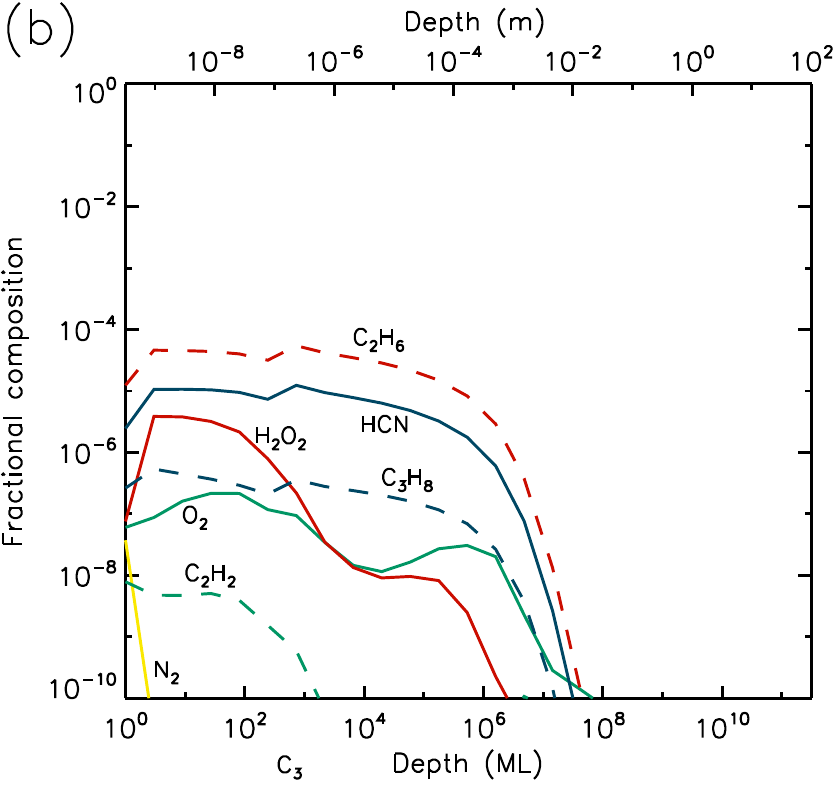}
\includegraphics[width=0.32\textwidth]{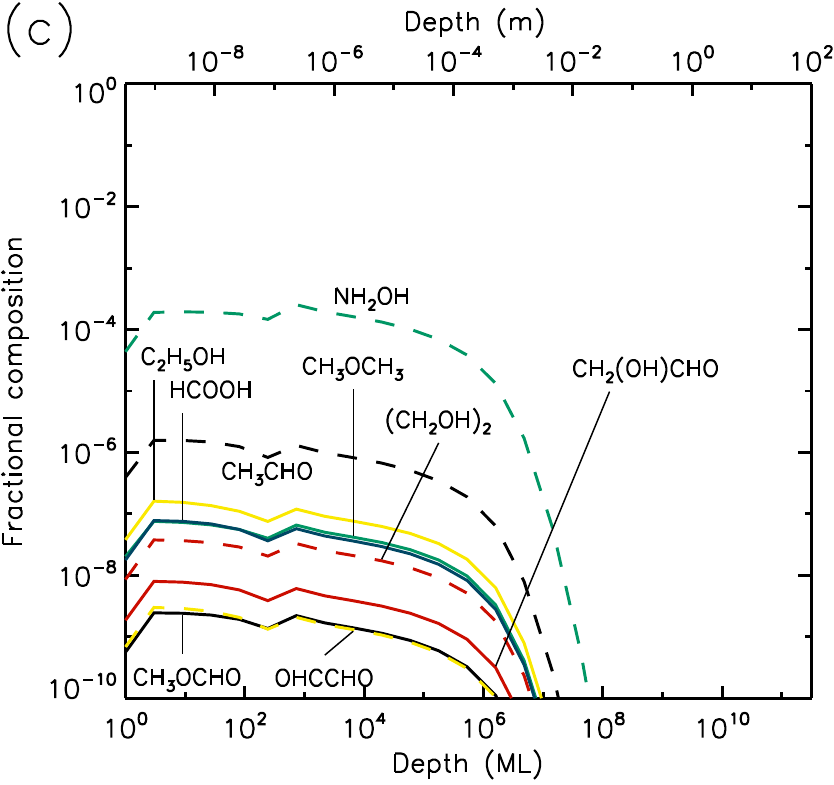}
\includegraphics[width=0.32\textwidth]{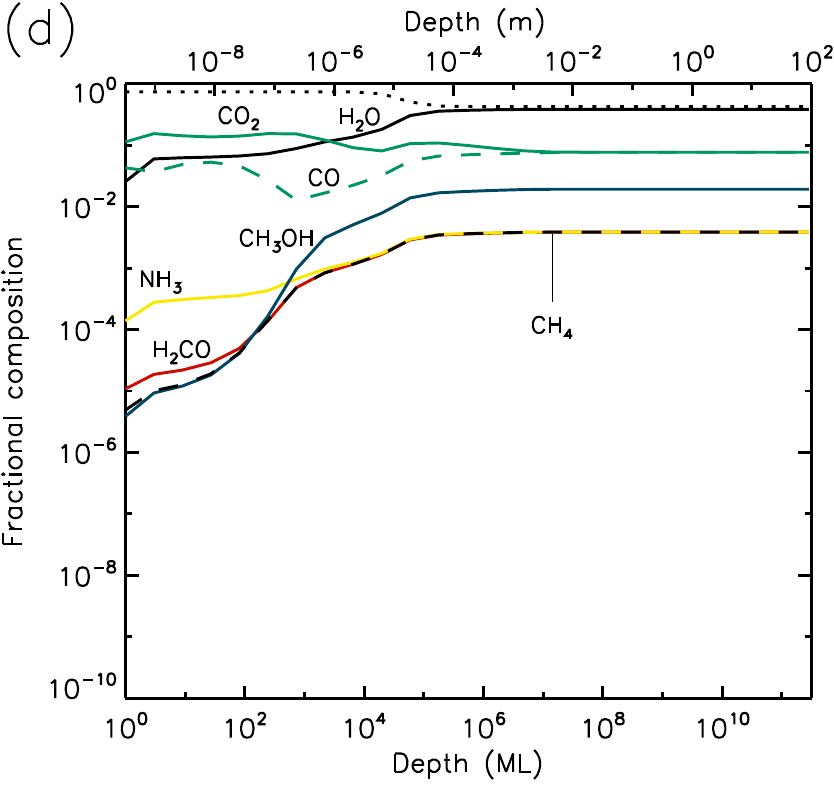}
\includegraphics[width=0.32\textwidth]{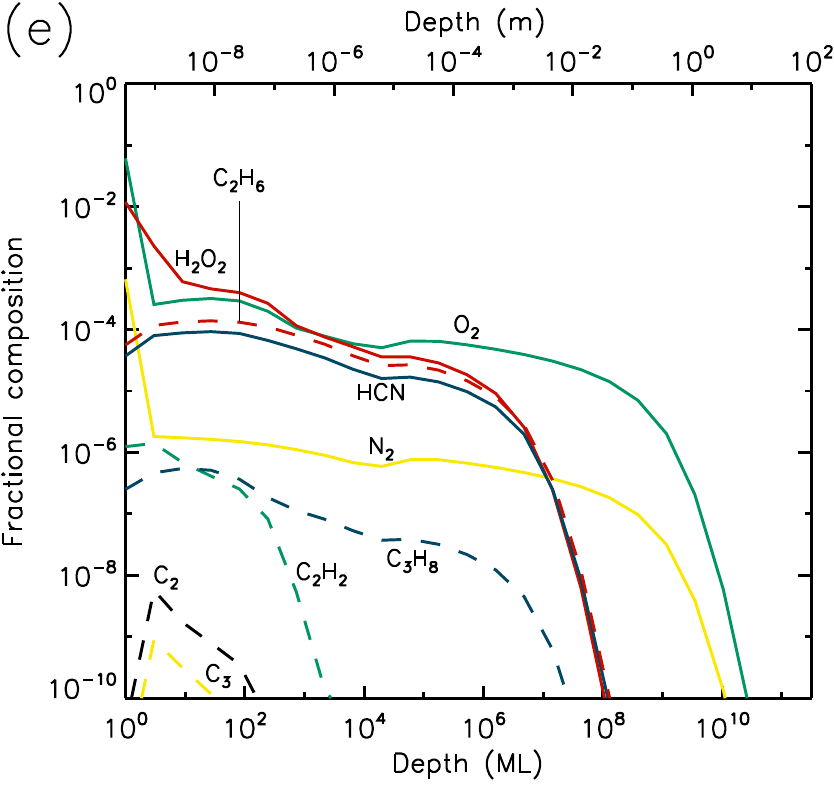}
\includegraphics[width=0.32\textwidth]{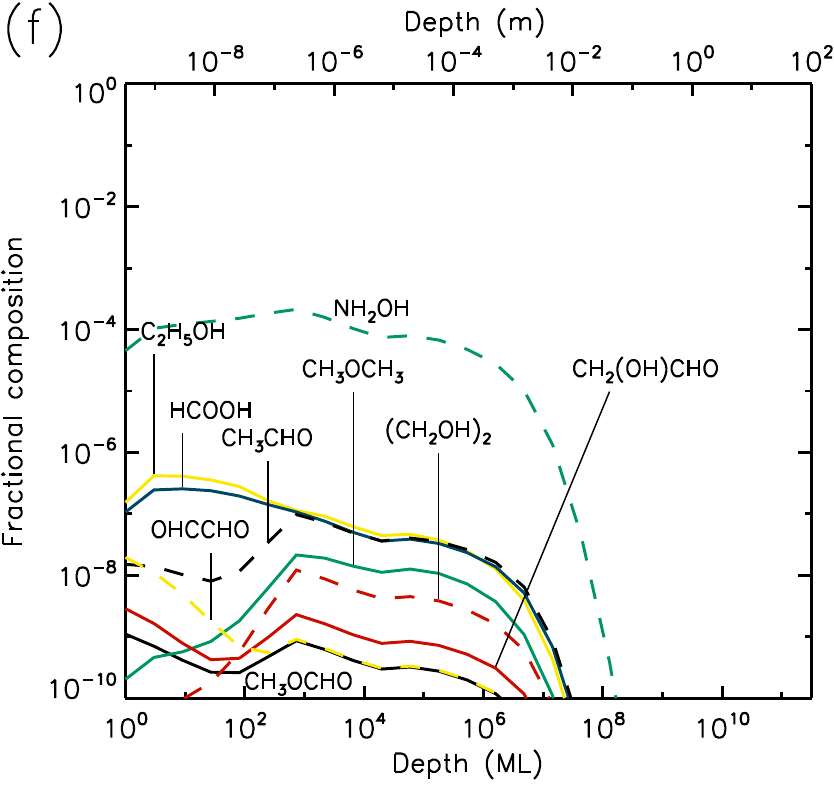}
\end{center}
\caption{\label{fig-20K-early} 20 K models at time $t=10^{6}$ yr. Abundances of the initial ice components are shown in the left panels, with dust shown as a dotted line. Results for two model setups are shown: low UV (upper panels) and high UV (lower panels).}
\end{figure*}

\begin{figure*}
\begin{center}
\includegraphics[width=0.32\textwidth]{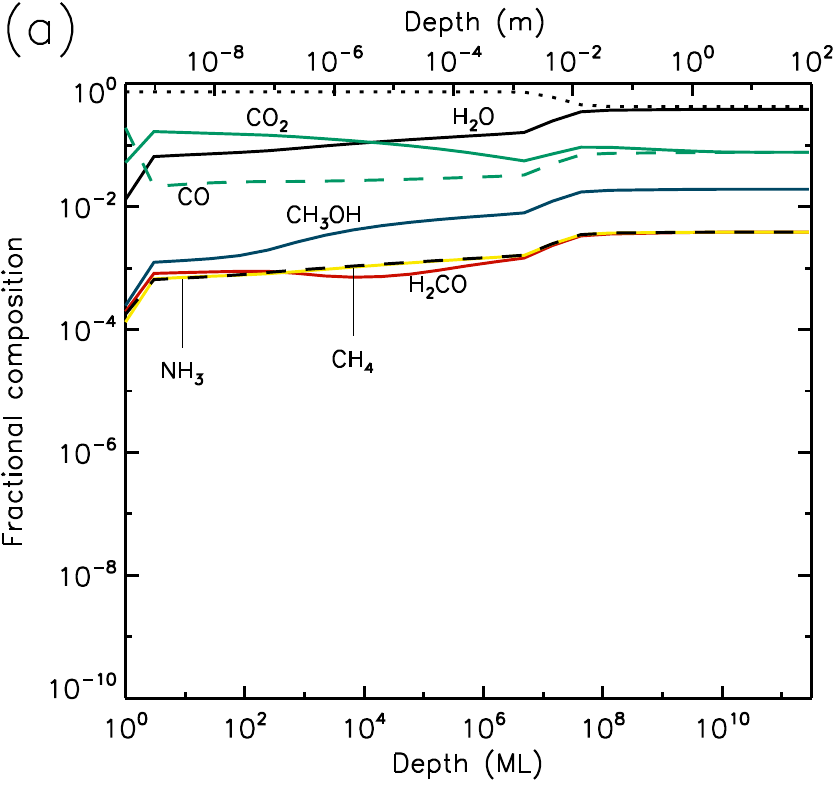}
\includegraphics[width=0.32\textwidth]{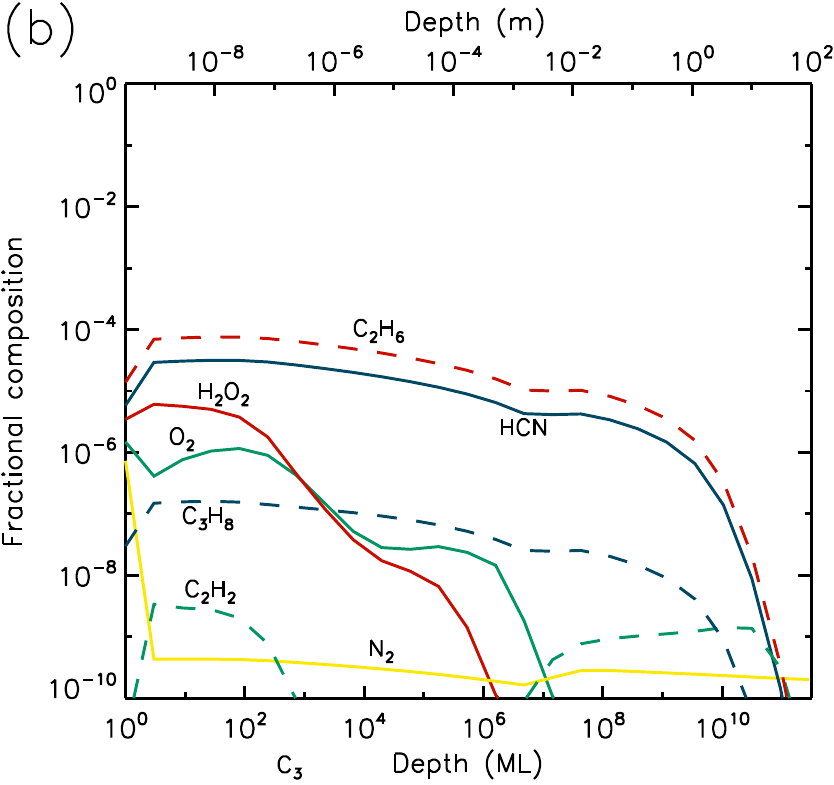}
\includegraphics[width=0.32\textwidth]{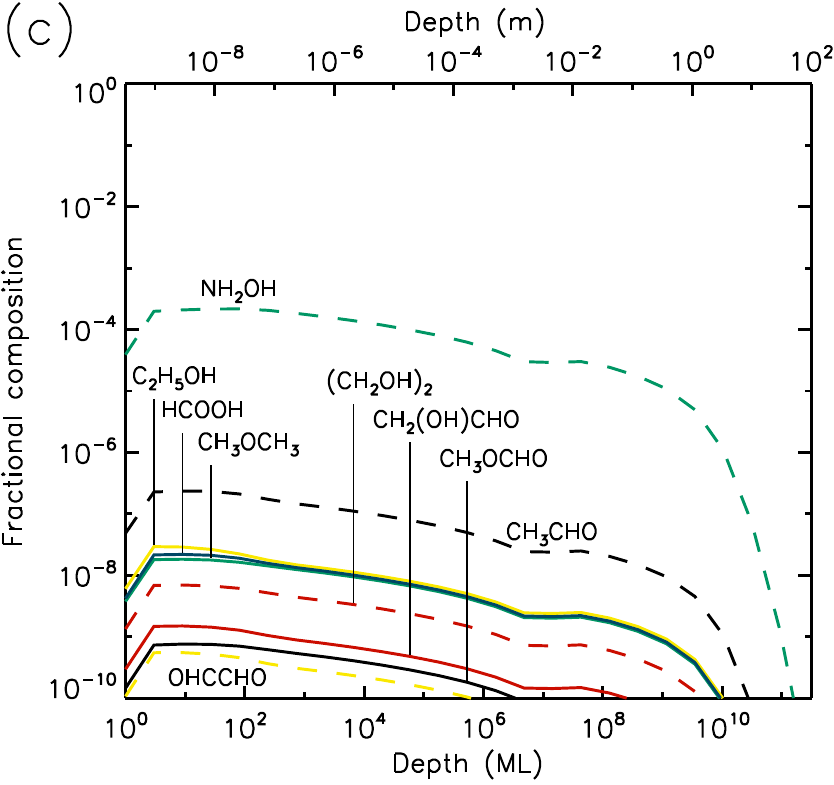}
\includegraphics[width=0.32\textwidth]{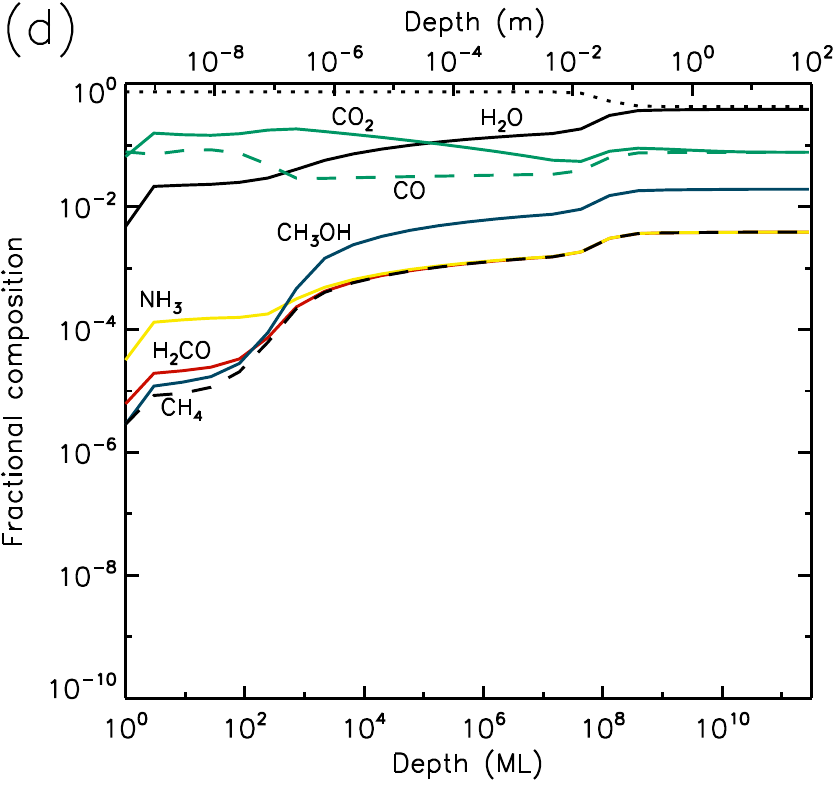}
\includegraphics[width=0.32\textwidth]{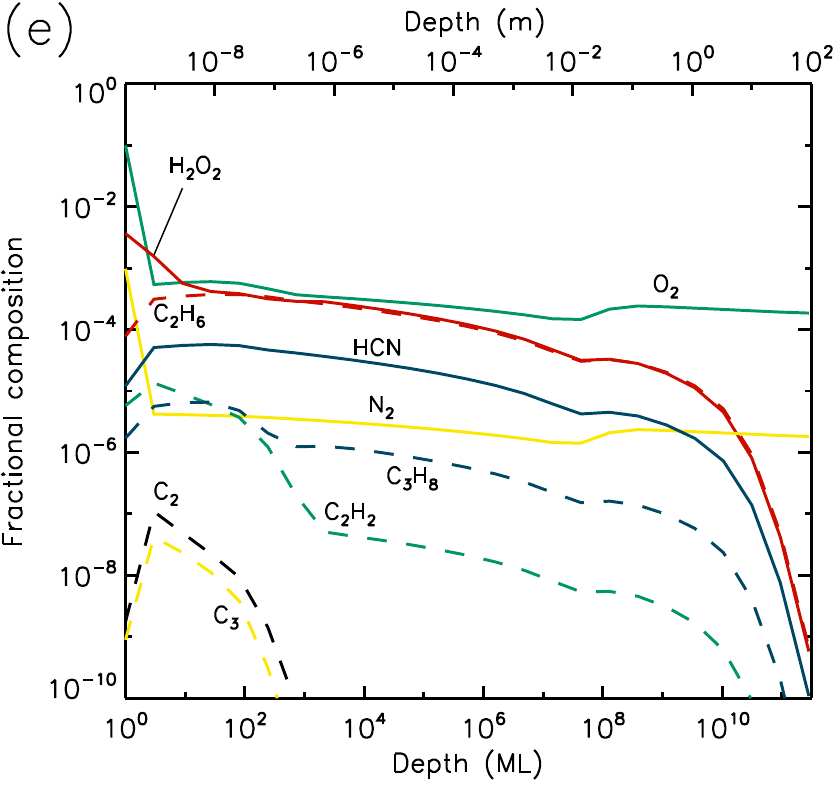}
\includegraphics[width=0.32\textwidth]{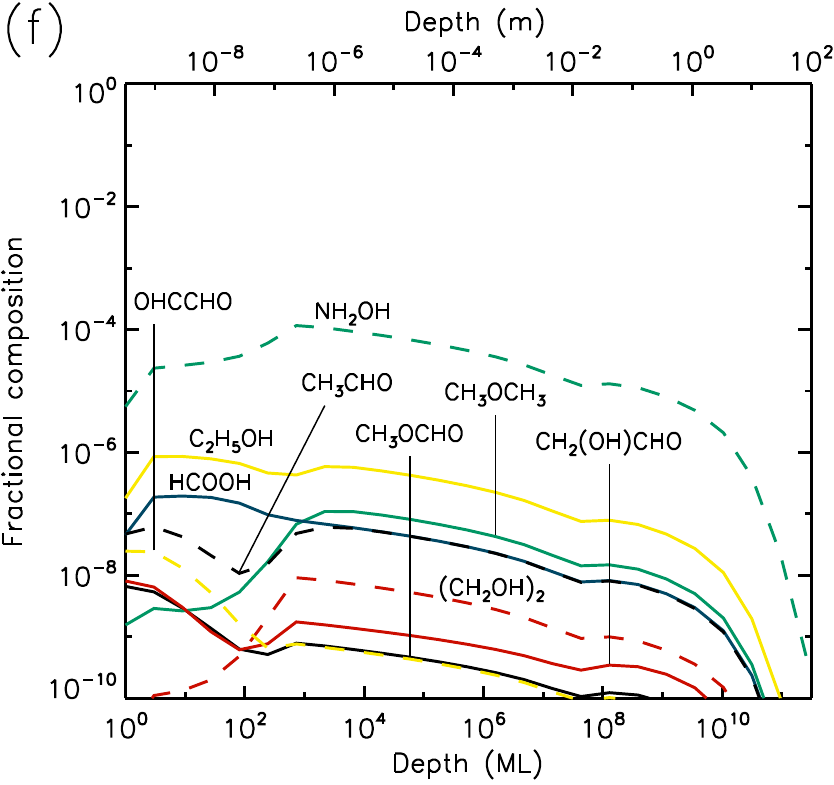}
\end{center}
\caption{\label{fig-20K-late} 20 K models at time $t=5 \times 10^{9}$ yr. Abundances of the initial ice components are shown in the left panels, with dust shown as a dotted line. Results for two model setups are shown: low UV (upper panels) and high UV (lower panels).}
\end{figure*}

Figures \ref{fig-5K-early} and \ref{fig-5K-late} show the 5~K model results at $t$=10$^6$ yr and $5\times10^9$ yr, respectively. It may be seen that the most drastic chemical changes occur in the upper 1~$\mu$m of the ice, with the initial ice constituents dropping significantly over time, except in the very surface layer. In Fig. \ref{fig-5K-early}, very similar behavior is seen between the low and high UV cases. Conversely, by the end of the model runs (Fig. \ref{fig-5K-late}), the abundance of dust has reached its maximum value, to a depth of approximately 10~$\mu$m for the low UV case, 100~$\mu$m for the medium UV case, and 1~mm for the high UV case. The increase in dust fraction produces a commensurate drop in the main ice components, which may be more easily distinguished from the photo-destruction of these species as the dust enrichment extends to greater depths over time. The concentration of dust in the outer layers is driven by the loss of material through desorption from the surface layer. At 5~K, this is caused exclusively by photo-desorption, with water loss making up the largest individual component. The total loss rate from the surface at the end of the 5~K low-UV model run is $6.3\times10^{-5}$ ML yr$^{-1}$, of which water makes up $\sim$55\%. Other such rates, along with the total amount of ice lost in each model and the depth to which the dust reaches peak concentration, are shown in Table \ref{tab-rates}.

The photo-dissociation of the initial ice components results in the production of new molecules from the radical/atomic fragments. The abundance behavior of  these products is similar across the various UV flux models. The large abundance of H$_2$, which here is the dominant molecule in the ice to a depth of a fraction of a micron, is mainly the product of the photo-dissociation of water (H$_2$O + $h\nu$ $\rightarrow$ OH + H) followed by immediate reaction between the newly-dissociated H-atom and a contiguous H-atom already present in the ice. The influence of thermal bulk-ice diffusion on reaction rates is negligible at 5~K compared to this immediate reaction process, even for H atoms. A signficant amount of H and O (around 10\% and 8\%, respectively, for the medium-UV model) is also present in the upper sub-surface layers, and these abundances are mostly constant to a depth of $\sim$1~$\mu$m. In the surface layer itself, the abundances of hydrides that are initially present in the ice remain relatively large throughout the model runs, due to the lower barriers to thermal diffusion, allowing photo-fragments to find each other and react, with hydrogen dominating the diffusion and reaction process. 

O$_2$ is very abundant in the upper layers, as seen in the middle panels of the figures. Its production is mostly the result of the liberation and immediate reaction of oxygen atoms via the dissociation of CO$_2$ and OH (produced mainly from water), although, once formed, O$_2$ may also be photo-dissociated. The abundance of O$_2$ produced through this immediate photo-dissociation/reaction mechanism is around 1\% of the total dust/ice composition, although in the region where O$_2$ is abundant, water itself is strongly depleted, so that the O$_2$/H$_2$O ratio is approximately unity. H$_2$O$_2$, N$_2$, and a range of hydrocarbons are also produced through similar mechanisms, via the destruction of the initial ice components. Oxygen atoms produced by photo-dissociation may also react with O$_2$, producing ozone (O$_3$, not shown in figures). 

At relatively early times ($10^{6}$ yr), product molecules are seen to reach a peak at their greatest depth (around 1~$\mu$m), before strongly dropping off in the deeper layers. However, at the end-time of the models, the product abundance profiles are generally flatter down to a depth of 1~$\mu$m. The deep-peaking effect at early times is caused by the maintenence of high abundances of parent species deeper into the ice, where photo-dissociation rates are lower and therefore chemical timescales are longer. Later in the models, the abundances of parent species are reduced, and the chemistry reaches stability, balancing between photo-destruction of parent and product species, and the formation/re-formation of both from the resulting radicals.

The right-hand panels show the abundances of a selection of complex organic molecules, either detected in cometary environments, observed in the gas phase toward star-forming sources, or considered in interstellar chemical models. These species show the same early-peaking effect as the simpler products. The majority of complex species plotted show abundances of around $10^{-8}$--$10^{-4}$ to a depth of 1$\mu$m. The most abundant of these is formic acid (HCOOH), which is formed through the reaction HCO + OH $\rightarrow$ HCOOH. The HCO radical is produced by reactions of liberated H atoms with CO already present in the ice; again, no thermal diffusion of H is required in this case, due to the production of H-atoms in the presence of the abundant CO molecule. Photo-dissociation of water then produces OH that reacts with the HCO. The large abundances of both CO and H$_2$O therefore result in significant formic acid production.

Many of the other complex organic molecules plotted in the figures are products of CH$_3$O or CH$_2$OH radicals. In interstellar ices, these radicals are usually thought to be formed either through H-atom addition to formaldehyde (H$_2$CO) on dust-grain surfaces, or via the abstraction of a hydrogen atom from methanol (CH$_3$OH) on grains. In the 5~K comet models, the reactions CH$_3$ + O $\rightarrow$ CH$_3$O and CH$_2$ + OH $\rightarrow$ CH$_2$OH are the main formation routes. The relatively low abundances of H$_2$CO and CH$_3$OH (as compared with CO), combined with the barriers involved in either H addition or abstraction, make the interstellar routes less favored at 5~K. Reactions of the CH$_3$O and CH$_2$OH radicals with HCO tend to be somewhat more efficient than reactions with CH$_3$, due to the rapid H + CO formation routes for HCO. 

Abundances of complex organics are similar between different UV-field models, especially toward the end of the model runs. This is because, under the conditions at 5~K in which diffusion between layers is generally minimal, the chemistry of those molecules reaches a balance between formation and destruction that is dependent almost exclusively on photo-dissociation (in the upper layers where COMs are more abundant); photo-dissociation of smaller molecules produces COMs, while those products are also destroyed by photo-dissociation. Both formation and destruction therefore scale in the same proportion as the UV field.

Hydrocarbons and many of the complex organics show a strong tendency to increase in the outer surface layer, due to the greater ability of e.g. methyl radicals to diffuse on the surface versus within the bulk ice. The gradual concentration of C$_2$ in the surface layer (due to desorption of more volatile species, following its formation in deeper layers, also results in higher abundances of its hydrogenation products such as C$_2$H$_2$ (see middle panels).

Glycine (NH$_2$CH$_2$COOH) is also plotted, although it barely reaches fractional abundances above $10^{-10}$. The reactions by which it may form in the current network all involve the addition of radicals, and are described in more detail by Garrod (2013). Here, production occurs mainly through the reaction NH$_2$ + CH$_2$COOH $\rightarrow$ NH$_2$CH$_2$COOH. The amino radical is formed through direct dissociation of ammonia, as well as through the repetitive addition of H to atomic N (ultimately sourced from NH$_3$). The radical CH$_2$COOH is a dissociation product of acetic acid, whose formation requires the production of the CH$_3$CO radical. The latter is formed by the barrier-mediated reaction of CH$_3$ with CO.

\subsection{10~K models}

Results for the 10~K models are plotted in Figures 3 \& 4. By the end of the model runs, the concentration of dust is similar to that of the 5~K models, as most of the molecular material is still not thermally mobile within the bulk ice. Thus, water photo-desorption still dominates the loss of surface material at rates determined purely by the UV field strength. However, aside from the effects of the concentration of dust in the upper layers, the abundance of water now remains close to its maximum value throughout the models. This is due to the thermal diffusion of H atoms in the bulk ice, which react with O and OH to re-form water. The mobility of atomic H means that an individual H atom has a short lifetime in the ice, disfavoring the reaction H + H $\rightarrow$ H$_2$, which depends on high fractional abundances of H. Hence the H$_2$ abundance in the bulk ice drops to low values at temperatures of 10~K and above. The H$_2$ abundance produced increases with the UV-field strength, and in the high-UV plot it can be seen that at 10~K this molecule is mobile enough in the ice to have thermally diffused into the deepest ice layers by the end of the model.

At 10~K, O$_2$ production in the ice is somewhat lessened, versus the 5~K models, due to H atoms reacting more frequently with atomic O and reducing its abundance in the ice. Production of O$_2$ still requires immediate reaction between photo-dissociation products. Hydrocarbon production is more pronounced at 10~K, with molecules such as C$_2$H$_2$, C$_2$H$_6$ and C$_3$H$_8$ reaching abundances in the upper micron that are 2--3 orders of magnitude greater at the end time of the model, versus the 5~K results. C$_2$ and C$_3$ also show more modest increases.

While the abundances of product molecules are still restricted to layers in the upper 1~$\mu$m, by the end of the 10~K model runs these species have been able to diffuse gradually to somewhat greater depths, or their radical or atomic precursors have done so, allowing the product species to be formed in the deeper layers to a small extent. 

At 10~K, the abundances of complex molecules are noticeably higher in the upper layers, versus 5~K, with the fractional abundances of many species reaching around $10^{-4}$ with respect to total ice and dust content. The survival of greater abundances of formaldehyde and methanol in the upper layers of the ice allows production of CH$_3$O/CH$_2$OH through addition or abstraction of H to be maintained, resulting in more complex organic molecule production. Glycine is also produced in greater abundance -- again through NH$_2$ addition to CH$_2$COOH, although the latter radical is formed more easily in the 10~K models through the abstraction of a hydrogen atom from acetic acid, which itself is formed as a result of H-abstraction from acetaldehyde followed by OH addition. Hydroxylamine, NH$_2$OH, suggested by some to be a possible precursor to glycine, is strongly produced through the addition of NH$_2$ and OH radicals, as well as by repetitive hydrogenation of NO by H atoms. The final abundance of NH$_2$OH continues to be substantial for all model temperatures, up to some limiting depth.

\begin{figure*}
\begin{center}
\includegraphics[width=0.32\textwidth]{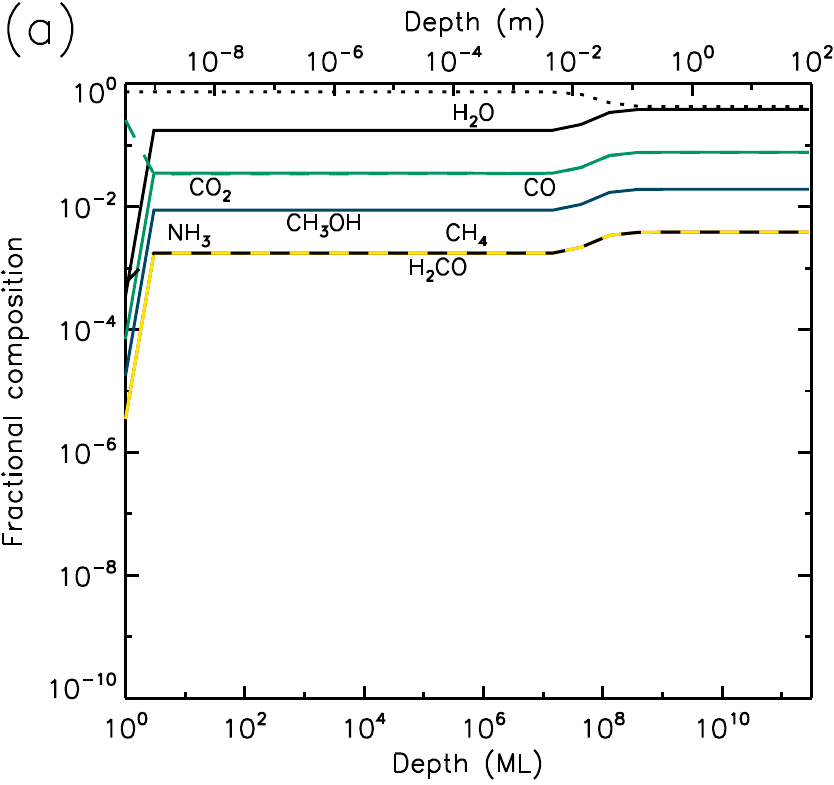}
\includegraphics[width=0.32\textwidth]{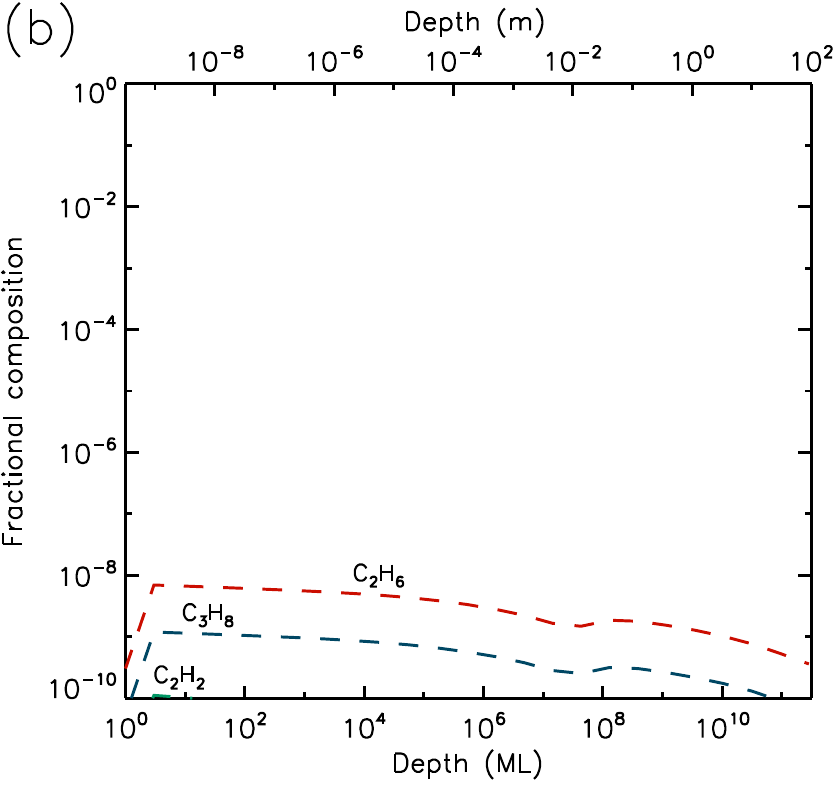}
\includegraphics[width=0.32\textwidth]{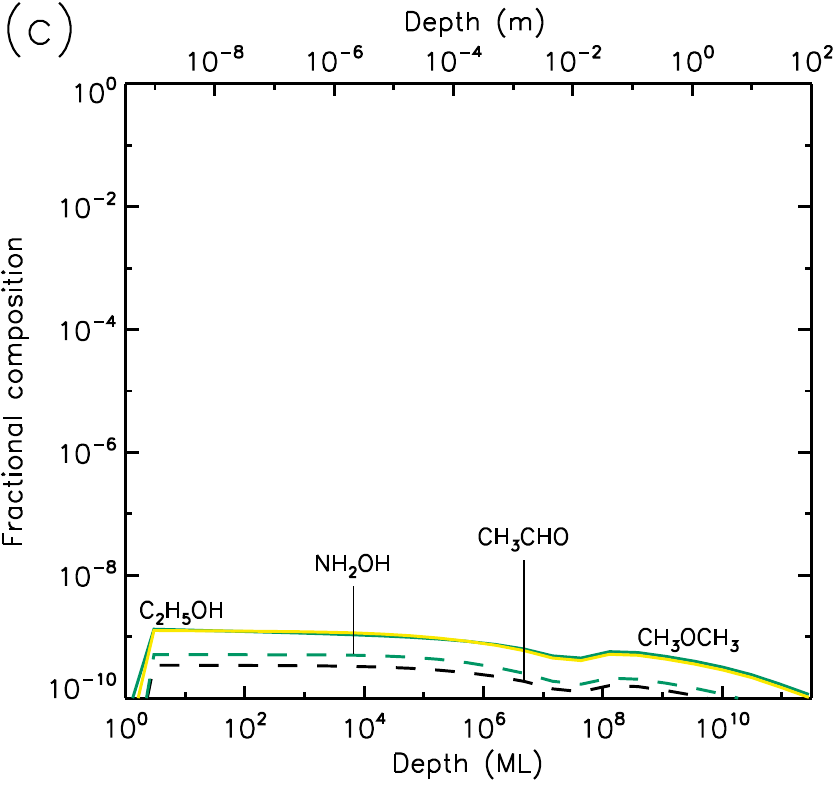}
\includegraphics[width=0.32\textwidth]{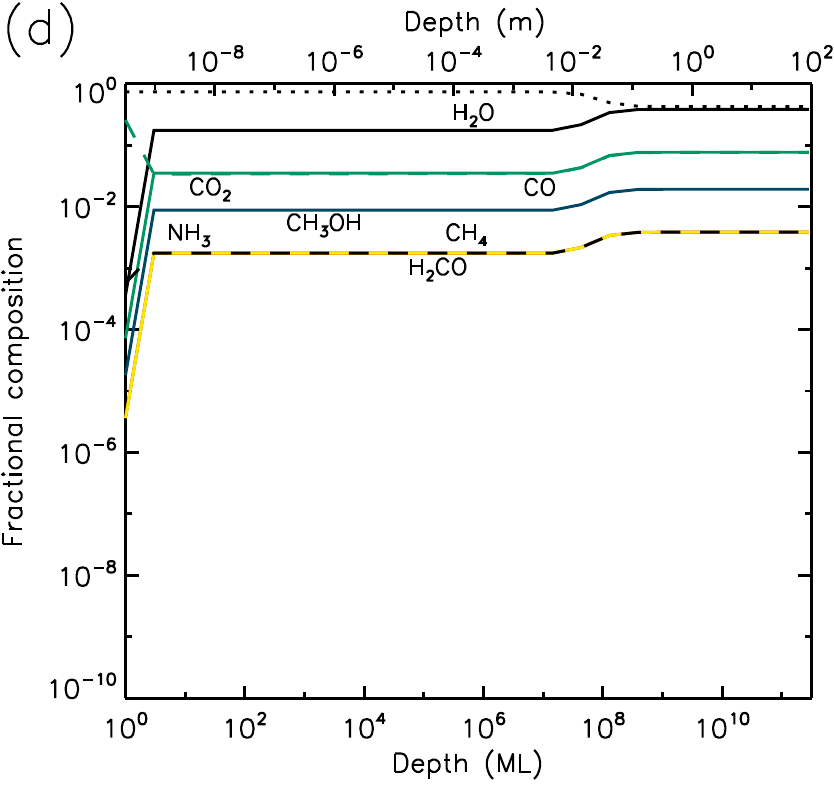}
\includegraphics[width=0.32\textwidth]{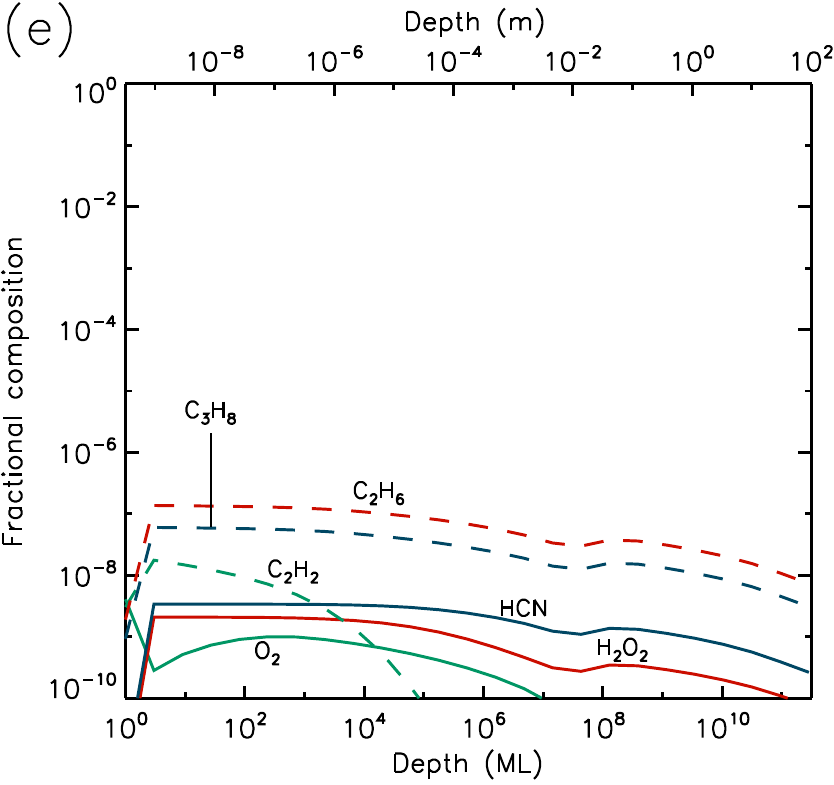}
\includegraphics[width=0.32\textwidth]{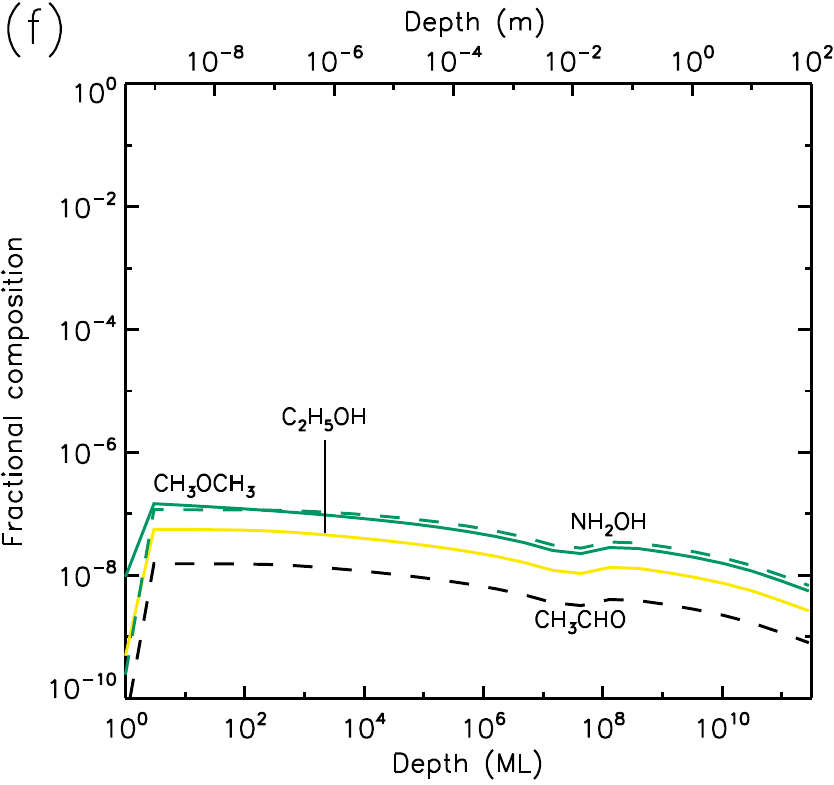}
\end{center}
\caption{\label{fig-30K-early} 30 K models at time $t=10^{6}$ yr. Abundances of the initial ice components are shown in the left panels, with dust shown as a dotted line. Results for two model setups are shown: low UV (upper panels) and high UV (lower panels).}
\end{figure*}

\begin{figure*}
\begin{center}
\includegraphics[width=0.32\textwidth]{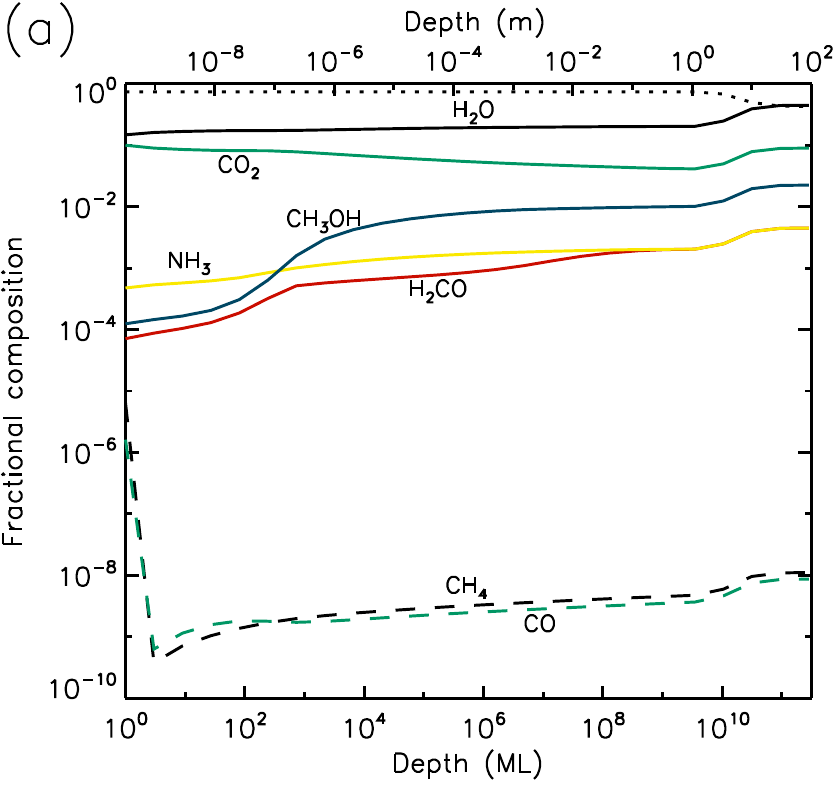}
\includegraphics[width=0.32\textwidth]{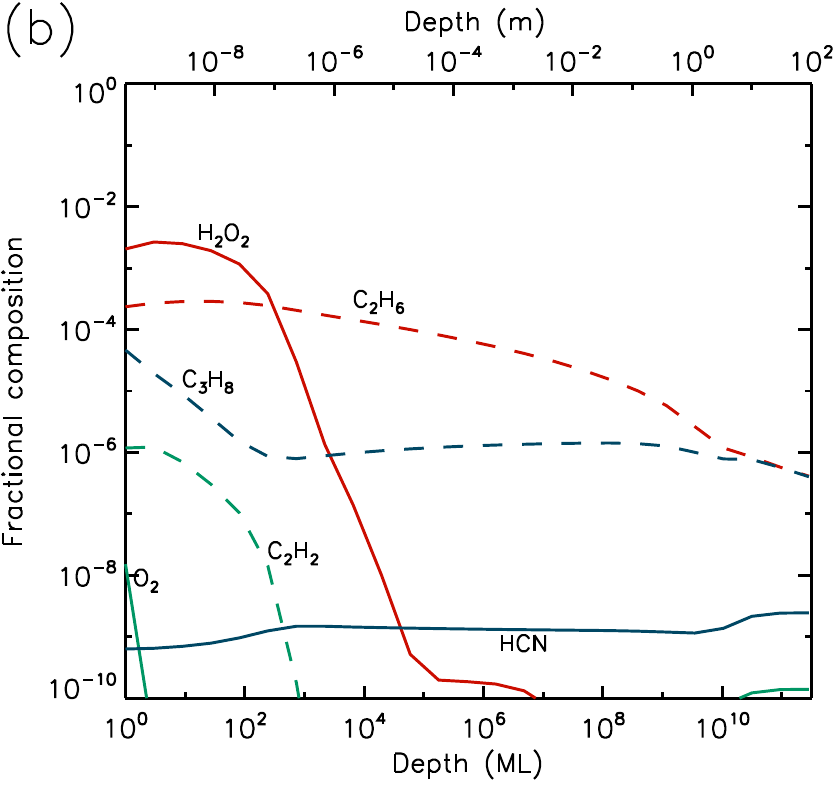}
\includegraphics[width=0.32\textwidth]{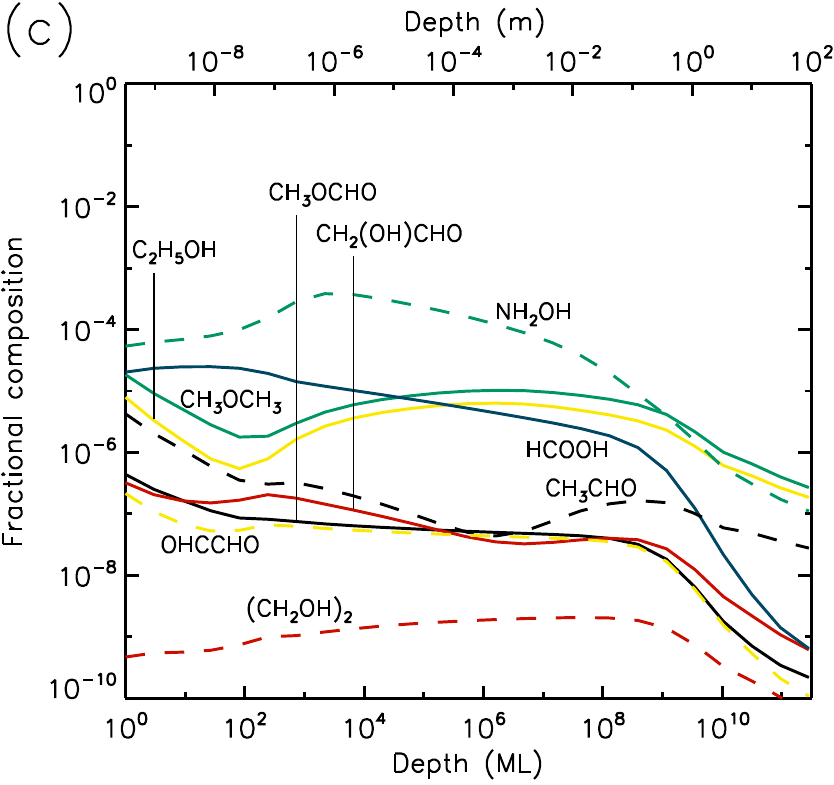}
\includegraphics[width=0.32\textwidth]{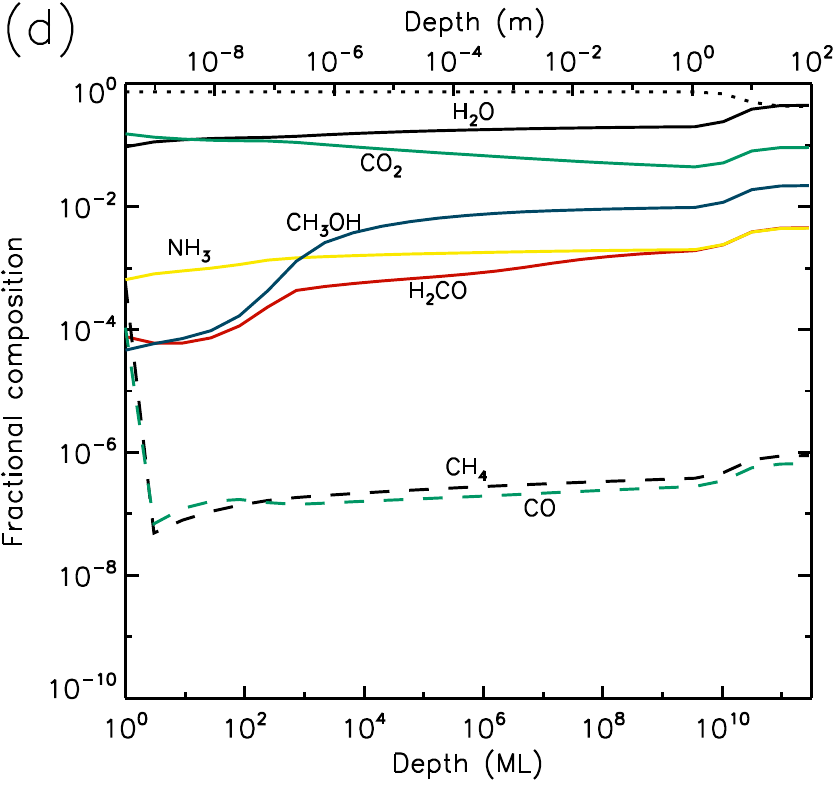}
\includegraphics[width=0.32\textwidth]{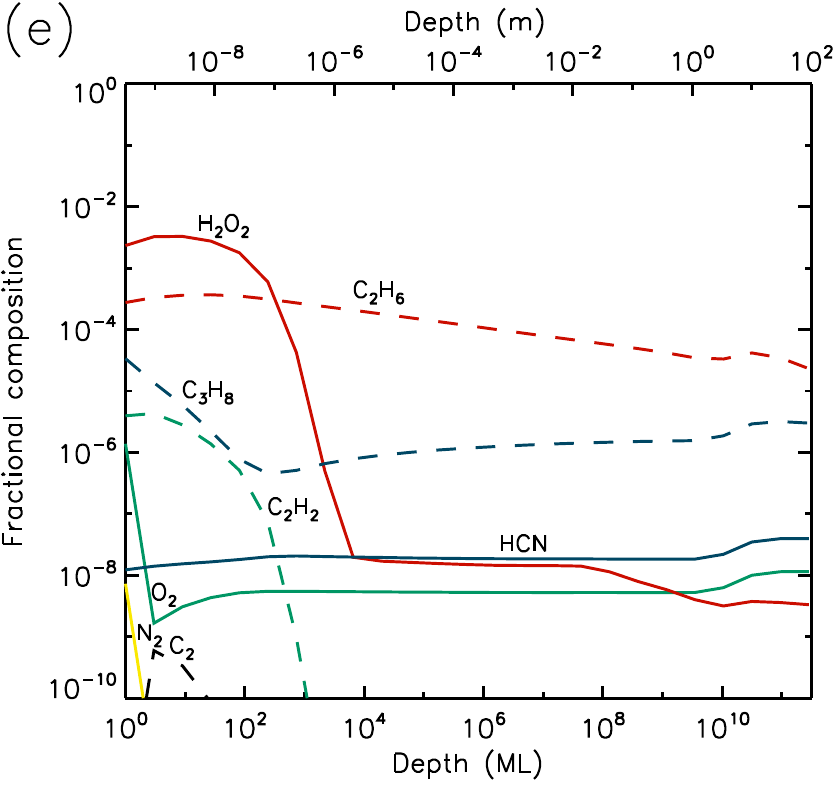}
\includegraphics[width=0.32\textwidth]{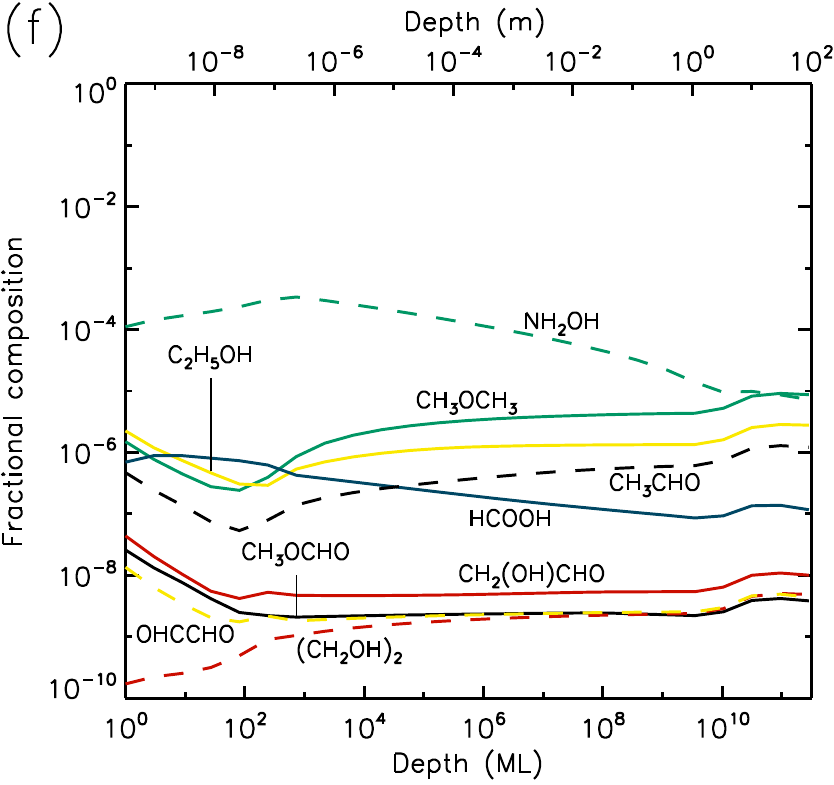}
\end{center}
\caption{\label{fig-30K-late} 30 K models at time $t=5 \times 10^{9}$ yr. Abundances of the initial ice components are shown in the left panels, with dust shown as a dotted line. Results for two model setups are shown: low UV (upper panels) and high UV (lower panels).}
\end{figure*}

\subsection{20~K models}

At 20~K (see Figs. \ref{fig-20K-early} \& \ref{fig-20K-late}), the effects of thermal mobility within the bulk ice are more noticeable. The region of maximum dust concentration extends much deeper and more uniformly for all UV-field strengths, with high dust abundances to depths of 1--10~mm at the end of the models. Thermal diffusion of CO within the ice becomes sufficiently rapid to allow it gradually to reach the surface layer, where it may be photo-desorbed; {\em thermal} desorption of CO is still less rapid than the corresponding photo-process by several orders of magnitude. However, for H$_2$, which is also able to reach the surface via diffusion, thermal desorption is indeed the dominant rate, and the production of H$_2$ within the bulk ice is still significant enough at 20~K for H$_2$ desorption to provide the greatest rate of loss of material from the surface (see Table \ref{tab-rates}).

Product molecules, both simple and complex, achieve relatively low abundances at 20~K, as compared with lower temperatures, because these molecules spread to much greater depths through either direct diffusion of their own, or by bulk-ice swapping with highly diffusive species. O$_2$ and N$_2$ in particular are able to maintain modest abundances throughout the ice by the end of the high-UV run. H$_2$O$_2$ and HCN are also abundant to depths of around 1~m.

Ethane (C$_2$H$_6$) becomes the dominant hydrocarbon in the ice at these temperatures, with a fractional abundance around 10$^{-4}$ near to the surface layer; this value is nevertheless somewhat less than the $\sim$10$^{-3}$ -- 10$^{-2}$ detected in comets. The C$_2$H$_6$ abundance appears fairly insensitive to the UV-field strength. The abundances of complex molecules like methyl formate (CH$_3$OCHO) are significantly lower than at 10~K, due to their spread to greater depths. The production of complex molecules is still occurring mainly in the upper 1$\mu$m of the ice, with the products then gradually diffusing to deeper layers. However, diffusion rates are high enough that production of new molecules through thermal diffusion of radicals in the ice has begun to dominate the chemistry (rather than immediate reaction with contiguous species, following photo-dissociation). The barrier against bulk diffusion for the CH$_3$ radical (823~K) in these models is somewhat lower than that for HCO (1120~K); the effect of this can be seen in the abundance ratios of CH$_3$- versus HCO-related product molecules such as CH$_3$OCHO and CH$_3$OCH$_3$, the latter of which attains abundances several times higher than the former. The achievement of these ratios is directly attributable to the change-over to a thermal diffusion-dominated bulk-ice chemistry. For such reasons, glycine abundance falls off at 20~K, as the radicals involved in its formation are not very mobile at these temperatures, and do not become mobile below around 50--60~K. 

It may be noted also that at 20~K, in the high-UV case, the production of CH$_3$O and CH$_2$OH in the upper 1~$\mu$m is dominated by direct photo-dissociation of methanol, rather than the addition/abstraction of an H-atom to/from H$_2$CO/CH$_3$OH. Production of HCO is also dominated by photo-dissociation of H$_2$CO, rather than the addition of H to CO, which delivers negligible amounts of HCO, due to the high thermal diffusion rate of atomic H competing with quantum tunneling through the activation energy barrier of the H + CO $\rightarrow$ HCO reaction.

\subsection{30~K models}

The 30~K models (Figs. \ref{fig-30K-early} \& \ref{fig-30K-late}) achieve strong dust enrichment to depths greater than 1~m by the end of the runs, and achieve maximum dust abundances to depths of 1~cm within 10$^6$ yr. Thermal desorption of small atoms and molecules such as H, H$_2$, O, and O$_2$ dominate the losses from the surface, but the precise values are strongly dependent on the UV-field strength of the model. Table \ref{tab-rates} shows that at a sustained temperature of  30~K, after 5 Gyr the comet has lost $\sim$13~m of ice. 

The strong degree of diffusion at these temperatures again results in low abundances of many product species, especially at early times. Some molecules, such as acetylene (C$_2$H$_2$), tend to survive only weakly at greater depths and late times, due to destruction by H atoms, which are highly mobile; acetylene is ultimately hydrogenated all the way to ethane (C$_2$H$_6$). On the other hand, hydrogen peroxide (H$_2$O$_2$) is not very diffusive, due to similar binding and diffusion characteristics to water itself. This means that H$_2$O$_2$ has its highest abundances in the upper 1$\mu$m where UV irradiation is strong. The more volatile initial ice constituents, especially CO and CH$_4$, desorb strongly at 30~K (although they do not dominate the total desorption rate), while CO may also react rapidly with OH to form CO$_2$ + H; as a result, CO$_2$ tends toward higher abundances in the shallower layers, over long timescales.

Although the local abundances of many molecules appear somewhat low in the 30~K models, their aggregated abundances through the ice are substantial, and are significantly higher for most species than in the 20~K models, as they reach to deeper layers of the cometary material.

\begin{figure*}
\begin{center}
\includegraphics[width=0.32\textwidth]{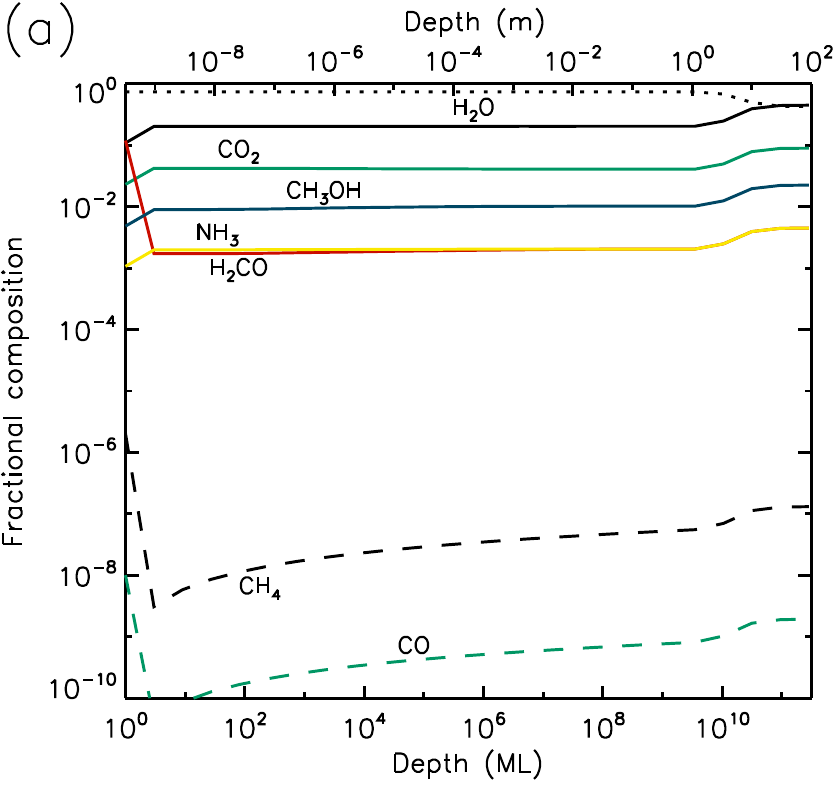}
\includegraphics[width=0.32\textwidth]{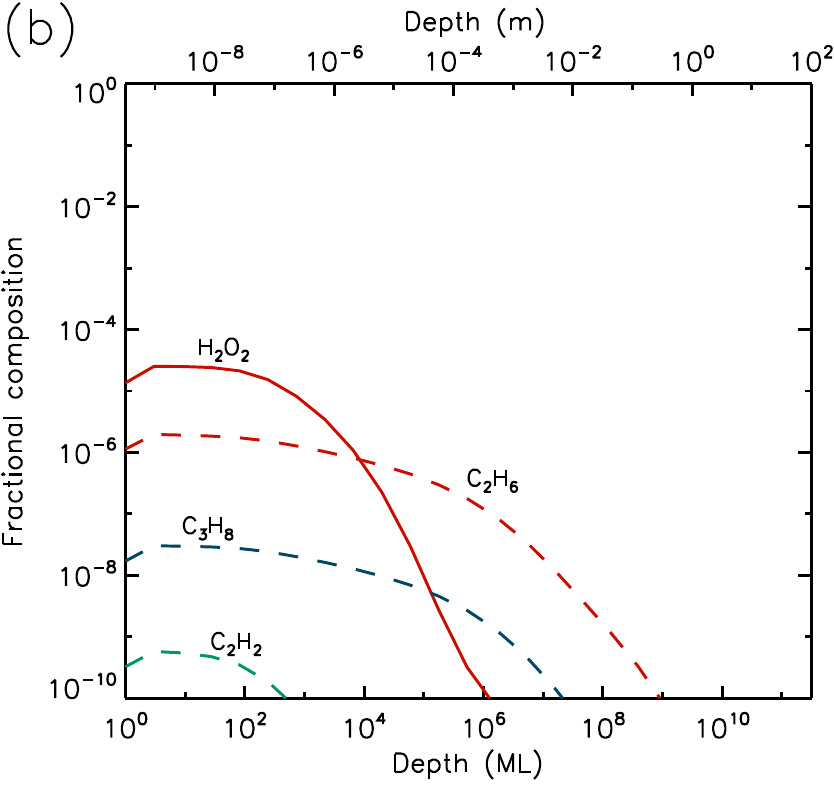}
\includegraphics[width=0.32\textwidth]{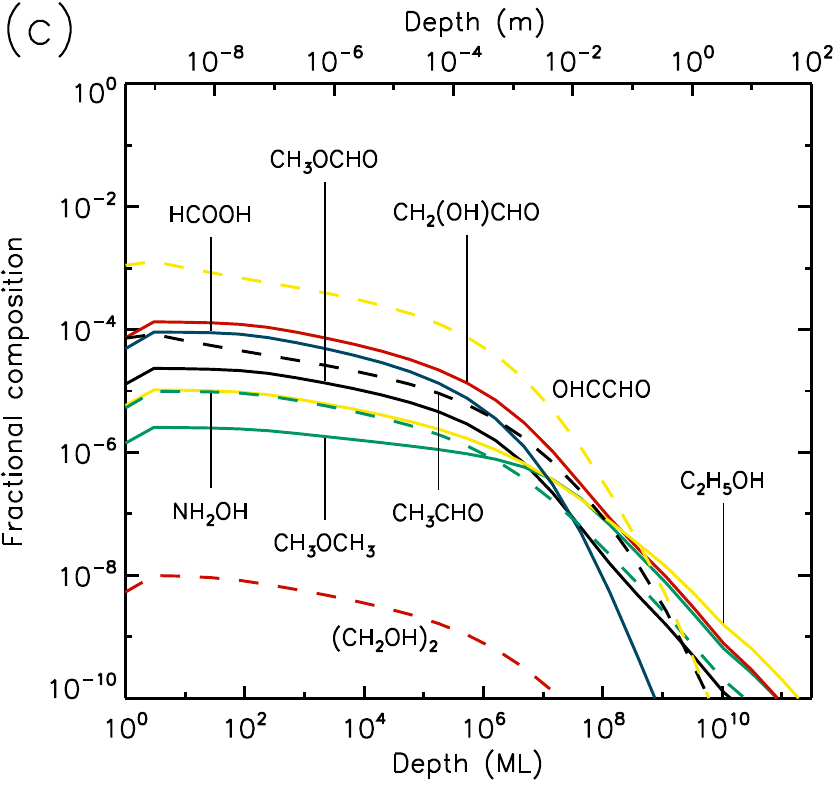}
\includegraphics[width=0.32\textwidth]{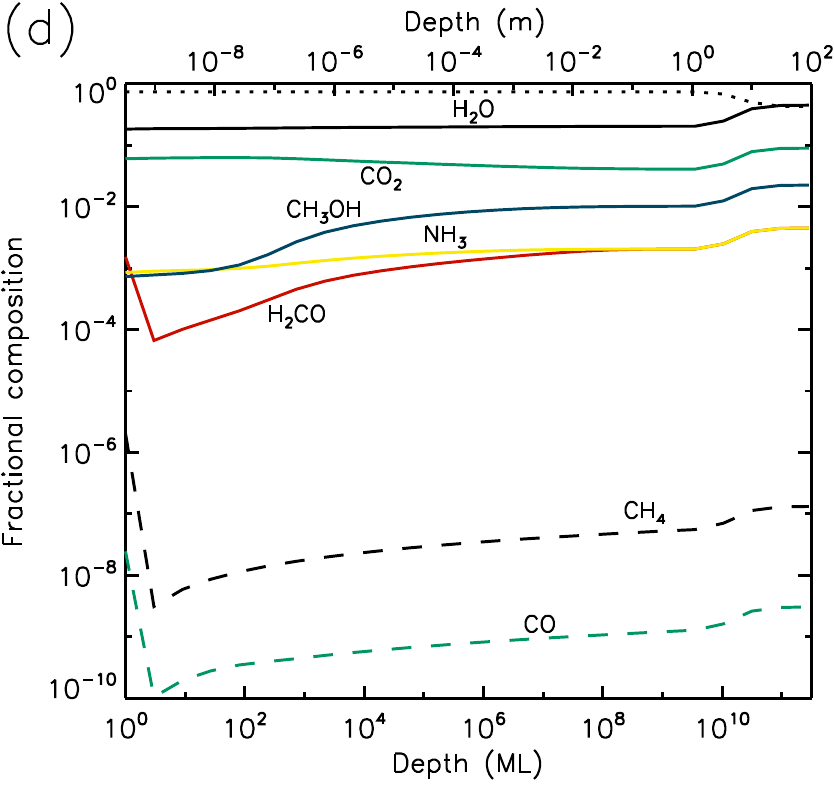}
\includegraphics[width=0.32\textwidth]{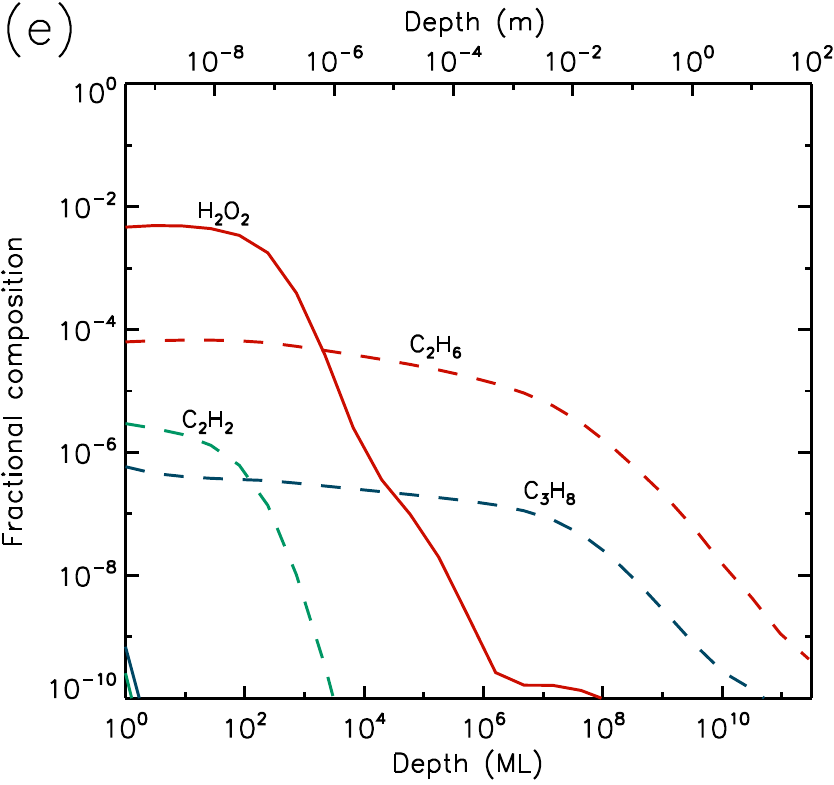}
\includegraphics[width=0.32\textwidth]{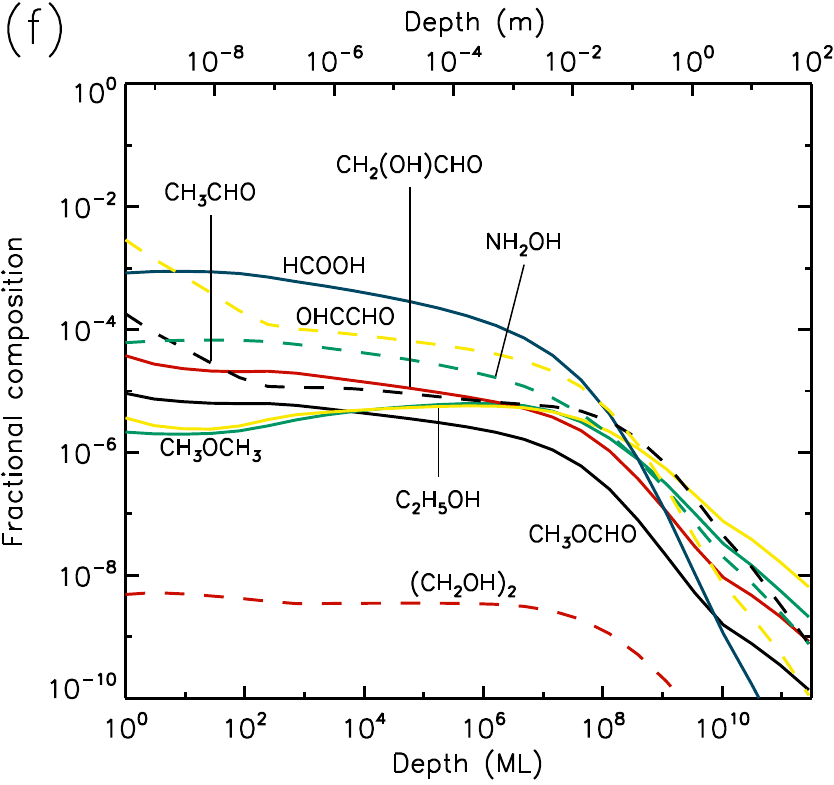}
\end{center}
\caption{\label{fig-40K-early} 40 K models at time $t=10^{6}$ yr. Abundances of the initial ice components are shown in the left panels, with dust shown as a dotted line. Results for two model setups are shown: low UV (upper panels) and high UV (lower panels).}
\end{figure*}

\begin{figure*}
\begin{center}
\includegraphics[width=0.32\textwidth]{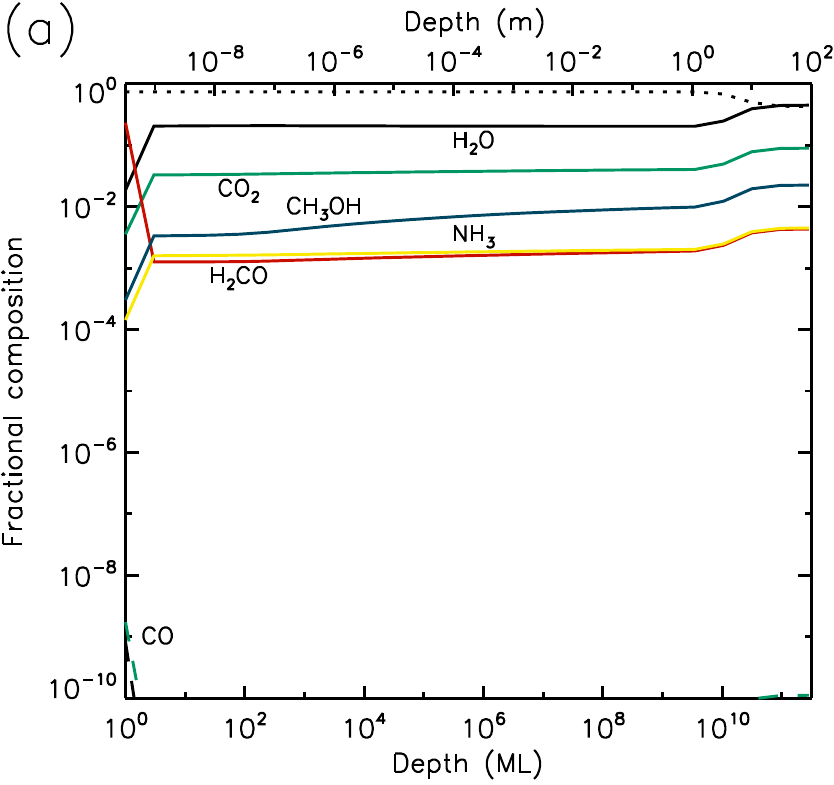}
\includegraphics[width=0.32\textwidth]{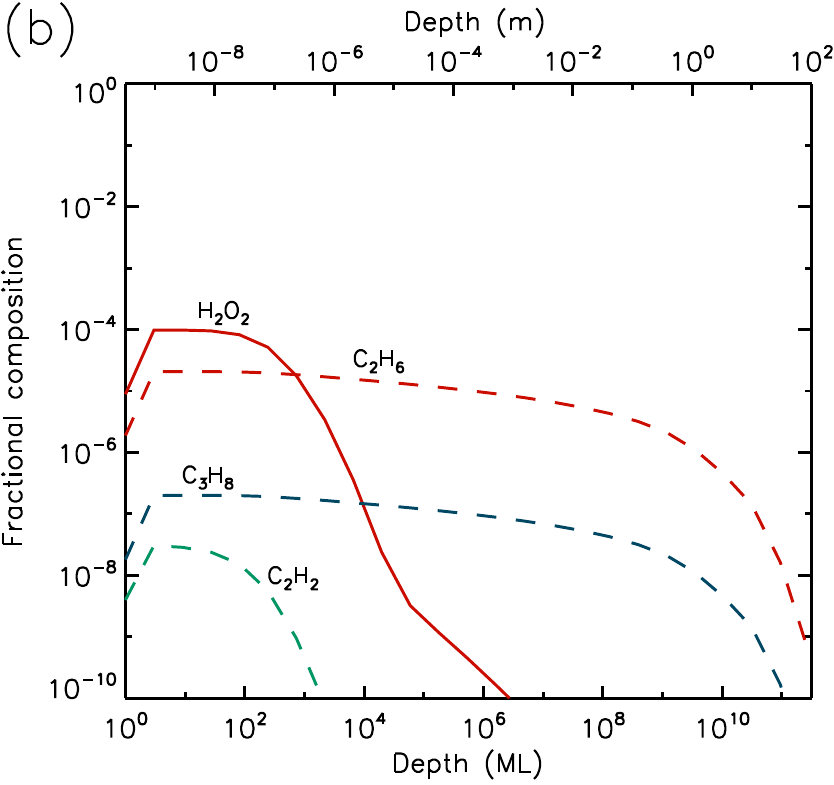}
\includegraphics[width=0.32\textwidth]{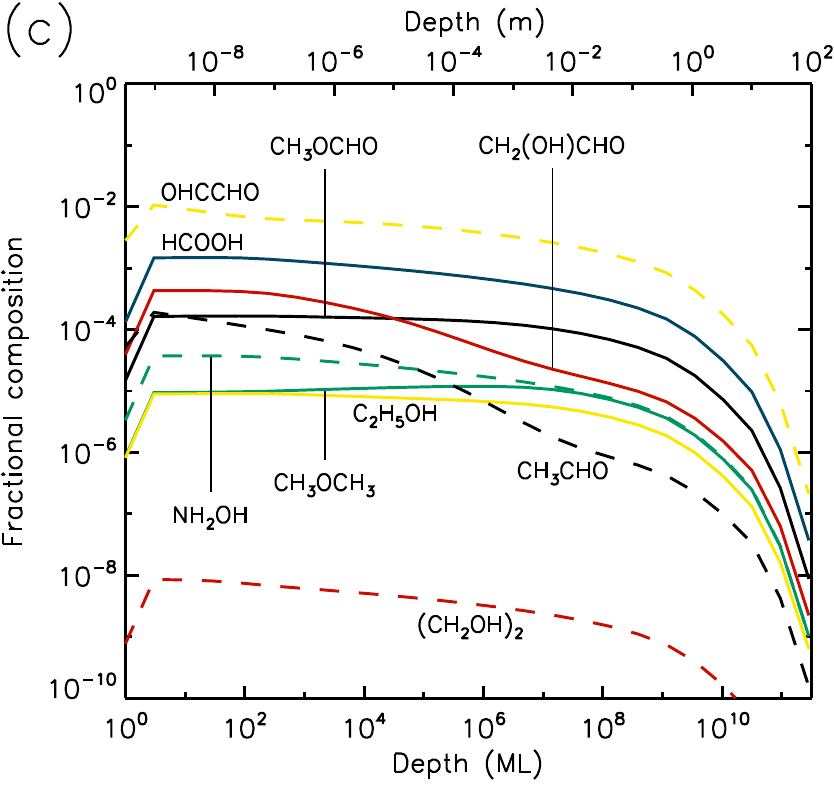}
\includegraphics[width=0.32\textwidth]{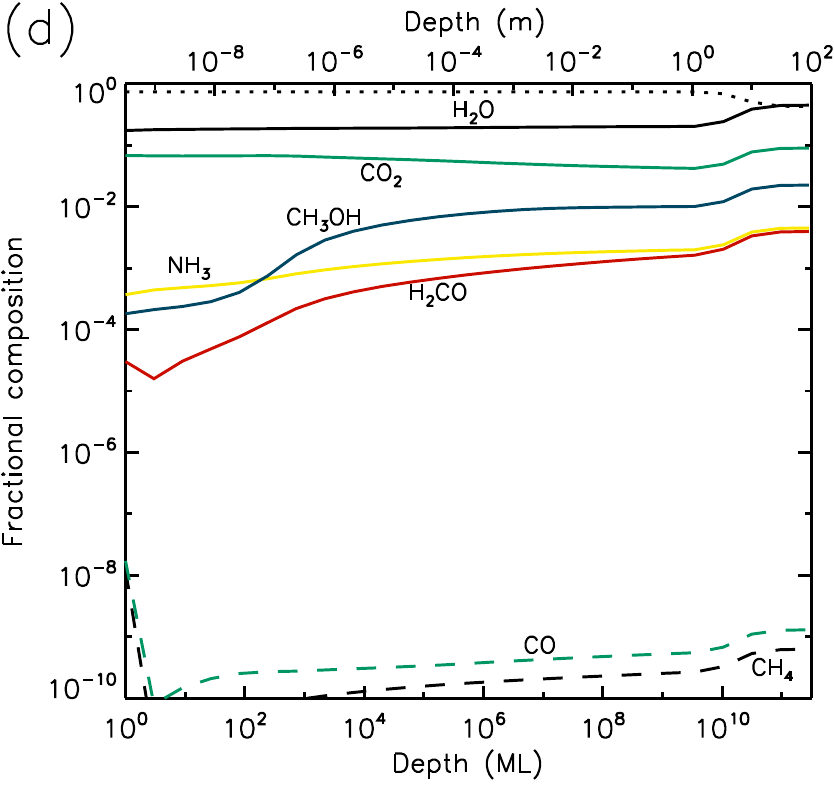}
\includegraphics[width=0.32\textwidth]{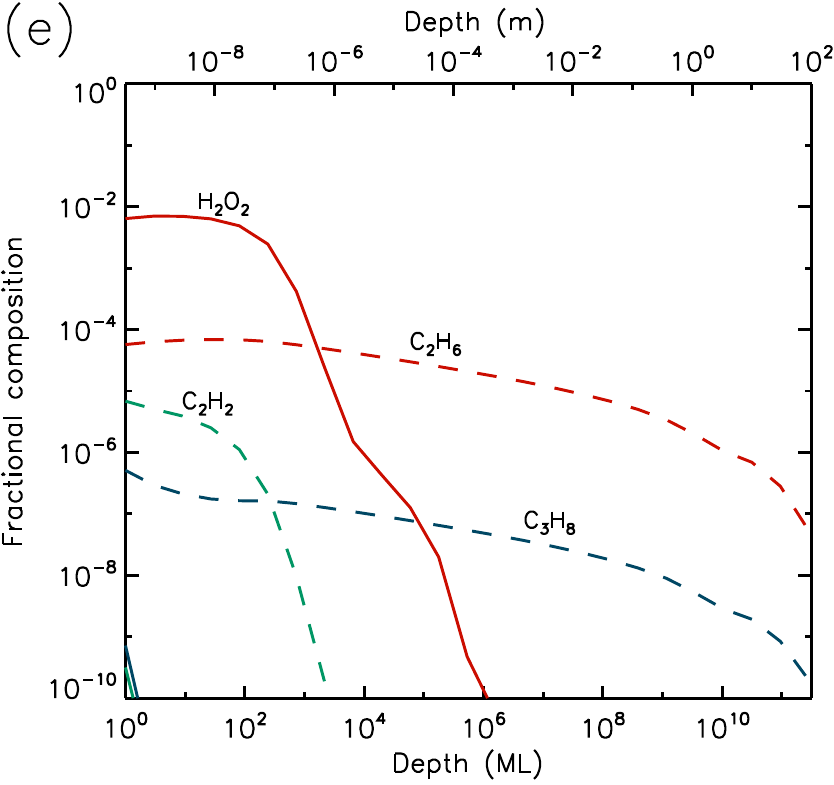}
\includegraphics[width=0.32\textwidth]{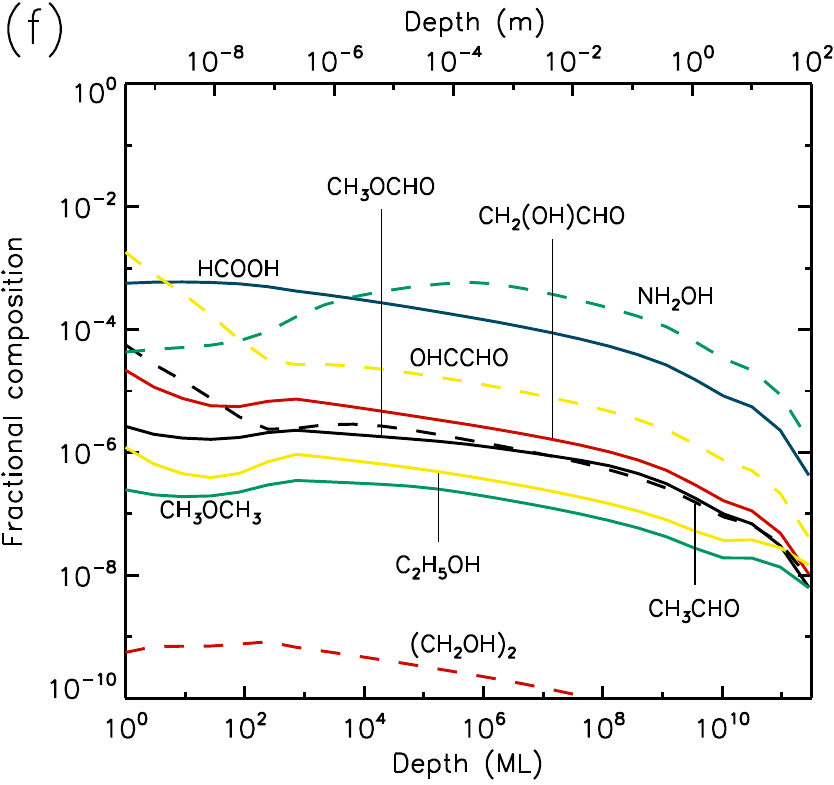}
\end{center}
\caption{\label{fig-40K-late} 40 K models at time $t=5 \times 10^{9}$ yr. Abundances of the initial ice components are shown in the left panels, with dust shown as a dotted line. Results for two model setups are shown: low UV (upper panels) and high UV (lower panels).}
\end{figure*}

\subsection{40--60~K models}

At 40 and 50~K, the loss of surface material is consistently dominated by H, H$_2$ and O desorption (Table \ref{tab-rates}). At 60~K, CO, CO$_2$, and H$_2$CO dominate. The total loss rates from the surface (corresponding to the {\em final} values) increase with temperature and UV-field strength. However, the overall loss of material over the course of each model is not substantially different for temperatures between 30--60~K. 

With increasing temperature, the abundances of CO, CH$_4$ and H$_2$CO become smaller due to diffusion, reaction, and surface desorption, so that by a time of 1 Myr in the 40~K models, CO and CH$_4$ abundances are negligible compared with the less volatile initial constituents. At a temperature of 60~K, all three of these species decrease substantially in abundance throughout the ice within the first 1000 years of model time.

The majority of complex organics are formed most efficiently around 40~K, although ethylene glycol, (CH$_2$OH)$_2$, is a notable exception; it is formed through the addition of CH$_2$OH radicals, which have a bulk-diffusion barrier in this model of 3556~K (see Garrod 2013). In the higher temperature models, the smaller radicals such as CH$_3$ and HCO become very diffusive and volatile, decreasing their abundances, as they are able to desorb from the nearby surface layer in preference to diffusing into the deeper layers. At 60~K, nevertheless, complex molecule abundances extend through every layer. Higher UV fields at high temperatures still have a noticeable effect on the ability to produce such molecules. At 60~K, glycine fractional abundance is around $10^{-8}$ for the high-UV case, while it is orders of magnitude lower for the low UV-field strength model.

It should be noted again that these high-temperature models, with fixed temperature, are run to a final time of 5 Gyr, for consistency with the other models. However, it is unlikely that stochastic heating events could produce these temperatures for such long timescales -- see Discussion.

\section{The influence of cosmic rays}

The results of the models for various temperatures and UV field strengths show that, at low temperatures, the effects of UV photo-dissociation can extend to depths of several hundred microns, firstly through the local UV-induced chemistry (in the upper 1~$\mu$m), and secondly as an indirect result of the loss of molecular material from the ice surface, which acts to enrich the dust content in the upper ice layers. Such effects would be important, for example, to the albedo of cold-storage comets unaffected by stochastic heating events. The higher ($\geq$20~K) temperature models produce results in which the higher mobility of bulk-ice species allows mixing of the chemical products into deeper layers in which the UV flux is weak.

However, comets will also be subject to a continuous flux of high-energy charged particles, either from interstellar space or directly from the Sun; the cosmic-ray energy distribution determined by Spitzer and Tomasko (1968) covers an energy range of 0.3 MeV to 100 GeV, with a peak value close to around 100 MeV. Oort Cloud bodies should be expected to experience the full spectrum of interstellar cosmic-ray fluxes. For Kuiper-belt objects, the spectrum of heliospheric protons is dominated by the interstellar contribution at particle energies greater than around 1 GeV (Cooper et al. 2003). The interaction cross-sections for such high-energy particles are small compared to UV photons, allowing them to penetrate to much greater depths into a comet. Unlike photons, cosmic rays can undergo multiple ionizing collisions with molecules in the target material, while giving up only a small fraction of their energy on each occasion. Experiments and models indicate that impingement of high-energy particles into both solids and liquids results in the ionization of the target material, producing energetic secondary electrons that can also cause many further ionizations (e.g. Johnson 1990). The effects of the interaction of cosmic rays with the ice (collectively referred to as radiolysis) will also include electronic/vibrational excitation, recombination of ions with liberated electrons, electron attachment to neutrals, and molecular dissociation resulting from each of these processes, as well as augmented reaction rates (Shingledecker, Le Gal \& Herbst, 2017).

Recent work by Shingledecker \& Herbst (2018), hereafter SH18, presented a methodology for the inclusion of ice radiolysis chemistry in standard interstellar gas-grain chemical models. That work was further explored in the context of cold interstellar clouds by Shingledecker et al. (2018). Crucially, those authors proposed in their model that cosmic ray-induced radiolysis could be treated accurately by assuming that any ionized species would rapidly recombine with electrons, to produce reactive dissociation products. Some products would be in an excited state, allowing fast reactions to occur with stable, neutral species already present in the ice, even where those reactions would usually be mediated by substantial activation energy barriers for reactants in the ground state. Unexcited dissociation products would be allowed to react thermally in the normal way with other radical species in the ice. The Shingledecker \& Herbst model presented rates averaged over all primary-ion and secondary-electron events in the ice, allowing time-averaged rates to be included within the framework of a typical rate equation-based chemical kinetics model. This method made use of a generic, excited form of each chemical species in the ice, to be simulated in parallel with the ground-state forms, and whose decay back to the ground state via quenching was assumed to be rapid. Shingledecker et al. (2018) found the radiolysis process to produce significant quantities of complex organics in dust-grain ice mantles over interstellar timecales.

While a full treatment of cosmic-ray impingement on cometary ices based on the approach of SH18 will be explored in detail in future work, an approximation to the resultant chemical processing has been made here by augmenting the rates of molecular {\em photo}-dissociation used in the model, to incorporate a contribution from cosmic ray-induced dissociation. Since the majority of the ice is composed of water, the calculation of cosmic ray-induced dissociation rates for all species is based on the detailed treatment for water molecules given below.

In order to include cosmic-ray processing in the comet model, the depth-dependence of the associated rates must be determined. The penetration depth of cosmic rays was not explicitly considered in the models of SH18 and Shingledecker et al. (2018), since the energy lost by penetration into a sub-micron-sized dust-grain ice mantle was assumed to be neglible. 

In the present models of comet chemistry, the results of a Monte Carlo simulation of cosmic-ray impingement into solid water by Atri (2016) are used to parameterize the depth-dependent dissociation rates. Those calculations were conducted using the GEANT4 Monte Carlo model, with a Galactic cosmic-ray proton energy spectrum reaching 1 TeV. Among several scenarios tested, Atri calculated the energy deposition rate at various depths into a planetary object that has no atmosphere and is composed of pure, solid water of density 1 g cm$^{-3}$. In fact, Jorda et al. (2016) determined the density of comet 67P to be $\sim$0.53 g~cm$^{-3}$, suggesting significant porosity. Some adjustment must therefore be made to scale the results of the energy-deposition simulations to a more appropriate density for a comet. Since penetration depths of high-energy particles can be scaled linearly with density\footnote{see e.g. https://physics.nist.gov/PhysRefData/Star/Text/PSTAR.html}, the depths corresponding to the energy-deposition rates obtained by Atri are simply scaled by a factor 1/0.53. Although the simulations of Atri (2016) assume a pure water ice, the presence of dust in the ice would be unlikely to change the result significantly, assuming the density were the same; penetration depths are mainly dependent on the overall density of the target, regardless of composition. (Note that in the associated chemical models that follow, the model retains the non-porous ice treatment of diffusion/reaction rates described in Sec. 2.1).

\begin{figure*}
\begin{center}
\includegraphics[width=0.8\textwidth]{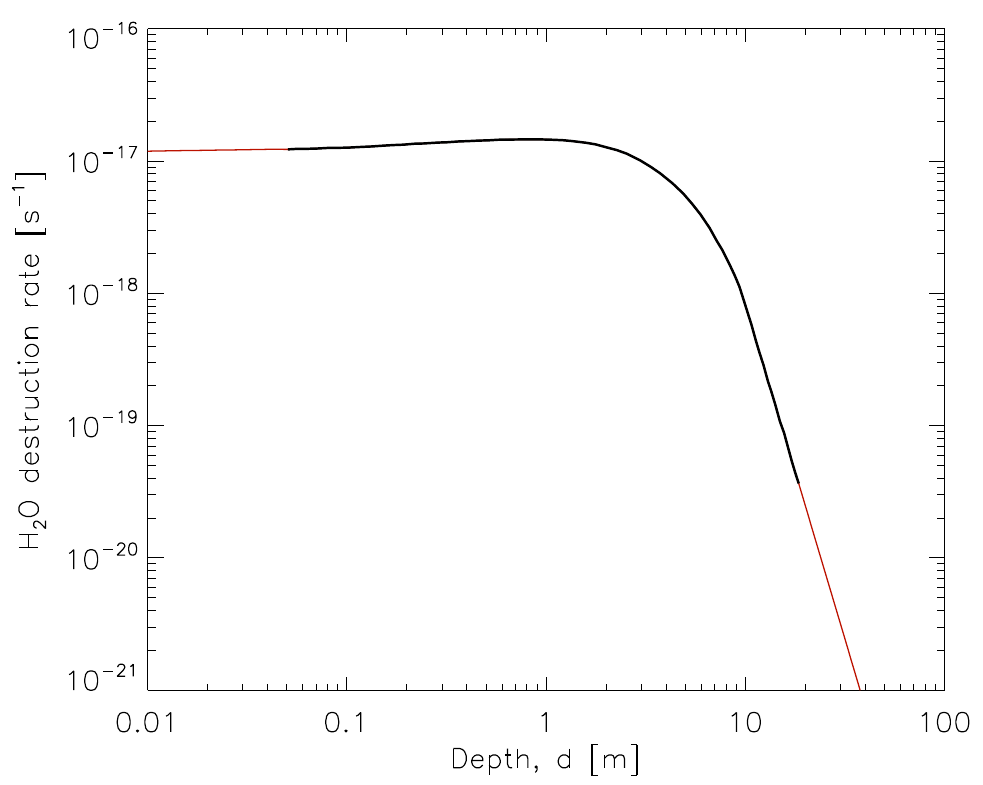}
\end{center}
\caption{\label{dest} Depth-dependent destruction rate of water molecules caused by cosmic ray impacts, as a function of depth, based on simulations by Atri (2016), scaled to an ice density of 0.53 g cm$^{-3}$. Extrapolated values are marked in red.}
\end{figure*}

To determine, using the adjusted energy-deposition rates, the water dissociation rates caused by cosmic rays at a given depth into the comet, it is here assumed that for every 27 eV lost by a cosmic ray to the target material (i.e. water ice), one water molecule is dissociated, following SH18 and Johnson (2011). The energy loss technically corresponds to the initial ionization process, but -- as per SH18 -- this is assumed to lead efficiently to recombination and thence dissociation. The resulting dissociation rate for water takes a value around $1.2 \times 10^{-17}$ s$^{-1}$ at a depth of 5 cm; the rate peaks at $1.6 \times 10^{-17}$ s$^{-1}$ at a depth of 85 cm, then falls off rapidly at greater depths. The depth-dependence of this profile at each extreme is well fit by a pair of power-law functions, which are used to extend the profile down to one monolayer depth, and up to 100~m (at which depth the dissociation rate is negligible). The resulting profile is plotted in Fig. \ref{dest}; the red line indicates the extrapolated portions.

For the purposes of the chemical model, each ice layer adopts the mean average value of the dissociation rate taken over its entire thickness. The average value used for the deepest ice layer, $4.7 \times 10^{-23}$, corresponds to the local value achieved at a depth of $\sim$81~m. For ice depths smaller than $\sim$1~m, the dissociation rates are all on the order of 10$^{-17}$ s$^{-1}$; the rate calculated in the outermost ice layer (of thickness 1 ML) is $8.0 \times 10^{-18}$ s$^{-1}$. 

The unattenuated UV photo-dissociation rate for water used in the chemical model (corresponding to H$_2$O + $h\nu$ $\rightarrow$ OH + H) is $3.28 \times 10^{-10}$ s$^{-1}$. The unattenuated cosmic ray-induced dissociation rate is thus a little over seven orders of magnitude smaller than for the equivalent UV process in the topmost layer of the ice (although the CR process maintains a substantial rate to much greater depths than the UV process). In order to apply the same general approach as for water to all molecular species in the model, every UV photo-dissociation rate in the network is given an additional CR-induced component based on its unattenuated UV rate and the calculated CR rate for H$_2$O, thus:

\begin{equation}
k_{\mathrm{CRD}}(i, d) = k_{\mathrm{CRD}}(\mathrm{H}_{2}\mathrm{O}, d)  \times \frac{ k_{\mathrm{UV}}(i, 0) }{ k_{\mathrm{UV}}(\mathrm{H}_{2}\mathrm{O}, 0) }
\end{equation}

\noindent where $k_{\mathrm{CRD}}(i, d)$ is the dissociation rate of species $i$ at depth $d$ into the comet.

It may be noted that simply modifying the rates of photo-dissociation to include a CR-induced component in the chemical model is equivalent to the assumption of all dissociation products being in the ground state, thus no supra-thermal reactions are explicitly considered. However, due to the inclusion (in the comet model) of immediate reactions following photo-dissociation (Sec. 2.5), something akin to the reaction of excited radicals with pre-existing ground-state radicals included by Shingledecker et al. (2018) may be achieved. Furthermore, unlike the present comet model, the model of Shingledecker et al. did not apparently include the possibility of immediate reaction between unexcited dissociation products and pre-existing ground-state radicals in the ice. This mechanism may provide a somewhat higher efficiency of reaction for radiolysis at low temperatures than suggested by those authors.

The applicability of branching ratios and relative production rates associated with photodissociation as a stand-in for explicit values relating to cosmic-ray impingement in this simple treatment is borne out to some degree by experimental work (Gerakines, Moore \& Hudson 2001), which suggests that the products of UV and CR processing do not vary widely, although modest differences exist in some cases, notably CO.

\begin{figure*}
\begin{center}
\includegraphics[width=0.32\textwidth]{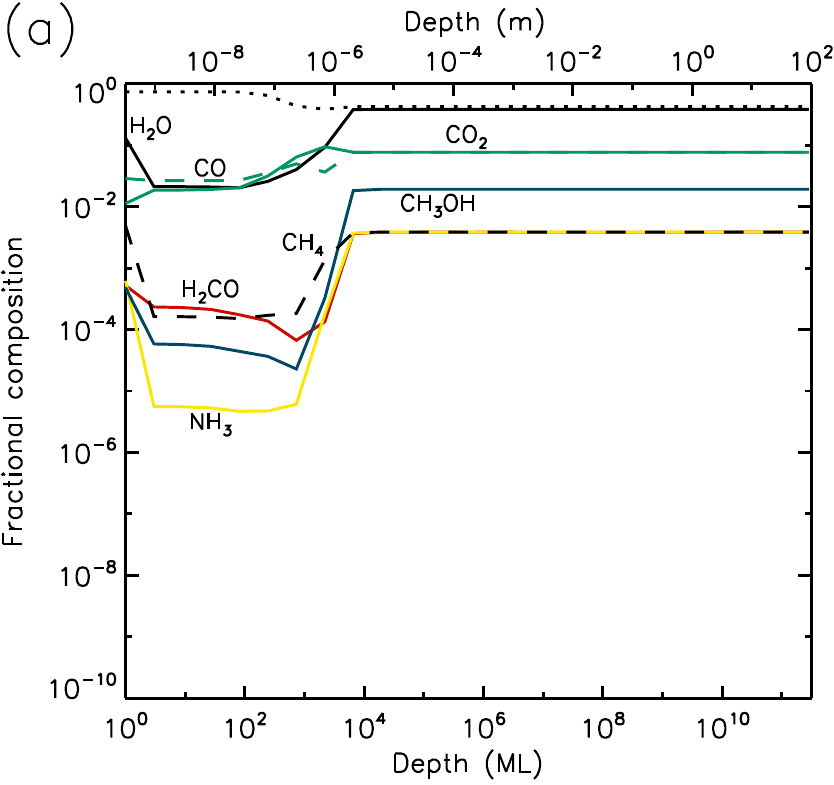}
\includegraphics[width=0.32\textwidth]{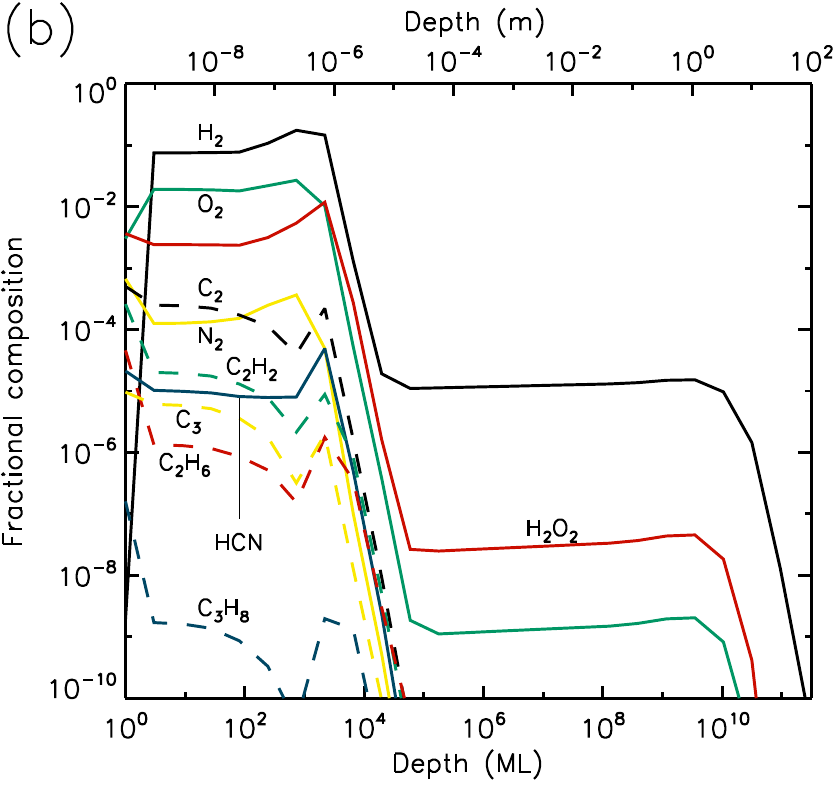}
\includegraphics[width=0.32\textwidth]{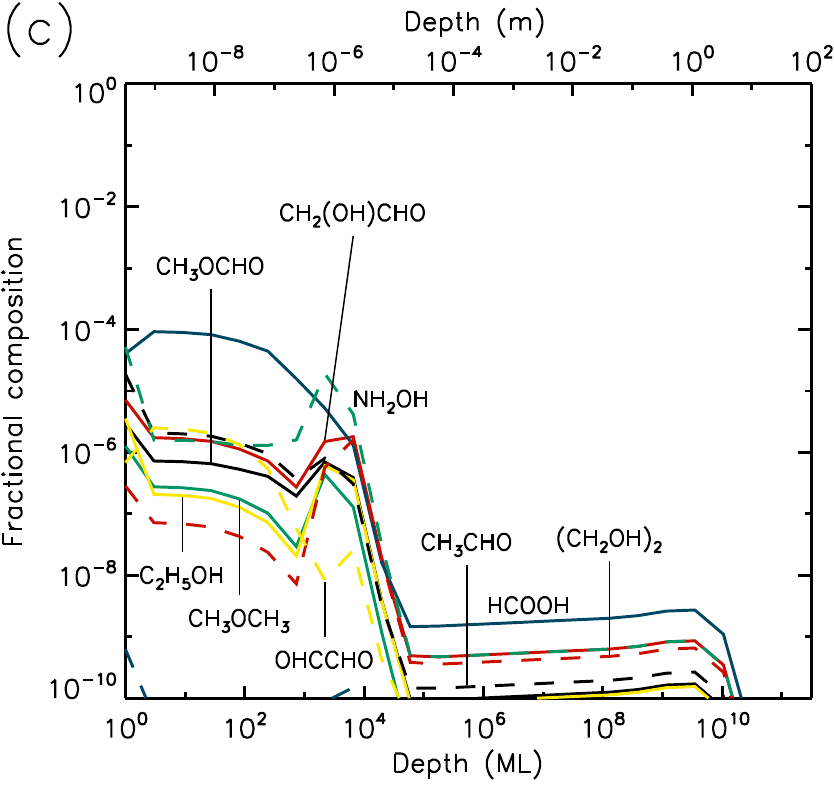}
\includegraphics[width=0.32\textwidth]{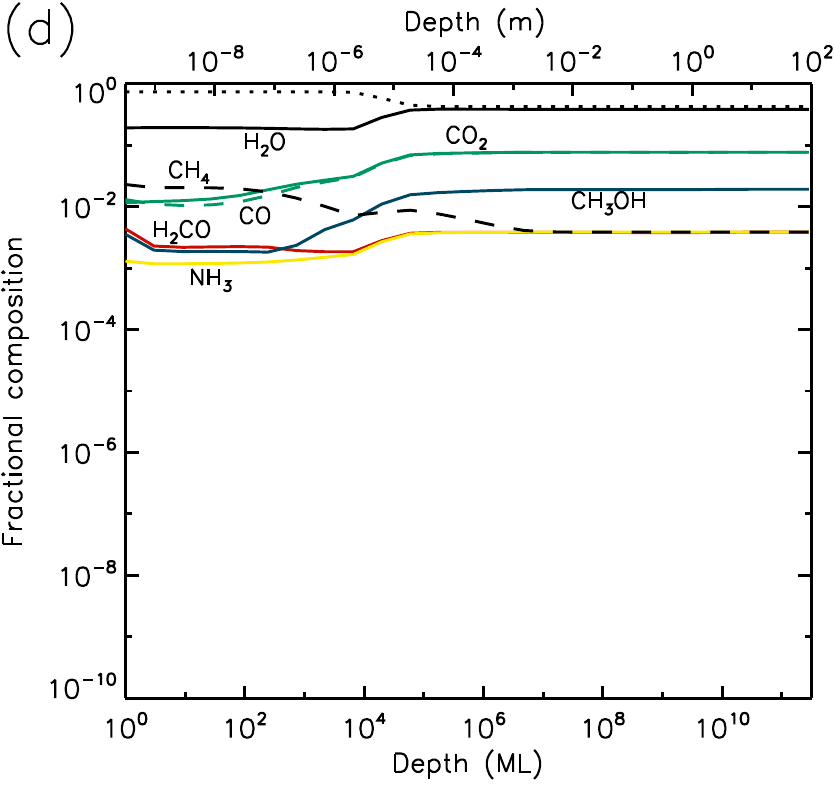}
\includegraphics[width=0.32\textwidth]{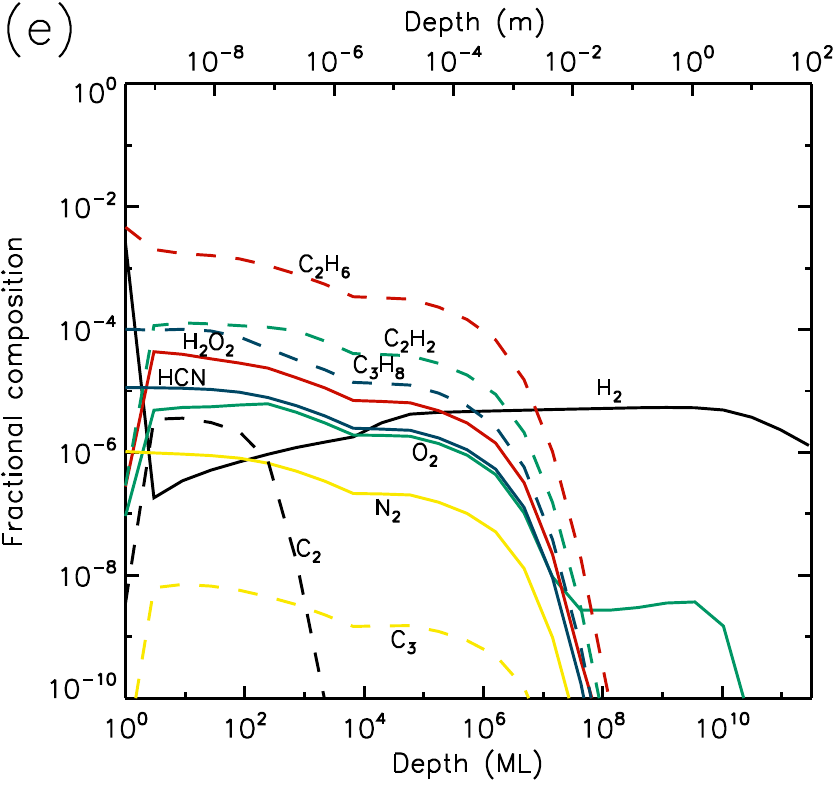}
\includegraphics[width=0.32\textwidth]{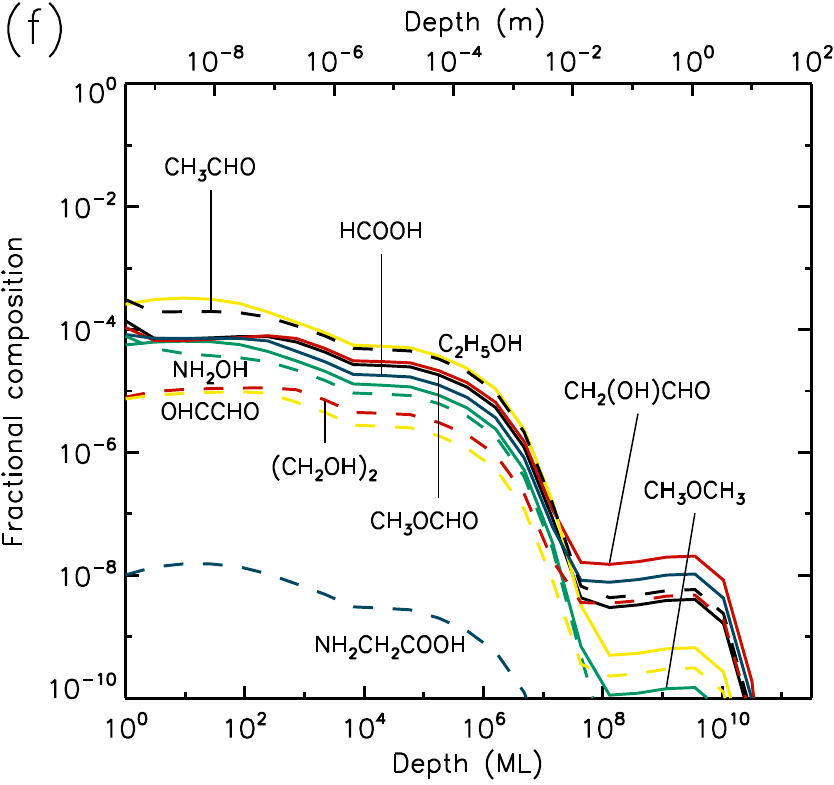}
\includegraphics[width=0.32\textwidth]{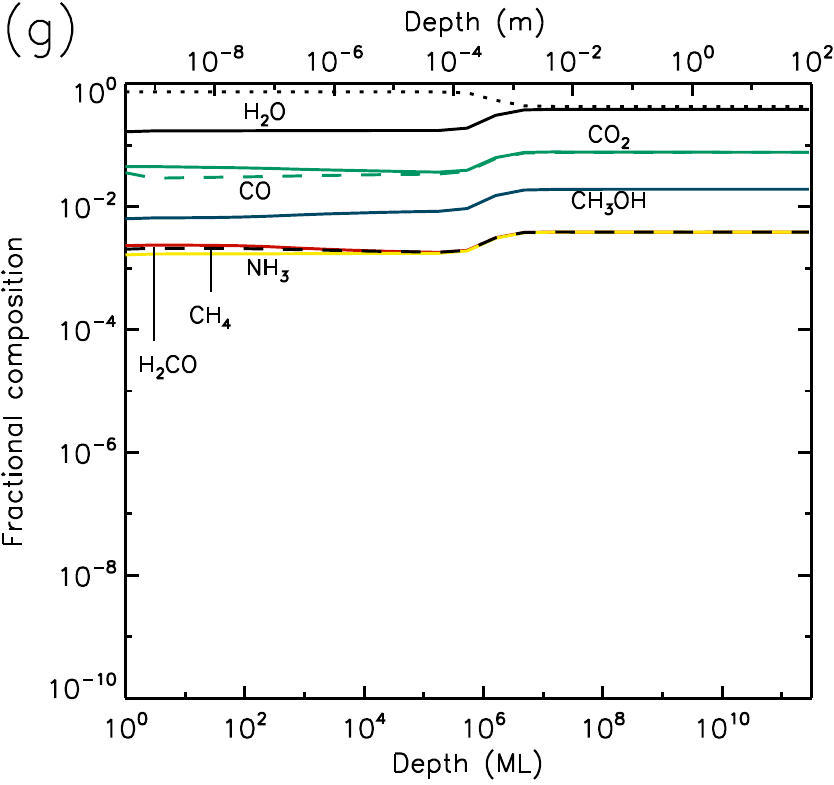}
\includegraphics[width=0.32\textwidth]{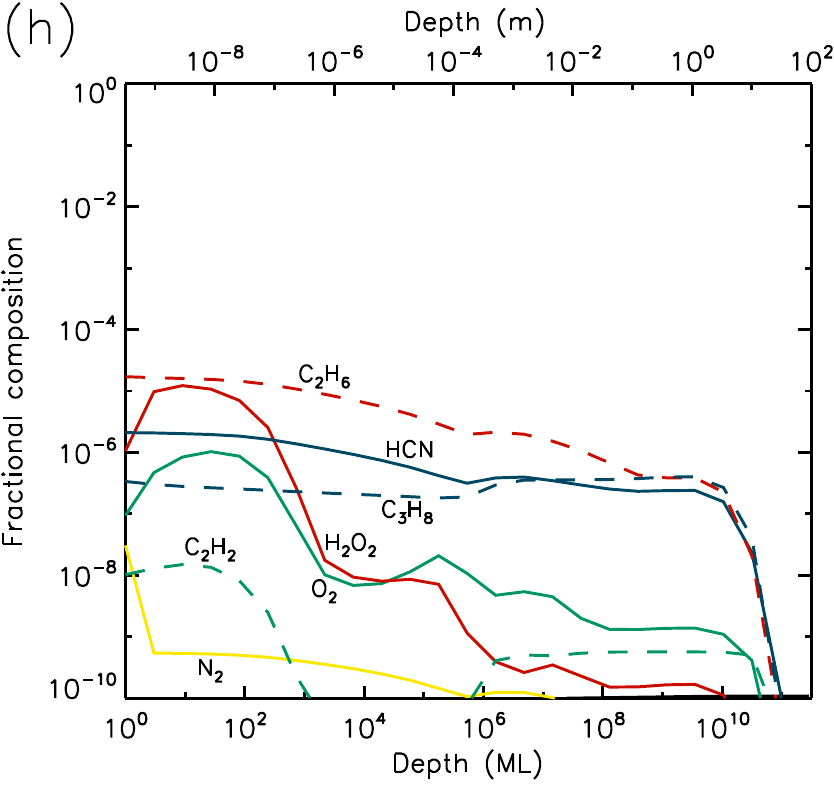}
\includegraphics[width=0.32\textwidth]{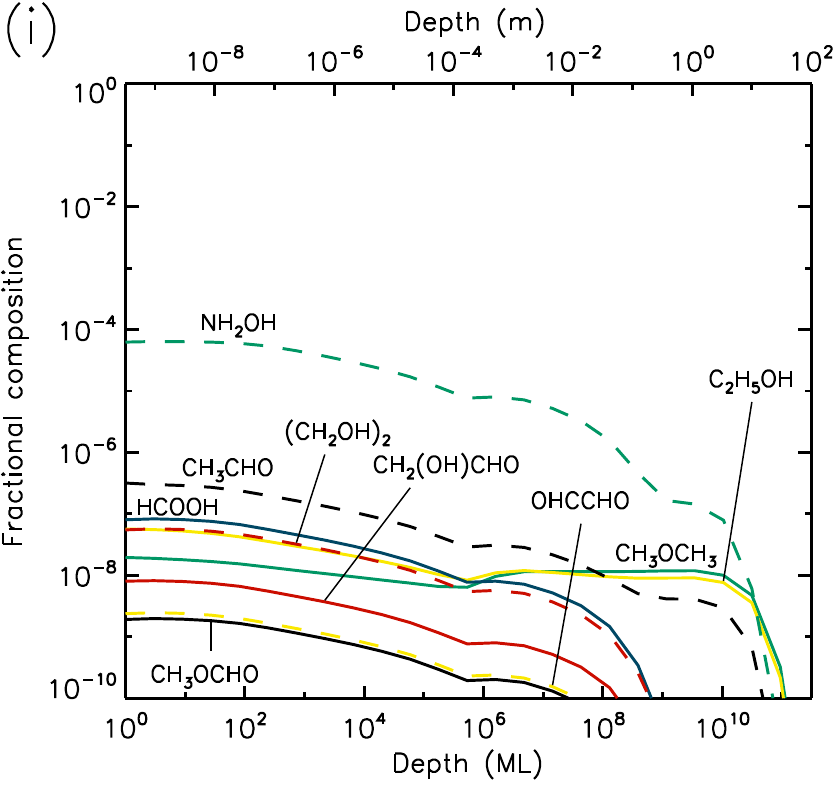}

\end{center}
\caption{\label{fig-CR-early} Chemical abundances for cosmic-ray dissociation models at time $t=10^{6}$ yr, assuming a temperature of 5~K (upper panels), 10~K (middle panels), and 20~K (lower panels). Abundances of the initial ice components are shown in the left panels, with dust shown as a dotted line.}
\end{figure*}

\begin{figure*}
\begin{center}
\includegraphics[width=0.32\textwidth]{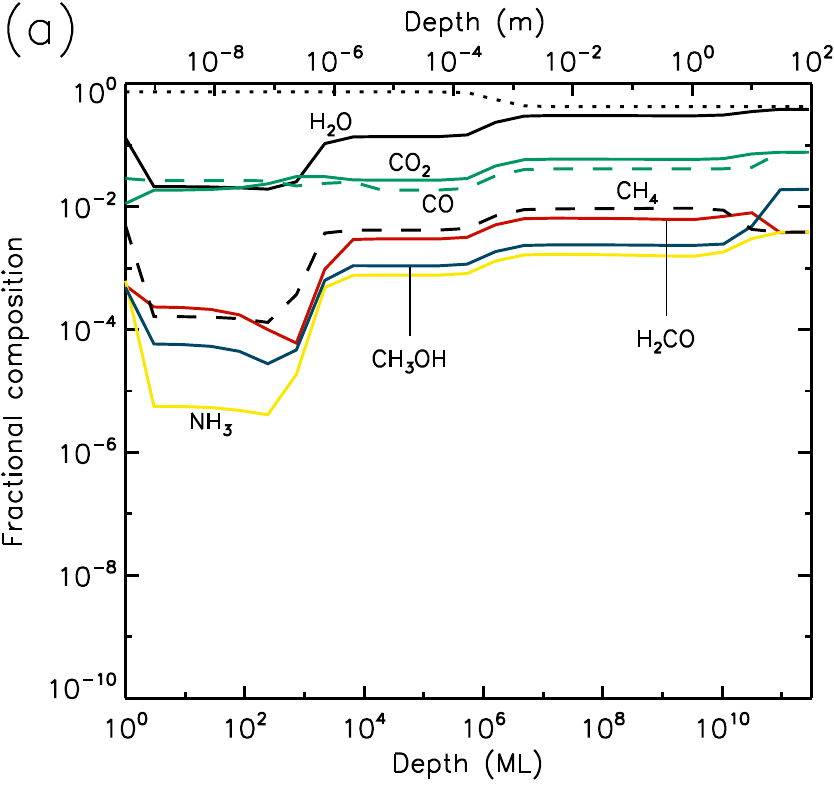}
\includegraphics[width=0.32\textwidth]{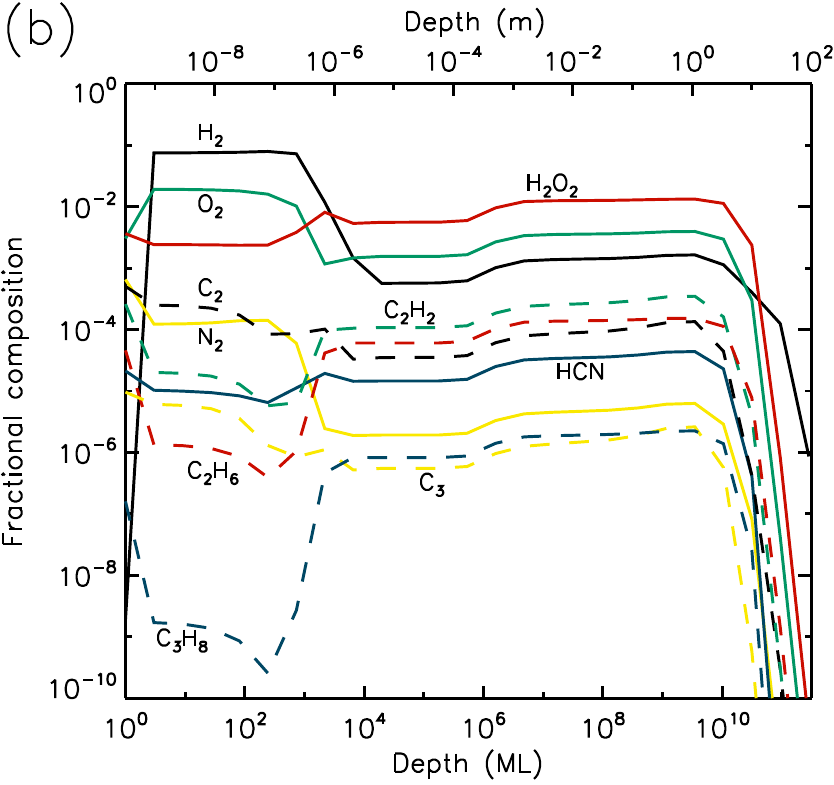}
\includegraphics[width=0.32\textwidth]{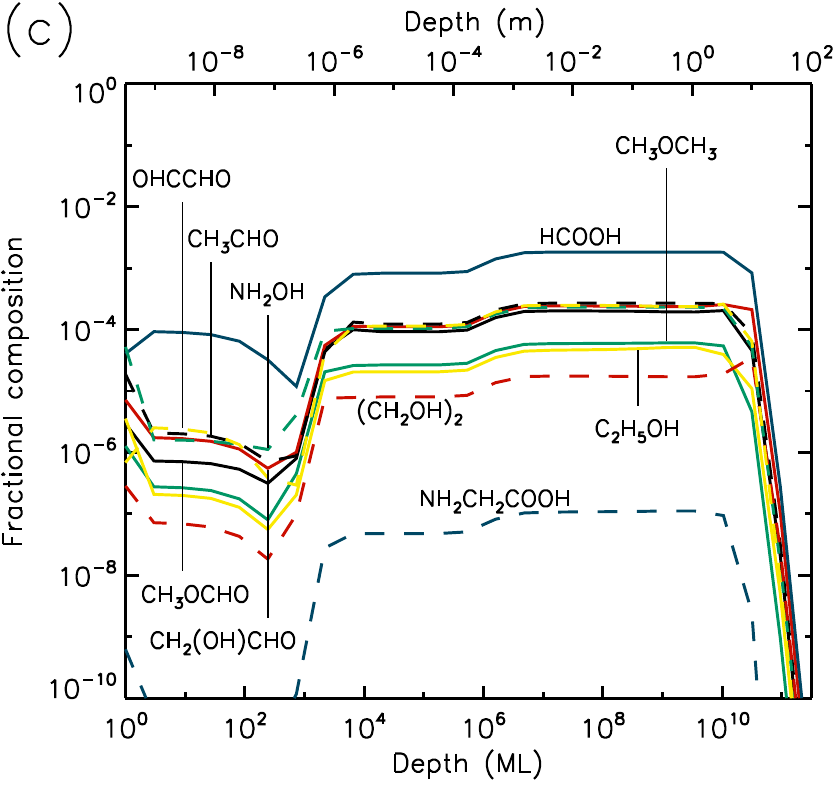}
\includegraphics[width=0.32\textwidth]{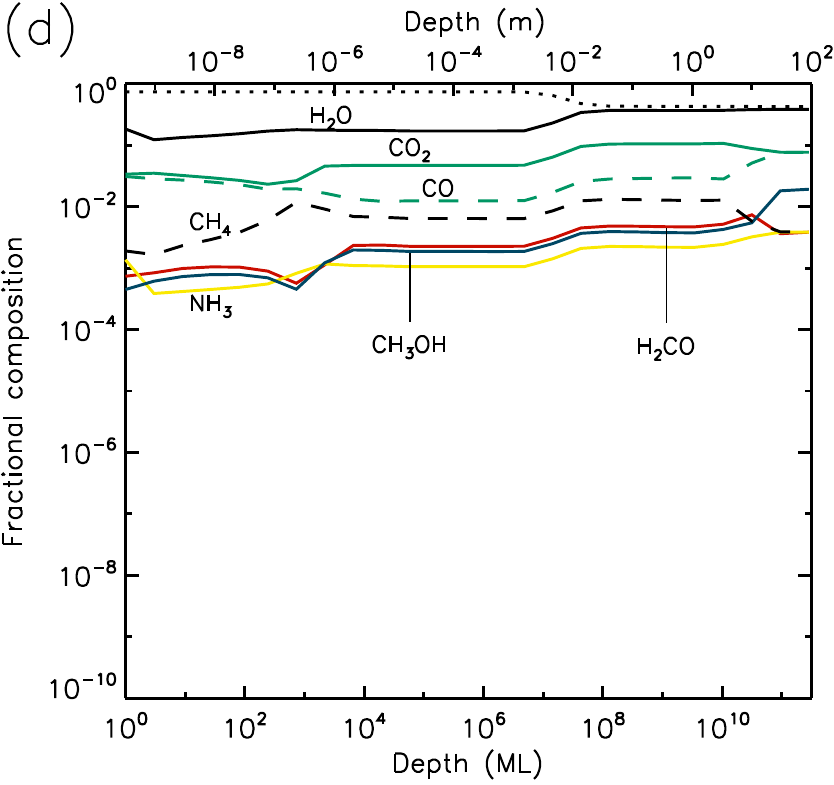}
\includegraphics[width=0.32\textwidth]{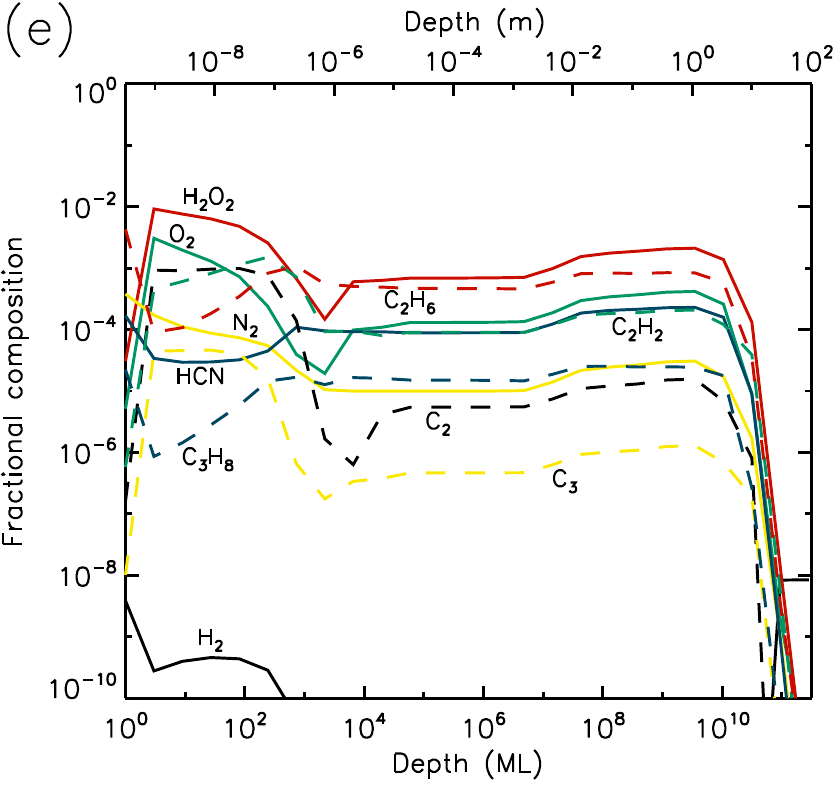}
\includegraphics[width=0.32\textwidth]{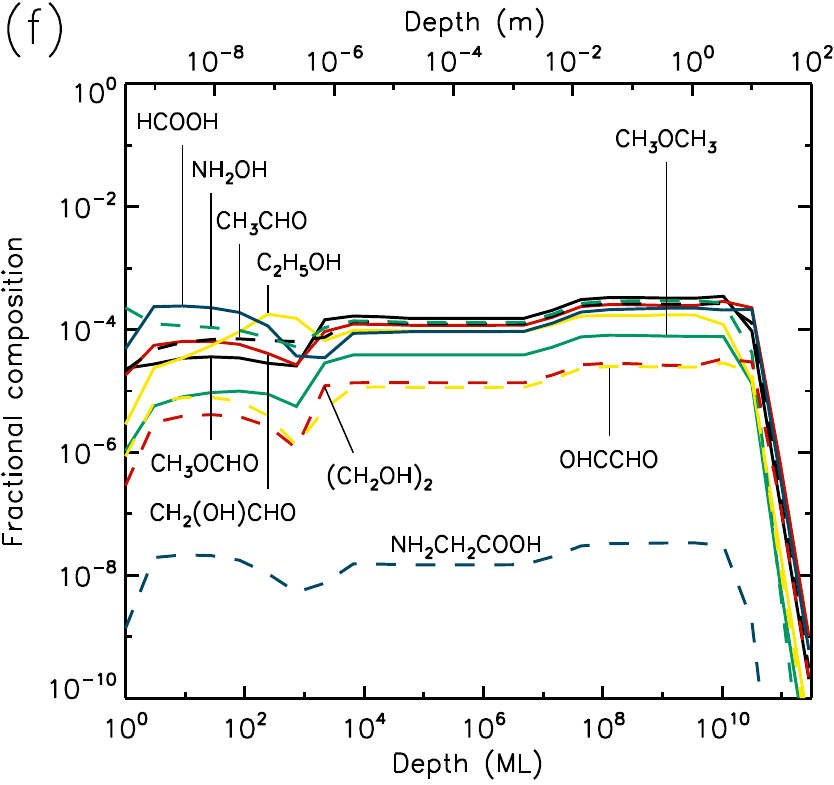}
\includegraphics[width=0.32\textwidth]{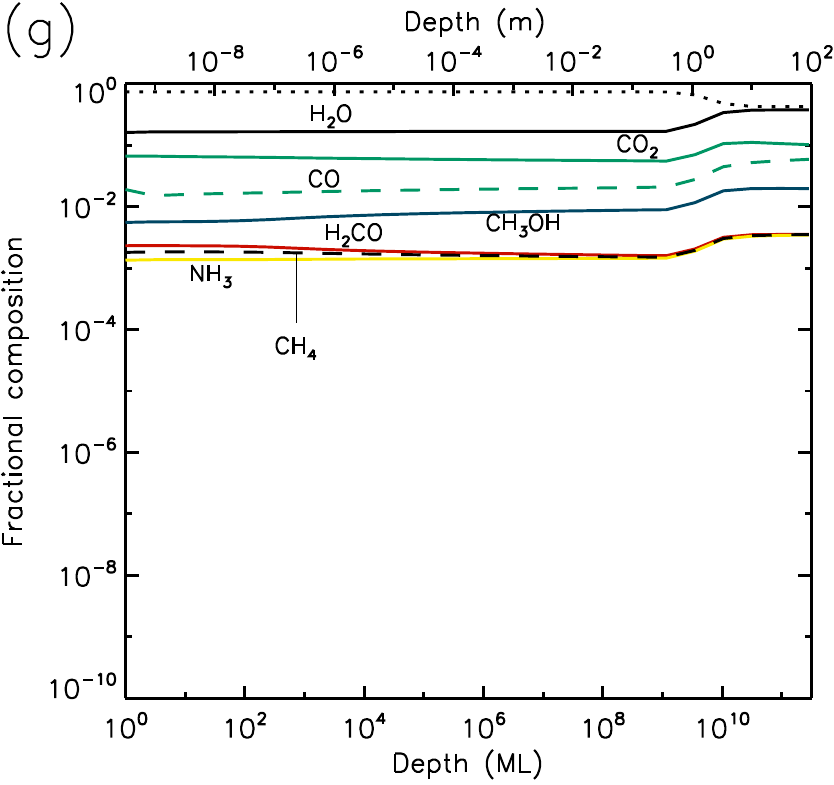}
\includegraphics[width=0.32\textwidth]{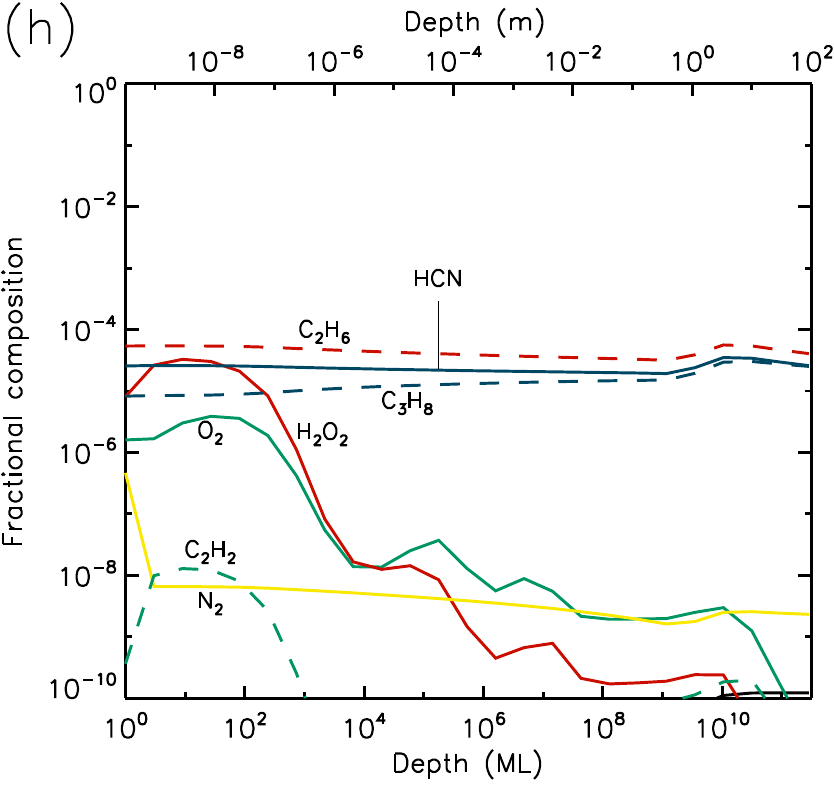}
\includegraphics[width=0.32\textwidth]{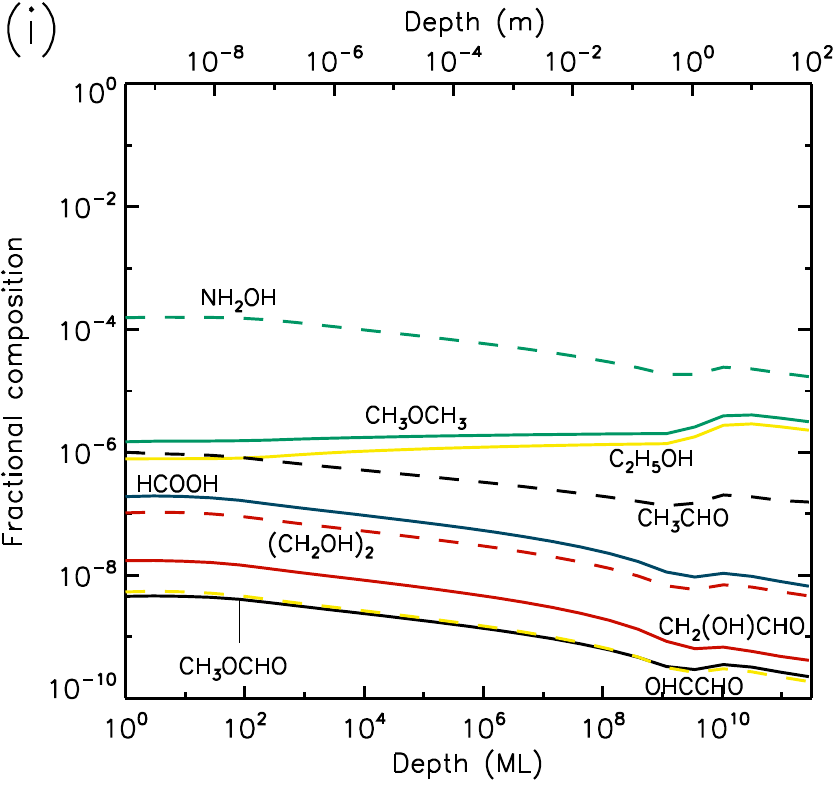}
\end{center}
\caption{\label{fig-CR-late} Chemical abundances for cosmic-ray dissociation models at time $t=5 \times 10^{9}$ yr, assuming a temperature of 5~K (upper panels), 10~K (middle panels), and 20~K (lower panels). Abundances of the initial ice components are shown in the left panels, with dust shown as a dotted line.}
\end{figure*}

\subsection{Results}

Figures \ref{fig-CR-early} and \ref{fig-CR-late} show results from the cosmic-ray models at times 10$^6$ yr and $5\times10^9$ yr, respectively. The upper row of panels (a--c) correspond to a temperature of 5~K, the middle panels (d--f) to 10~K, and the lower panels (g--i) to 20~K. 

 
The low temperatures in the cosmic-ray models mean that the rates of thermal diffusion are fairly small, especially at 5~K, thus reaction rates are dominated by the immediate reaction of radicals following their production via the dissociation of parent species (whether through UV or cosmic-ray processing). In the upper micron of material, the chemical behavior is similar to the models with only UV processing included;  for many molecules a step-like feature is seen, as the UV-induced dissociation drops off. However, at greater depths, various product species take on non-negligible abundances, even after as little as 1 Myr (Fig. \ref{fig-CR-early}), due to the weaker -- but more penetrating -- CR-induced dissociation. At 5~K, some complex organics reach fractional abundances of around $10^{-10}$ to depths on the order of 10~m at this time. In the 10~K model, complex organics have somewhat higher abundances at those depths, while their abundances become as high as $10^{-5}$ to depths of 1~mm. Such behavior is not seen in the UV-only models, either in the 5~K or the 10~K case. The enhancements down to $\sim$1~mm, and the weaker enhancement to $\sim$10~m, are the result of the cosmic-ray processing alone. At 20~K, the behavior is more comparable with the UV-only models to depths of around 1~mm, but the enhancements seen to depths $>$10~m are unique to the cosmic-ray model. At $10^6$ yr, this model produces complex organic molecule fractional abundances on the order of $10^{-8}$ to depths of around 10~m. 

Abundances of H$_2$, O$_2$ and H$_2$O$_2$ are also noticeably enhanced at a time of $10^6$ yr in the 5~K and 10~K models with cosmic-ray processing. In the 10~K and 20~K models, the abundances of hydrocarbons are enhanced at greater depths also, although the hydrocarbon abundances at depths less than 1~$\mu$m are suppressed versus the UV-only models.

At a time of 5 Gyr in the 5~K models, H$_2$O$_2$ is very abundant to depths of around 10~m or more, while O$_2$ maintains a fractional abundance greater than $10^{-3}$ to a depth of several meters, and greater than $10^{-4}$ to a depth of 10~m. H$_2$ is abundant throughout the ice, and maintains an abundance greater than $10^{-4}$ to several tens of meters. In the 10~K models, O$_2$ and H$_2$O$_2$ are somewhat lower in abundance, but retain the same general behavior, while H$_2$ becomes a very minor component of the ice, versus the 5~K model. All three species achieve only very small abundances at 20~K. Meanwhile, HCN and the saturated hydrocarbons C$_2$H$_6$ and C$_3$H$_8$ retain a consistent abundance around 10$^{-4}$--10$^{-5}$ at all three temperatures at the end of the cosmic-ray models. N$_2$ maintains an abundance of at least 10$^{-5}$ to a depth of a few meters in the 5~K and 10~K models, but is lower in the 20~K model.

The production of O$_2$ at large depths into the ice, where only cosmic-ray processing is important, occurs through the dissociation of O$_2$H (which is formed via the reaction between O and OH), and by direct addition of oxygen atoms. OH radicals are produced by the dissociation primarily of water, but also H$_2$O$_2$, while atomic oxygen is formed by dissociation of CO$_2$, OH, CO and several other species, including O$_2$ itself.

At 5 Gyr, many complex molecules are quite abundant throughout the 5 and 10~K ices, maintaining fractional abundances as high as 10$^{-8}$ down to around 10~m. Values as high as 10$^{-4}$ are achieved to these depths for many species. Glycine attains a maximum abundance around 10$^{-8}$, to a depth of a few meters. Local abundances of the most complex organics become significantly lower in the 20~K model, caused by the gradual thermal mixing of molecules into the deepest layers over the full 5 Gyr timescale; however, as a result, substantial abundances of these molecules are maintained to the very greatest depths sampled in the models. By the end-time of the cosmic-ray models, the larger hydrocarbon species and many other complex organics tend to achieve lower abundances in the upper ice layers than they do at depths greater than 1 micron; the combined UV and cosmic ray-induced destruction of these species is rapid in the upper layers.


\begin{deluxetable}{lrrrrrrrr}
\tabletypesize{\small}
\tablecaption{\label{tab-60K-hi} Abundances with respect to H$_2$O for UV-only 60K model with high UV}
\tablewidth{0pt}
\tablehead{ \colhead{Species}  & \colhead{$0$ yr} & \colhead{$10^3$ yr} & \colhead{$10^6$ yr} & \colhead{$10^9$ yr}  & \colhead{$5 \times 10^9$ yr}  & \colhead{Max.}  & \colhead{Hale-Bopp $^{a}$}  &  \colhead{67P/CG $^{b}$} }
\startdata
               Dust  &      1.1(+0)  &      1.7(+0)  &      1.7(+0)  &      1.8(+0)  &      1.9(+0)  &      1.9(+0)  &           --  &           -- \\
                  CO &      2.0(-1)  &      4.1(-8)  &      2.2(-9)  &     3.1(-11)  &     2.6(-11)  &      2.0(-1)  & 1.2--2.3(-1)  &      1.2(-2) \\
              CO$_2$ &      2.0(-1)  &      2.0(-1)  &      2.0(-1)  &      1.9(-1)  &      1.4(-1)  &      2.0(-1)  &        6(-2)  &           -- \\
              CH$_4$ &      1.0(-2)  &      8.4(-9)  &     4.6(-10)  &     6.8(-12)  &     5.1(-12)  &      1.0(-2)  &      1.5(-2)  &        5(-3) \\
             H$_2$CO &      1.0(-2)  &      9.8(-3)  &      5.0(-4)  &      4.9(-7)  &      3.2(-7)  &      1.0(-2)  &      1.1(-2)  &           -- \\
            CH$_3$OH &      5.0(-2)  &      5.0(-2)  &      5.0(-2)  &      4.3(-2)  &      2.0(-2)  &      5.0(-2)  &      2.4(-2)  &           -- \\
              NH$_3$ &      1.0(-2)  &      1.0(-2)  &      1.0(-2)  &      9.0(-3)  &      5.8(-3)  &      1.0(-2)  &        7(-3)  &           -- \\
\hline
               O$_2$ &           --  &           --  &           --  &     2.0(-14)  &     1.5(-14)  &     2.0(-14)  &           --  &  3.8(-2)$^c$ \\
          H$_2$O$_2$ &           --  &     1.3(-12)  &     2.6(-11)  &      1.6(-6)  &      8.2(-7)  &      1.6(-6)  &           --  &           -- \\
               O$_3$ &           --  &           --  &     5.5(-13)  &     3.7(-12)  &     5.4(-11)  &     1.4(-10)  &           --  &           -- \\
               N$_2$ &           --  &           --  &           --  &     7.5(-14)  &     5.5(-14)  &     7.7(-14)  &           --  &           -- \\
          C$_2$H$_2$ &           --  &           --  &           --  &     2.8(-10)  &     6.4(-10)  &     6.4(-10)  &     1--3(-3)  &           -- \\
          C$_2$H$_6$ &           --  &     3.1(-13)  &      2.0(-9)  &      9.9(-6)  &      8.5(-5)  &      8.5(-5)  &        6(-3)  &           -- \\
          C$_3$H$_8$ &           --  &     4.0(-15)  &     3.5(-11)  &      2.9(-7)  &      2.6(-6)  &      2.6(-6)  &           --  &           -- \\
               HCOOH &           --  &     7.4(-12)  &      5.5(-8)  &      3.2(-6)  &      2.2(-5)  &      2.2(-5)  &        9(-4)  &           -- \\
           CH$_3$CHO &           --  &     9.7(-11)  &      5.5(-8)  &      4.7(-8)  &      2.9(-7)  &      2.9(-7)  &        2(-4)  &        5(-3) \\
              OHCCHO &           --  &      8.8(-9)  &      5.5(-6)  &      3.3(-6)  &      3.6(-5)  &      3.6(-5)  &           --  &           -- \\
          CH$_3$OCHO &           --  &      3.2(-9)  &      4.1(-7)  &      1.0(-5)  &      3.0(-5)  &      3.0(-5)  &        8(-4)  &           -- \\
       CH$_3$OCH$_3$ &           --  &     3.1(-10)  &      2.4(-8)  &      1.8(-7)  &      3.5(-7)  &      3.5(-7)  &     $<$5(-3)  &           -- \\
        C$_2$H$_5$OH &           --  &     2.8(-10)  &      2.2(-7)  &      1.2(-4)  &      3.1(-4)  &      3.1(-4)  &           --  &           -- \\
       CH$_2$(OH)CHO &           --  &      4.1(-9)  &      2.3(-6)  &      1.6(-4)  &      7.0(-4)  &      7.0(-4)  &           --  &        4(-3) \\
      (CH$_2$OH)$_2$ &           --  &     6.4(-14)  &     7.7(-11)  &      2.3(-7)  &      3.1(-7)  &      3.8(-7)  &      2.5(-3)  &        2(-3) \\
       C$_2$H$_5$CHO &           --  &     1.3(-13)  &     5.3(-10)  &      2.9(-7)  &      4.9(-6)  &      4.9(-6)  &           --  &        1(-3) \\
      CH$_3$COCH$_3$ &           --  &     1.5(-12)  &      3.8(-9)  &      2.1(-9)  &      1.2(-8)  &      1.2(-8)  &           --  &        3(-3) \\
                 HCN &           --  &           --  &           --  &           --  &           --  &           --  &      2.5(-3)  &        9(-3) \\
                 HNC &           --  &     9.0(-14)  &     3.8(-15)  &           --  &           --  &     9.9(-14)  &        4(-4)  &           -- \\
                HNCO &           --  &     1.2(-12)  &     3.0(-12)  &     1.4(-11)  &     1.6(-11)  &     1.9(-11)  &        1(-3)  &        3(-3) \\
            CH$_3$CN &           --  &           --  &           --  &           --  &           --  &           --  &        2(-4)  &        3(-3) \\
             HC$_3$N &           --  &           --  &           --  &           --  &           --  &           --  &        2(-4)  &           -- \\
           NH$_2$CHO &           --  &      1.0(-9)  &      6.1(-7)  &      2.7(-5)  &      1.5(-4)  &      1.5(-4)  &      1.5(-4)  &      1.8(-2) \\
    C$_2$H$_5$NH$_2$ &           --  &     3.6(-14)  &     9.1(-12)  &      2.7(-7)  &      1.9(-6)  &      1.9(-6)  &           --  &        3(-3) \\
      CH$_3$CONH$_2$ &           --  &     5.6(-13)  &     8.2(-10)  &      1.2(-7)  &      1.5(-6)  &      1.5(-6)  &           --  &        7(-3) \\
            NH$_2$OH &           --  &     1.4(-10)  &      9.3(-8)  &      2.2(-4)  &      9.2(-4)  &      9.2(-4)  &           --  &           -- \\
    NH$_2$CH$_2$COOH &           --  &           --  &     2.0(-15)  &     9.7(-10)  &      2.5(-8)  &      2.5(-8)  &           --  &   $<$2.5(-3) \\
\enddata
\tablecomments{$A(B)=A \times 10^B$. Fractional abundances are mean values calculated to a depth of 15~m.}
\tablenotetext{a}{Hale-Bopp fractional abundances: Mumma \& Charnley (2011).}
\tablenotetext{b}{Comet 67P/CG abundances: Goesmann et al. (2015); glycine value: Altwegg et al. (2016).}
\tablenotetext{c}{Coma measurement of O$_2$: Bieler et al. (2015)}
\end{deluxetable}
\clearpage

\begin{deluxetable}{lrrrrrrrr}
\tabletypesize{\small}
\tablecaption{\label{tab-5K-CR} Abundances with respect to H$_2$O for 5~K model with cosmic-ray processing}
\tablewidth{0pt}
\tablehead{ \colhead{Species}  & \colhead{$0$ yr} & \colhead{$10^3$ yr} & \colhead{$10^6$ yr} & \colhead{$10^9$ yr}  & \colhead{$5 \times 10^9$ yr}  & \colhead{Max.}  & \colhead{Hale-Bopp $^{a}$}  &  \colhead{67P/CG $^{b}$} }
\startdata
               Dust  &      1.1(+0)  &      1.1(+0)  &      1.1(+0)  &      1.2(+0)  &      1.3(+0)  &      1.3(+0)  &           --  &           -- \\
                  CO &      2.0(-1)  &      2.0(-1)  &      2.0(-1)  &      1.6(-1)  &      1.3(-1)  &      2.0(-1)  & 1.2--2.3(-1)  &      1.2(-2) \\
              CO$_2$ &      2.0(-1)  &      2.0(-1)  &      2.0(-1)  &      2.0(-1)  &      2.0(-1)  &      2.1(-1)  &        6(-2)  &           -- \\
              CH$_4$ &      1.0(-2)  &      1.0(-2)  &      1.0(-2)  &      1.1(-2)  &      1.7(-2)  &      1.7(-2)  &      1.5(-2)  &        5(-3) \\
             H$_2$CO &      1.0(-2)  &      1.0(-2)  &      1.0(-2)  &      1.3(-2)  &      2.2(-2)  &      2.2(-2)  &      1.1(-2)  &           -- \\
            CH$_3$OH &      5.0(-2)  &      5.0(-2)  &      5.0(-2)  &      2.8(-2)  &      1.2(-2)  &      5.0(-2)  &      2.4(-2)  &           -- \\
              NH$_3$ &      1.0(-2)  &      1.0(-2)  &      1.0(-2)  &      9.0(-3)  &      7.6(-3)  &      1.0(-2)  &        7(-3)  &           -- \\
\hline
               O$_2$ &           --  &     7.0(-10)  &      3.9(-9)  &      6.8(-4)  &      3.8(-3)  &      3.8(-3)  &           --  &  3.8(-2)$^c$ \\
          H$_2$O$_2$ &           --  &     6.5(-10)  &      2.6(-8)  &      4.5(-3)  &      1.7(-2)  &      1.7(-2)  &           --  &           -- \\
               O$_3$ &           --  &     6.2(-11)  &     2.9(-10)  &      1.2(-5)  &      1.3(-4)  &      1.3(-4)  &           --  &           -- \\
               N$_2$ &           --  &     6.0(-13)  &     2.5(-11)  &      2.1(-7)  &      4.1(-6)  &      4.1(-6)  &           --  &           -- \\
          C$_2$H$_2$ &           --  &     2.8(-12)  &     9.5(-12)  &      1.1(-5)  &      2.3(-4)  &      2.3(-4)  &     1--3(-3)  &           -- \\
          C$_2$H$_6$ &           --  &     7.4(-13)  &     3.8(-11)  &      1.4(-5)  &      1.4(-4)  &      1.4(-4)  &        6(-3)  &           -- \\
          C$_3$H$_8$ &           --  &     3.2(-15)  &           --  &      6.4(-8)  &      1.7(-6)  &      1.7(-6)  &           --  &           -- \\
               HCOOH &           --  &     1.4(-11)  &      1.4(-9)  &      1.1(-3)  &      3.4(-3)  &      3.4(-3)  &        9(-4)  &           -- \\
           CH$_3$CHO &           --  &     7.8(-13)  &     1.4(-10)  &      1.1(-4)  &      4.3(-4)  &      4.3(-4)  &        2(-4)  &        5(-3) \\
              OHCCHO &           --  &     2.8(-13)  &     4.0(-11)  &      8.0(-5)  &      3.8(-4)  &      3.8(-4)  &           --  &           -- \\
          CH$_3$OCHO &           --  &     6.2(-13)  &     9.1(-11)  &      6.2(-5)  &      2.9(-4)  &      2.9(-4)  &        8(-4)  &           -- \\
       CH$_3$OCH$_3$ &           --  &     2.1(-13)  &     1.5(-11)  &      6.4(-6)  &      6.5(-5)  &      6.5(-5)  &     $<$5(-3)  &           -- \\
        C$_2$H$_5$OH &           --  &     4.4(-13)  &     8.2(-11)  &      1.5(-5)  &      6.4(-5)  &      6.4(-5)  &           --  &           -- \\
       CH$_2$(OH)CHO &           --  &     3.1(-12)  &     4.6(-10)  &      2.8(-4)  &      6.6(-4)  &      6.6(-4)  &           --  &        4(-3) \\
      (CH$_2$OH)$_2$ &           --  &     1.3(-12)  &     3.5(-10)  &      4.6(-5)  &      8.7(-5)  &      8.7(-5)  &      2.5(-3)  &        2(-3) \\
       C$_2$H$_5$CHO &           --  &     3.4(-15)  &           --  &      4.1(-7)  &      3.8(-6)  &      3.8(-6)  &           --  &        1(-3) \\
      CH$_3$COCH$_3$ &           --  &     2.6(-14)  &     2.5(-12)  &      7.0(-7)  &      7.5(-6)  &      7.5(-6)  &           --  &        3(-3) \\
                 HCN &           --  &     1.4(-12)  &     6.7(-12)  &      1.4(-6)  &      3.0(-5)  &      3.0(-5)  &      2.5(-3)  &        9(-3) \\
                 HNC &           --  &     3.8(-13)  &     1.1(-11)  &      2.3(-6)  &      8.6(-6)  &      8.6(-6)  &        4(-4)  &           -- \\
                HNCO &           --  &     8.0(-11)  &      7.6(-8)  &      3.4(-5)  &      6.4(-5)  &      6.4(-5)  &        1(-3)  &        3(-3) \\
            CH$_3$CN &           --  &     2.1(-15)  &     4.3(-15)  &      2.5(-9)  &      1.3(-7)  &      1.3(-7)  &        2(-4)  &        3(-3) \\
             HC$_3$N &           --  &           --  &           --  &     5.1(-13)  &     4.0(-10)  &     4.0(-10)  &        2(-4)  &           -- \\
           NH$_2$CHO &           --  &     4.5(-13)  &     4.2(-11)  &      3.5(-5)  &      9.6(-5)  &      9.6(-5)  &      1.5(-4)  &      1.8(-2) \\
    C$_2$H$_5$NH$_2$ &           --  &           --  &           --  &      1.2(-8)  &      1.4(-7)  &      1.4(-7)  &           --  &        3(-3) \\
      CH$_3$CONH$_2$ &           --  &     1.5(-14)  &     2.1(-12)  &      4.5(-7)  &      2.0(-6)  &      2.0(-6)  &           --  &        7(-3) \\
            NH$_2$OH &           --  &     7.6(-12)  &     4.6(-10)  &      9.6(-5)  &      3.3(-4)  &      3.3(-4)  &           --  &           -- \\
    NH$_2$CH$_2$COOH &           --  &           --  &           --  &      7.4(-9)  &      1.0(-7)  &      1.0(-7)  &           --  &   $<$2.5(-3) \\
\enddata
\tablecomments{See Table 2}
\end{deluxetable}
\clearpage

\begin{deluxetable}{lrrrrrrrr}
\tabletypesize{\small}
\tablecaption{\label{tab-10K-CR} Abundances with respect to H$_2$O for 10~K model with cosmic-ray processing}
\tablewidth{0pt}
\tablehead{ \colhead{Species}  & \colhead{$0$ yr} & \colhead{$10^3$ yr} & \colhead{$10^6$ yr} & \colhead{$10^9$ yr}  & \colhead{$5 \times 10^9$ yr}  & \colhead{Max.}  & \colhead{Hale-Bopp $^{a}$}  &  \colhead{67P/CG $^{b}$} }
\startdata
               Dust  &      1.1(+0)  &      1.1(+0)  &      1.1(+0)  &      1.1(+0)  &      1.1(+0)  &      1.1(+0)  &           --  &           -- \\
                  CO &      2.0(-1)  &      2.0(-1)  &      2.0(-1)  &      1.6(-1)  &      1.2(-1)  &      2.0(-1)  & 1.2--2.3(-1)  &      1.2(-2) \\
              CO$_2$ &      2.0(-1)  &      2.0(-1)  &      2.0(-1)  &      2.2(-1)  &      2.5(-1)  &      2.5(-1)  &        6(-2)  &           -- \\
              CH$_4$ &      1.0(-2)  &      1.0(-2)  &      1.0(-2)  &      1.5(-2)  &      2.2(-2)  &      2.2(-2)  &      1.5(-2)  &        5(-3) \\
             H$_2$CO &      1.0(-2)  &      1.0(-2)  &      1.0(-2)  &      1.5(-2)  &      1.7(-2)  &      1.8(-2)  &      1.1(-2)  &           -- \\
            CH$_3$OH &      5.0(-2)  &      5.0(-2)  &      5.0(-2)  &      2.0(-2)  &      1.3(-2)  &      5.0(-2)  &      2.4(-2)  &           -- \\
              NH$_3$ &      1.0(-2)  &      1.0(-2)  &      1.0(-2)  &      9.3(-3)  &      7.8(-3)  &      1.0(-2)  &        7(-3)  &           -- \\
\hline
               O$_2$ &           --  &     3.7(-11)  &      2.1(-9)  &      2.5(-5)  &      2.9(-4)  &      2.9(-4)  &           --  &  3.8(-2)$^c$ \\
          H$_2$O$_2$ &           --  &     1.7(-10)  &     4.0(-10)  &      2.4(-4)  &      1.7(-3)  &      1.7(-3)  &           --  &           -- \\
               O$_3$ &           --  &     9.7(-13)  &     6.7(-13)  &      1.7(-6)  &      3.9(-5)  &      3.9(-5)  &           --  &           -- \\
               N$_2$ &           --  &     2.9(-13)  &     1.5(-11)  &      2.4(-6)  &      2.2(-5)  &      2.2(-5)  &           --  &           -- \\
          C$_2$H$_2$ &           --  &     1.7(-11)  &      2.5(-9)  &      3.1(-5)  &      2.0(-4)  &      2.0(-4)  &     1--3(-3)  &           -- \\
          C$_2$H$_6$ &           --  &     7.9(-12)  &      1.9(-8)  &      5.0(-5)  &      6.6(-4)  &      6.6(-4)  &        6(-3)  &           -- \\
          C$_3$H$_8$ &           --  &     1.0(-13)  &     7.4(-10)  &      5.6(-7)  &      1.8(-5)  &      1.8(-5)  &           --  &           -- \\
               HCOOH &           --  &     1.2(-11)  &      6.6(-9)  &      3.0(-4)  &      5.6(-4)  &      5.6(-4)  &        9(-4)  &           -- \\
           CH$_3$CHO &           --  &     1.0(-11)  &      5.8(-9)  &      2.0(-4)  &      4.6(-4)  &      4.6(-4)  &        2(-4)  &        5(-3) \\
              OHCCHO &           --  &     1.0(-12)  &     3.2(-10)  &      2.7(-5)  &      5.9(-5)  &      5.9(-5)  &           --  &           -- \\
          CH$_3$OCHO &           --  &     6.4(-12)  &      3.7(-9)  &      1.6(-4)  &      4.7(-4)  &      4.7(-4)  &        8(-4)  &           -- \\
       CH$_3$OCH$_3$ &           --  &     1.1(-12)  &     7.6(-10)  &      2.4(-5)  &      9.2(-5)  &      9.2(-5)  &     $<$5(-3)  &           -- \\
        C$_2$H$_5$OH &           --  &     1.9(-12)  &      3.4(-9)  &      2.4(-5)  &      1.5(-4)  &      1.5(-4)  &           --  &           -- \\
       CH$_2$(OH)CHO &           --  &     1.3(-11)  &      1.3(-8)  &      3.5(-4)  &      6.4(-4)  &      6.4(-4)  &           --  &        4(-3) \\
      (CH$_2$OH)$_2$ &           --  &     1.3(-12)  &      2.8(-9)  &      5.0(-5)  &      8.0(-5)  &      8.0(-5)  &      2.5(-3)  &        2(-3) \\
       C$_2$H$_5$CHO &           --  &     6.9(-14)  &     5.0(-11)  &      4.0(-7)  &      2.8(-6)  &      2.8(-6)  &           --  &        1(-3) \\
      CH$_3$COCH$_3$ &           --  &     1.7(-13)  &     6.2(-11)  &      1.2(-6)  &      5.4(-6)  &      5.4(-6)  &           --  &        3(-3) \\
                 HCN &           --  &     1.7(-12)  &     1.5(-10)  &      9.5(-6)  &      1.8(-4)  &      1.8(-4)  &      2.5(-3)  &        9(-3) \\
                 HNC &           --  &     2.8(-13)  &     5.8(-12)  &      3.8(-5)  &      1.7(-4)  &      1.7(-4)  &        4(-4)  &           -- \\
                HNCO &           --  &     7.8(-11)  &      7.6(-8)  &      1.5(-5)  &      1.9(-5)  &      1.9(-5)  &        1(-3)  &        3(-3) \\
            CH$_3$CN &           --  &     1.0(-14)  &     4.4(-13)  &      2.7(-8)  &      7.2(-7)  &      7.2(-7)  &        2(-4)  &        3(-3) \\
             HC$_3$N &           --  &           --  &           --  &     7.1(-10)  &      8.8(-9)  &      8.8(-9)  &        2(-4)  &           -- \\
           NH$_2$CHO &           --  &     2.4(-12)  &     8.4(-10)  &      2.8(-5)  &      5.0(-5)  &      5.0(-5)  &      1.5(-4)  &      1.8(-2) \\
    C$_2$H$_5$NH$_2$ &           --  &     5.3(-15)  &     1.0(-12)  &      1.6(-8)  &      1.8(-7)  &      1.8(-7)  &           --  &        3(-3) \\
      CH$_3$CONH$_2$ &           --  &     5.9(-14)  &     1.0(-11)  &      5.0(-7)  &      1.4(-6)  &      1.4(-6)  &           --  &        7(-3) \\
            NH$_2$OH &           --  &     7.5(-12)  &     5.2(-10)  &      6.4(-5)  &      3.2(-4)  &      3.2(-4)  &           --  &           -- \\
    NH$_2$CH$_2$COOH &           --  &     1.4(-15)  &     1.6(-13)  &      4.9(-9)  &      3.1(-8)  &      3.1(-8)  &           --  &   $<$2.5(-3) \\
\enddata
\tablecomments{See Table 2}
\end{deluxetable}
\clearpage

\begin{deluxetable}{lrrrrrrrr}
\tabletypesize{\small}
\tablecaption{\label{tab-20K-CR} Abundances with respect to H$_2$O for 20~K model with cosmic-ray processing}
\tablewidth{0pt}
\tablehead{ \colhead{Species}  & \colhead{$0$ yr} & \colhead{$10^3$ yr} & \colhead{$10^6$ yr} & \colhead{$10^9$ yr}  & \colhead{$5 \times 10^9$ yr}  & \colhead{Max.}  & \colhead{Hale-Bopp $^{a}$}  &  \colhead{67P/CG $^{b}$} }
\startdata
               Dust  &      1.1(+0)  &      1.1(+0)  &      1.1(+0)  &      1.2(+0)  &      1.4(+0)  &      1.4(+0)  &           --  &           -- \\
                  CO &      2.0(-1)  &      2.0(-1)  &      2.0(-1)  &      1.7(-1)  &      1.4(-1)  &      2.0(-1)  & 1.2--2.3(-1)  &      1.2(-2) \\
              CO$_2$ &      2.0(-1)  &      2.0(-1)  &      2.0(-1)  &      2.4(-1)  &      3.0(-1)  &      3.0(-1)  &        6(-2)  &           -- \\
              CH$_4$ &      1.0(-2)  &      1.0(-2)  &      1.0(-2)  &      9.6(-3)  &      9.0(-3)  &      1.0(-2)  &      1.5(-2)  &        5(-3) \\
             H$_2$CO &      1.0(-2)  &      1.0(-2)  &      1.0(-2)  &      9.5(-3)  &      9.4(-3)  &      1.0(-2)  &      1.1(-2)  &           -- \\
            CH$_3$OH &      5.0(-2)  &      5.0(-2)  &      5.0(-2)  &      5.2(-2)  &      5.3(-2)  &      5.3(-2)  &      2.4(-2)  &           -- \\
              NH$_3$ &      1.0(-2)  &      1.0(-2)  &      1.0(-2)  &      9.5(-3)  &      8.8(-3)  &      1.0(-2)  &        7(-3)  &           -- \\
\hline
               O$_2$ &           --  &     4.4(-10)  &      1.7(-9)  &      3.2(-9)  &      5.0(-9)  &      5.0(-9)  &           --  &  3.8(-2)$^c$ \\
          H$_2$O$_2$ &           --  &     1.2(-11)  &     1.5(-10)  &     2.2(-10)  &     3.1(-10)  &     3.1(-10)  &           --  &           -- \\
               O$_3$ &           --  &      2.3(-9)  &      3.0(-8)  &      6.6(-8)  &      1.2(-7)  &      1.2(-7)  &           --  &           -- \\
               N$_2$ &           --  &     2.7(-15)  &     2.5(-12)  &      1.0(-9)  &      7.0(-9)  &      7.0(-9)  &           --  &           -- \\
          C$_2$H$_2$ &           --  &     1.6(-10)  &      1.3(-9)  &     8.8(-10)  &     5.3(-10)  &      1.4(-9)  &     1--3(-3)  &           -- \\
          C$_2$H$_6$ &           --  &     1.5(-11)  &      2.9(-7)  &      7.0(-5)  &      1.5(-4)  &      1.5(-4)  &        6(-3)  &           -- \\
          C$_3$H$_8$ &           --  &     2.2(-13)  &      3.4(-7)  &      5.3(-5)  &      8.3(-5)  &      8.3(-5)  &           --  &           -- \\
               HCOOH &           --  &     5.0(-14)  &     3.5(-11)  &      1.0(-8)  &      2.9(-8)  &      2.9(-8)  &        9(-4)  &           -- \\
           CH$_3$CHO &           --  &     4.0(-13)  &      4.2(-9)  &      4.3(-7)  &      5.4(-7)  &      5.4(-7)  &        2(-4)  &        5(-3) \\
              OHCCHO &           --  &     1.5(-15)  &     1.0(-12)  &     2.9(-10)  &     8.0(-10)  &     8.0(-10)  &           --  &           -- \\
          CH$_3$OCHO &           --  &     1.5(-15)  &     1.2(-12)  &     3.4(-10)  &     9.4(-10)  &     9.4(-10)  &        8(-4)  &           -- \\
       CH$_3$OCH$_3$ &           --  &     1.3(-12)  &      1.7(-8)  &      4.5(-6)  &      1.1(-5)  &      1.1(-5)  &     $<$5(-3)  &           -- \\
        C$_2$H$_5$OH &           --  &     9.5(-13)  &      1.3(-8)  &      3.4(-6)  &      7.9(-6)  &      7.9(-6)  &           --  &           -- \\
       CH$_2$(OH)CHO &           --  &     6.3(-15)  &     3.7(-12)  &     7.9(-10)  &      1.8(-9)  &      1.8(-9)  &           --  &        4(-3) \\
      (CH$_2$OH)$_2$ &           --  &     5.7(-14)  &     3.1(-11)  &      7.7(-9)  &      1.9(-8)  &      1.9(-8)  &      2.5(-3)  &        2(-3) \\
       C$_2$H$_5$CHO &           --  &           --  &     6.2(-11)  &      1.1(-8)  &      1.7(-8)  &      1.7(-8)  &           --  &        1(-3) \\
      CH$_3$COCH$_3$ &           --  &           --  &     1.0(-11)  &      3.7(-7)  &      1.1(-6)  &      1.1(-6)  &           --  &        3(-3) \\
                 HCN &           --  &     7.8(-11)  &      2.0(-7)  &      4.5(-5)  &      9.5(-5)  &      9.5(-5)  &      2.5(-3)  &        9(-3) \\
                 HNC &           --  &     2.8(-10)  &      3.0(-7)  &      5.0(-5)  &      8.7(-5)  &      8.7(-5)  &        4(-4)  &           -- \\
                HNCO &           --  &     9.9(-12)  &     4.3(-10)  &     3.8(-10)  &     3.3(-10)  &     4.3(-10)  &        1(-3)  &        3(-3) \\
            CH$_3$CN &           --  &     1.4(-14)  &     2.4(-10)  &      3.6(-8)  &      6.5(-8)  &      6.5(-8)  &        2(-4)  &        3(-3) \\
             HC$_3$N &           --  &     2.5(-14)  &     1.1(-12)  &     4.8(-13)  &     2.7(-13)  &     1.4(-12)  &        2(-4)  &           -- \\
           NH$_2$CHO &           --  &     3.6(-13)  &      2.8(-8)  &      7.1(-7)  &      7.2(-7)  &      7.4(-7)  &      1.5(-4)  &      1.8(-2) \\
    C$_2$H$_5$NH$_2$ &           --  &           --  &     3.3(-11)  &      6.4(-9)  &      1.2(-8)  &      1.2(-8)  &           --  &        3(-3) \\
      CH$_3$CONH$_2$ &           --  &           --  &     1.8(-11)  &      6.8(-9)  &      1.4(-8)  &      1.4(-8)  &           --  &        7(-3) \\
            NH$_2$OH &           --  &     1.1(-10)  &      1.3(-7)  &      3.1(-5)  &      6.6(-5)  &      6.6(-5)  &           --  &           -- \\
    NH$_2$CH$_2$COOH &           --  &           --  &           --  &           --  &     2.2(-15)  &     2.2(-15)  &           --  &   $<$2.5(-3) \\
\enddata
\tablecomments{See Table 2}
\end{deluxetable}
\clearpage

\section{Discussion}

The first set of models presented here focuses on the effects of the UV irradiation of comets by interstellar UV photons, at various fixed comet temperatures. This physical treatment is intended to approximate conditions during ``cold storage'' in the Oort cloud. Under extremely low-temperature conditions (5--10~K), the effects of the UV photo-chemistry are restricted to approximately the upper 1$\mu$m (although this is dependent on the dust model and UV penetration treatment described in Secs. 2.3 \& 2.4, which presently assume only a single, representative particle size). Both simple and complex organic molecules are formed at these depths. In parallel, the dust concentration builds up in the outer layers, dependent on the rates of loss from the comet surface, largely due to UV photo-desorption. For this reason, the elevated dust concentration extends to greater depths than the effects of UV. Under low temperature (i.e. low mobility) conditions, the dust becomes maximally concentrated to depths of less than 1~mm, but at 20~K this rises to around 1~cm. At temperatures greater than this, a maximum concentrated dust depth of around 1~m is reached, although more than 10~m of ice is lost from the surface in these models, over a period of 5 Gyr. The mobility of volatile species such as H and H$_2$ within the ice at these temperatures allows them to diffuse toward the surface, where they are thermally desorbed. 

The ability of diffusive species to reach the surface is dependent on the open structure of the dust component that is assumed in this model, via the imposition of a maximum dust abundance per monolayer ($\sim$74\%), which is based on close-packing of identical spheres. The presence of some minimum amount of icy material in the mixture thus allows a path to the outer surface via bulk diffusion. However, if refractory material, either formed in the outer ice layers or included in the ice at its formation, were to reach sufficient abundance in the outer layers, then the ability of volatiles to reach the surface could be significantly hindered. It is unclear in such a case whether hydrogen atoms and/or H$_2$ molecules (which in the present model are responsible for much of the loss from the surface) would be capable of passing through. In the model presented here, the mobility of diffusive species via bulk diffusion is not affected by the build-up of immobile species, while the complexity of the molecules produced in the chemical network is also limited. 

In any case, while the present models suggest that a substantial thickening of the dust component would result indirectly from volatile loss from the surface, both the degree and depth to which this occurs are dependent on the assumed maximum dust fraction, and these quantities cannot therefore be considered to be well constrained by the models. However, if the model picture is correct, then the outer layers of the comet would consist of closely packed dust-grain material, loosely bound together by a small amount (on the order of say 10--20\% by volume) of a mixture of water and other molecules, including some organic species of considerable complexity. The models indicate that these complex organics are formed in the outer layers, in which UV processing occurs, even at extremely low temperatures (although the precise depths to which UV alone may have an effect is dependent both on the dust concentration and on the degree of attenutation caused by that dust, for which the models adopt only a simple treatment). 

Through the inclusion of an explicit mechanism for immediate reaction between newly formed photo-dissociation products and existing atoms and radicals in the ices, the models demonstrate that reactions that form complex organic molecules need not be mediated by diffusion at low temperatures. Indeed, preliminary versions of the comet models (not presented here) that did not include this immediate reaction process suffered from the build-up of implausibly large quantities of reactive species, unable to react due to low thermal diffusion rates. 

It is worth noting that the interstellar chemistry model upon which the comet models are based did not include the immediate reaction mechanism either, nor do any other interstellar models to the author's knowledge. This omission must lead to a bias against complex molecule formation in models of cold, dense objects such as pre-stellar cores, or during equivalent low-temperature regimes appropriate to high-mass star formation. Indeed, the detection of modest quantities of complex organics in the gas phase in cold, dense cores by Bacmann et al. (2012) suggest that low-temperature mechanisms for the production of these molecules must exist. A more careful consideration of low-temperature photochemistry may thus provide some explanation.

In the comet models with temperatures of 20~K and above, the diffusion of reactive radicals -- and even their more complex products -- leads to significant abundances of complex organics appearing even in the deeper layers beyond the reach of UV photons. The achievement of such temperatures in Oort cloud comets may require stochastic heating events, and these would last for a much shorter period ($\sim$10$^4$ yr; Stern \& Shull 1988) than the maximum lifetime of a comet as modeled here. The temperature fluctuation would also reach to only a limited depth. A short period of heating might, however, be sufficient to allow the diffusion of complex organics to greater depths, even if the period of elevated temperature were too short to have a strong effect on the total degree of chemical processing. The latter is limited by total UV or cosmic-ray fluence, although one or other of these values, as well as the comet's temperature, would be raised by the close approach of a hot star or a nearby supernova event. Such a stochastic heating event would produce only a small increase in the temperature of a Kuiper belt object ($\sim$1~K; Stern 2003), although these may generally maintain temperatures as high as 60~K without such heating; this could lead to a generally more enhanced chemistry than in the Oort cloud, dependent on UV intensity. There is also evidence from models of early internal heating by the radionuclide $^{26}$Al (Prialnik \& Podolak 1999) that gas-flows may exist within the ice during early evolution, which cannot easily be modeled here.

It should be noted that the use of a fixed temperature in the preliminary models is unlikely to demonstrate the most effective complex organic molecule (COM) production, nor the most appropriate relative enhancements between different molecules; in the case of star-forming cores, it is the gradual ramping up through key temperature bands (at which certain molecules or radicals become mobile) that allows the full spectrum of molecules to be produced (e.g. Garrod et al. 2008). The inclusion in the model of time- and depth-dependent temperature changes would allow these effects to be assessed.

\subsection{Comparison with observed cometary molecular abundances}

Tables \ref{tab-60K-hi} -- \ref{tab-20K-CR} show aggregated abundances to a depth of 15~m, presented as a fraction of water in that same depth, for the 60~K high-UV (UV-only) model and for each of the cosmic-ray processing models at various times. The last two columns in the tables show abundances of various molecules for comet Hale-Bopp (C/1995 O1), determined via remote observations, and for Churyumov-Gerasimenko (67P), determined by the {\em COSAC} and {\em ROSINA} mass spectrometers. 

The 15~m depth over which the aggregated abundances are calculated is appropriate to demonstrate the overall effect on abundances of, in particular, cosmic-ray processing (Tables \ref{tab-5K-CR} -- \ref{tab-20K-CR}). However, beyond this depth, the resolution of the models is insufficient to provide accurate aggregated abundances, as the depth at which cosmic-ray processing falls off (and thus molecular abundances fall sharply) lies somewhere within the next deepest layer, which has a thickness of $\sim$30~m, making the integration quite imprecise. Future models will adopt higher depth resolution around the threshold region.

In order to be able to compare directly between the abundances produced by the models and those obtained from observations of active phase comets, consideration must be given as to how much material has been lost by those bodies since their entry into the inner solar system. The loss of surface material during a comet's active phase would mean that at least some fraction of the high molecular abundances residing in the upper layers (based on the present models) would likely be lost. Typical dynamical lifetimes for comets are a few hundred thousand years, so in the case of Jupiter family comets (JFCs), for example, there may have been thousands of orbits (stripping away thousands of meters of the surface) prior to observation. The present models are therefore most applicable to relatively dynamically new comets (those which have only entered the inner solar system a few tens of times at most), or dynamically new comets (those which are entering the inner solar system for the first time). 

It is useful, then, to consider the depth to which the modeled abundances must extend in order to be directly comparable with the {\em Rosetta} values, any further chemical processing during the active phase notwithstanding. The erosion rate of comet 67P has been calculated to be approximately 1.0~m per orbit ($\pm$50\%) by Bertaux (2015), while its orbital period is determined to be 6.44 yr, providing an average rate of 0.16 m/yr. Assuming no significant erosion occurred before its 1840 encounter with Jupiter that brought it into a closer orbit (although its earlier orbital activity is uncertain; Kr{\'o}likowska 2003), this corresponds to $\sim$13--41~m of material lost during the comet's active phase. For the solid-phase chemistry simulated in the model to be important to the recently-observed chemical abundances of 67P, such chemical processing should therefore extend to such a depth, or greater. The 15~m aggregated abundances calculated from the models thus satisfy at least the lower limit of this range. Substantial abundances of molecules of interest could extend beyond 15~m; however, as mentioned above, the present models lack the depth resolution to determine this.

The other comet with which the modeled abundances are compared is Hale-Bopp; prior to 1997, it is estimated to have made its last perihelion in 2215~BC, which is also been speculated to have been its first (Marsden 1997). Based on submillimeter measurements of particle mass loss during its most recent solar approach, Jewitt \& Matthews (1999) calculated a mean value of around 10~m of material lost per orbit. This would suggest, assuming only one previous perihelion with similar mass-loss characteristics, that only around 10~m of material should have been lost prior to the 1997 perihelion, and that at least 5~m of native material within the upper 15~m or so would have survived to be released during that well-observed event. Direct comparison of the aggregated values presented in Tables \ref{tab-60K-hi} -- \ref{tab-20K-CR} with the observed molecular abundances determined for Hale-Bopp therefore seems justified. It should be noted, however, that the penetration depths of cosmic rays used in the models correspond to the density of material in 67P in particular, which could differ from that of Hale-Bopp.

Although some UV-only models show significant abundance enhancements of COMs to a depth of around 10~m or more, for most temperatures they cannot reproduce values detected in comets 67P or Hale-Bopp (or not to appropriate depth, see below), and are thus not presented in the tables. The 60~K UV-only model is perhaps the most successful such model at reproducing the observed values, and is thus included to demonstrate the maximum likely contribution from UV surface processing.
Unlike the other UV-only models, the 60~K models maintain substantial abundances of COMs throughout all layers as a result of efficient diffusion. At this temperature, the aggregated abundances begin to approach the measured values for 67P or Hale-Bopp in some cases, but do not actually reach them, aside from formamide (NH$_2$CHO), with formic acid (HCOOH) and methyl formate (CH$_3$OCHO) falling only a little short. In general, it appears that none of the UV-only models is capable of reproducing either set of observed values at appropriate depths. However, for depths less than the 15~m used in the tables, fractional abundances of complex organics produced over the 5 Gyr lifetime are comparable to those observed in the gas phase in star-forming cores (with the presumption that many of these molecules may have formed in dust-grain ice mantles). This in itself would suggest that UV processing of cometary ices at moderately high temperatures could allow a comparable degree of COM production in post-formation cometary material as might be expected to be incorporated from the interstellar or proto-nebular stage (however, see also Sec. 5.3).

\subsection{Cosmic ray-induced radiolysis}

In contrast to the UV-only models, those including cosmic-ray processing of the ice sustain COM production to great depths, regardless of temperature. Crucially, the comparison of aggregated abundances with measured values for Hale-Bopp and 67P is much better for these models. Indeed, in some cases, certain abundances are a little higher than the observed values by the end of the models, in particular formic acid (HCOOH). However, it is only in the 5 and 10~K models of CR processing that a good match is provided for both the simple and complex species; diffusion into the deepest layer of the ice at 20~K diminishes abundances throughout. Weaker bulk-diffusion effects as discussed in Sec. 5.3, below, would raise the temperature at which the models could reproduce observational abundances.

At 5 and 10~K, COMs including formic acid, acetaldehyde (CH$_3$CHO), methyl formate (CH$_3$OCHO), and formamide (NH$_2$CHO) achieve aggregated abundances that are in reasonably good agreement with abundances observed for Hale-Bopp, to within a factor of 2--3 above or below the observations, while the abundance of dimethyl ether (CH$_3$OCH$_3$) is also in line with its observed upper limit. Ethylene glycol, (CH$_2$OH)$_2$, is underproduced with respect to Hale-Bopp and 67P abundances, although its only formation mechanism included in the network -- the addition two CH$_2$OH radicals -- also does a poor job in interstellar models. An alternative production mechanism, such as the barrier-mediated hydrogenation of glycolaldehyde by H atoms, may be required to explain its abundance. Glycolaldehyde (CH$_2$(OH)CHO), uniquely among the COMs, comes close to the detected abundance in 67P. Other complex organics broadly fall around 1--2 orders of magnitude below thier detected abundances in 67P. 

The success in reproducing the abundances of COMs in Hale-Bopp in particular may give credence to the idea that its chemistry is indicative of a young and relatively uneroded surface, whose outer layers from the cold-storage phase survived, allowing their complex molecular content to be observed in its coma.

Interestingly, the 5~K model produces aggregate abundances of O$_2$ that are comparable to the observed value for 67P.  Again, the agreement is not so good at 10~K, and very poor at 20~K. However, the results would indicate that cosmic-ray processing of water over several Gyr may be sufficient to produce the observed fractional abundances of this molecule in comet 67P; fractional abundances of at least 10$^{-3}$ with respect to water are maintained to a depth of around 10~m, although at the depth resolution used in this model, the precise depth is rather uncertain. Resolution on the order of 1~m or so at depths of 1~m and beyond may be necessary to investigate fully the cosmic-ray processing of cometary ices.

The production of significant amounts of O$_2$ to large depths in the cosmic-ray processing models is dependent on several factors, including the high porosity/low density of the cometary material assumed here (based on estimates for comet 67P), which determines how deeply the cosmic rays may penetrate. The precise origin of this porosity, and whether it was present at the earliest times in the evolution of 67P, or whether it was itself the result of some kind of processing, will determine whether the density estimate is appropriate for use in these models.

It should also be noted, however, that the rate and/or efficiency of water dissociation (and of other molecules) at various depths by galactic cosmic rays is not well constrained, and may be a more important parameter. The estimate of the water dissociation rate used in the present models (discussed in Sec. 4) is around 50 times lower than the value implied by the calculations of SH18, for the case where the cosmic-ray flux is unchanged by penetration into the material (i.e. the zero-depth rate). 

Bieler et al. (2015) measured an average O$_2$ abundance with respect to water in comet 67P of $3.80 \pm 0.85$ \%, with individual measurements ranging from 1-10 \%. If O$_2$ can be formed and maintained during the cold-storage phase at an abundance around 1\% with respect to water, to a depth on the order of ten meters, this may help to explain the mysteriously high abundance of this molecule in 67P. If O$_2$ is indeed formed this way, within the comet itself during cold storage, its production at large depths would be dependent on cosmic rays of energies $>$10~GeV; such high-energy particles are likely to be of similar flux both in the Oort cloud and within the heliosphere, thus all comet families would be expected to be similarly affected (prior to any surface ablation). The 5~K models presented here produce appropriate O$_2$ abundances to a depth of around 10~m, which is shy of the depth required if 67P has lost 27~m of ice in its recent history, but close to the lower limit on this estimate (13~m). However, an increase in the estimated cosmic ray-induced dissociation rate of water by even a factor of a few would push the efficient production of O$_2$ to a depth comfortably consistent with the required range. An increase by a factor 50, to be consistent with SH18, would be expected to give values well in agreement with observations. However, the relatively low depth-resolution of the present models in this depth range means that the precise depth of O$_2$ production is poorly constrained.

The failure of the cosmic-ray processing model at 20~K model to provide O$_2$ in significant abundance in the same way as the 5~K model is largely due to the thermal mobility of the O$_2$ molecule, while at temperatures much greater than 20~K, O$_2$ might be expected to leave the ice entirely. Mousis et al. (2016) suggest that the retention of O$_2$ to high temperatures in 67P itself is due to its clathration with surrounding H$_2$O molecules when it is formed via radiolysis, although those authors suggest that the formation process occurs in dust-grain ice mantles in the proto-solar nebula, prior to incorporation into a comet. Nevertheless, if clathration is indeed the mechanism that allows O$_2$ to remain trapped up to the high temperatures of an active-phase comet, high-energy cosmic-ray processing of the comet itself would also appear to be a plausible mechanism for its formation, even for a Jupiter family comet such as 67P, based on the present models. It is also likely that the deepest layers of a comet would retain a very low temperature (e.g. Guilbert-Lepoutre et al. 2016), even during the active phase, which would minimize thermal mobility over long periods. As discussed below, the back-diffusion effect may also reduce somewhat the net mixing rate, although this would be a relatively small effect if O$_2$ retains a large abundance. The formation of a dust-rich layer on the comet surface could also help to retain any mobile O$_2$.

Bieler et al. (2015) also measured hydrogen peroxide abundance in 67P, obtaining a value H$_2$O$_2$/O$_2$ $\simeq$$6 \times 10^{-4}$. The comet chemical models actually produce H$_2$O$_2$ more abundantly than O$_2$ in the deeper ice layers. Due to their related formation mechanisms, and the possibility for O$_2$ to be directly hydrogenated to H$_2$O$_2$, it is possible that the production of hydrogen peroxide is too efficient in the models. O$_2$ might therefore be expected to attain higher abundances than the models suggest, at the expense of H$_2$O$_2$. The depth attained by O$_2$ in the ice would also be greater as a result, more similar to that achieved by H$_2$O$_2$ in the models. The sum of these two molecules down to the aggregated depth of 15~m is certainly sufficient to agree adequately with the observed O$_2$ abundance, based on the 5~K model. The 10~K model also is not far from the observed value.

Rubin et al. (2015) detected N$_2$ emanating from 67P, although, since the ratio with H$_2$O is uncertain, its abundance is not included for comparison in the tables in this paper. Rubin et al. measured an N$_2$:CO ratio of $5.7(\pm 0.66) \times 10^{-3}$. The CR-processing models presented here produce aggregated 
abundance ratios for N$_2$:CO of $2.1 \times 10^{-5}$ and $1.1 \times 10^{-4}$ for the 5 and 10~K models, respectively, while the ratio for the 20~K model falls short by a larger margin.
All of the cosmic-ray processing model values thus fall short of the measurements; also, the depth reached by N$_2$ in the 5 and 10~K models is a little less than is achieved for O$_2$. 
The precise values of the modeled ratios for these, as for all other molecules, will depend also on the initial abundances assumed for them and/or their parent molecules, i.e. water, ammonia, CO and CO$_2$. It is also possible that N$_2$ and other small molecules may be formed in the gas phase, so the solid-phase model abundances may not bear a direct comparison.

\subsection{The influence of thermal diffusion}

The achievement of significant COM abundances to depths of tens of meters in the models at 60~K depends on the long-range mobility of either the complex organics themselves or their radical or atomic precursors, over periods on the order of a billion years. The models assume bulk diffusion is reasonably rapid, and there is some laboratory evidence for such behavior for certain molecules mixed with water (e.g. CO$_2$; Fayolle et al. 2011). Laboratory experiments of this kind typically involve large quantities of the diffusing species in relatively thin ices (on the order of 1~$\mu$m, often much less), meaning that the effects of back-diffusion, i.e. random walk, on the diffusion timescale are small. However, the extension to ice many meters in thickness, with diffusers of very low fractional abundance, could well produce different behavior. The interstellar gas-grain chemistry model of Garrod et al. (2017) applied a back-diffusion treatment for transfer between ice layers that was based on the parameterization of the results from a simple three-dimensional Monte Carlo diffusion model. Here, the number of bulk diffusion events, $N_{\mathrm{move}}$, required for a single diffuser (placed at an arbitrary starting position) to reach the surface of an ice of finite thickness, $N_{\mathrm{thick}}$ monolayers, followed the expression $N_{\mathrm{move}} = 2( N_{\mathrm{thick}} + 1/2)^{2}$. For transfer of low-abundance species between the deepest (i.e. thickest) layers in the comet model, this would suggest diffusion rates many orders of magnitude lower than assumed in the present models. However, in the other extreme, diffusers that make up a fraction of the total ice on the order of unity will show a negligible {\em net} back-diffusion effect, regardless of ice thickness. Back-diffusion is therefore highly sensitive to the abundance of the diffuser, and requires a more careful calculation, as per Garrod et al. (2017), when dealing with very thick ices. For this reason, the Garrod et al. expression for low-abundance diffusers (given above) was not used in the comet models.  Furthermore, its blanket application in fact would violate the necessary equality of diffusion rates across an arbitrary boundary for high-abundance species, due to the mismatch in layer thicknesses employed in these models. 

It is also likely that the degree (or even the possibility) of bulk diffusion will be dependent on the specific composition of the ice. The degree of diffusion allowed to non-water molecules in a highly water-dominated ice may be negligible, due to the build-up of crystalline structure. Ices composed of more balanced mixtures of water, CO, CO$_2$ and other commonly-detected solid-phase molecules may make possible the degree of COM diffusion required by the UV-only models to produce significant COM abundances at large depths into the ice. The degree of mixing exhibited in these models for various chemical species -- complex or otherwise -- into the deeper layers in particular may therefore represent something of an upper limit.

While it has been assumed here that the diffusion within the ices occurs through bulk diffusion, cometary ices are expected to be somewhat porous; indeed Jorda et al. (2016) determined the density of comet 67P to be $\sim$0.53 g~cm$^{-3}$, suggesting significant porosity. This value was used to determine the cosmic-ray penetration in the models presented here, based on a pure-water ice. A porous ice structure would provide an internal surface that could allow relatively rapid diffusion compared with bulk processes. This would have the local effect of increasing the rates of reactions that are moderated by diffusion, as well as making mixing between layers more rapid, assuming that the porous structures are well connected and not enclosed. Again, surface random walk would likely have a significant influence on the mixing rates of low-abundance species through such a mechanism. Willis \& Garrod (2017) studied the influence of two-dimensional random walk on the reaction rates of diffusive species on interstellar dust-grain surfaces, taking into consideration the abundance of the diffuser with respect to the available number of surface sites. While the effect was small in the case of discrete microscopic dust grains, on the order of a factor of a few, when extended to comet scales, the random-walk effect could arguably slow down diffusion by a factor of around 10-100 for a molecule of fractional abundance 10$^{-8}$ (depending on the available surface area and its geometry). 

The chemical and/or cosmic-ray processing of the ice could plausibly result in the gradual production of porous structures over time, even starting from a relatively non-porous initial ice. More detailed models that explicitly include pores/cavities, and chemistry upon their internal surfaces, would be necessary to confirm such effects.

\begin{figure*}
\begin{center}
\includegraphics[width=0.32\textwidth]{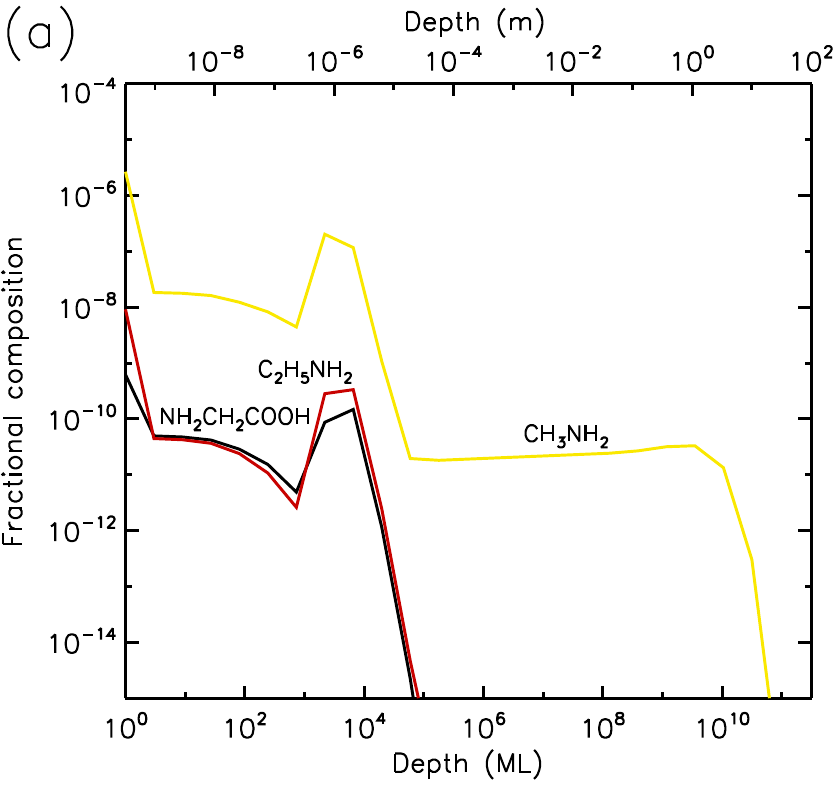}
\includegraphics[width=0.32\textwidth]{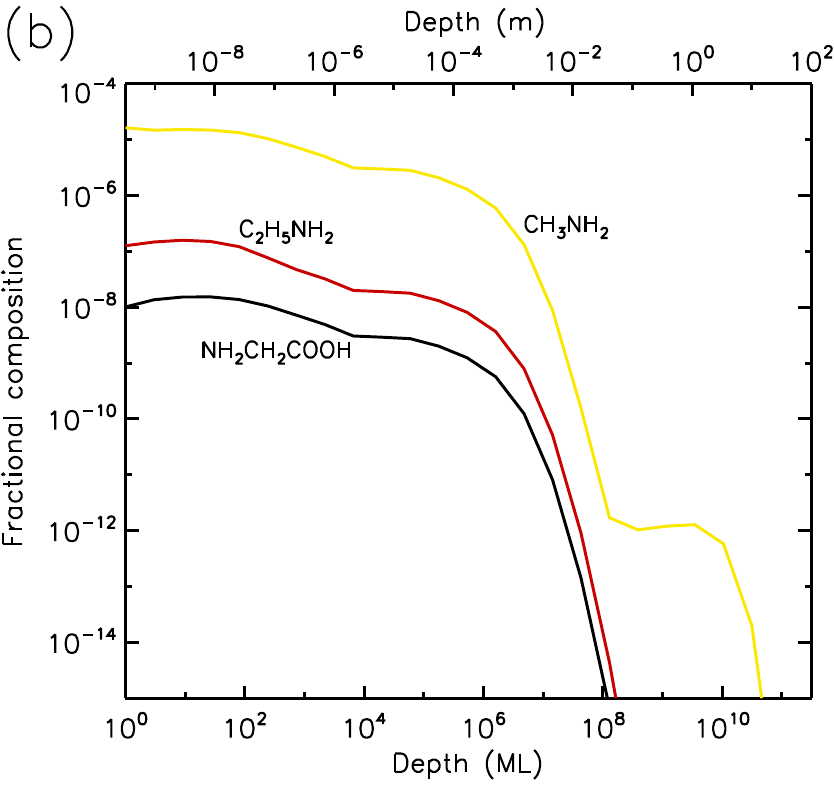}
\includegraphics[width=0.32\textwidth]{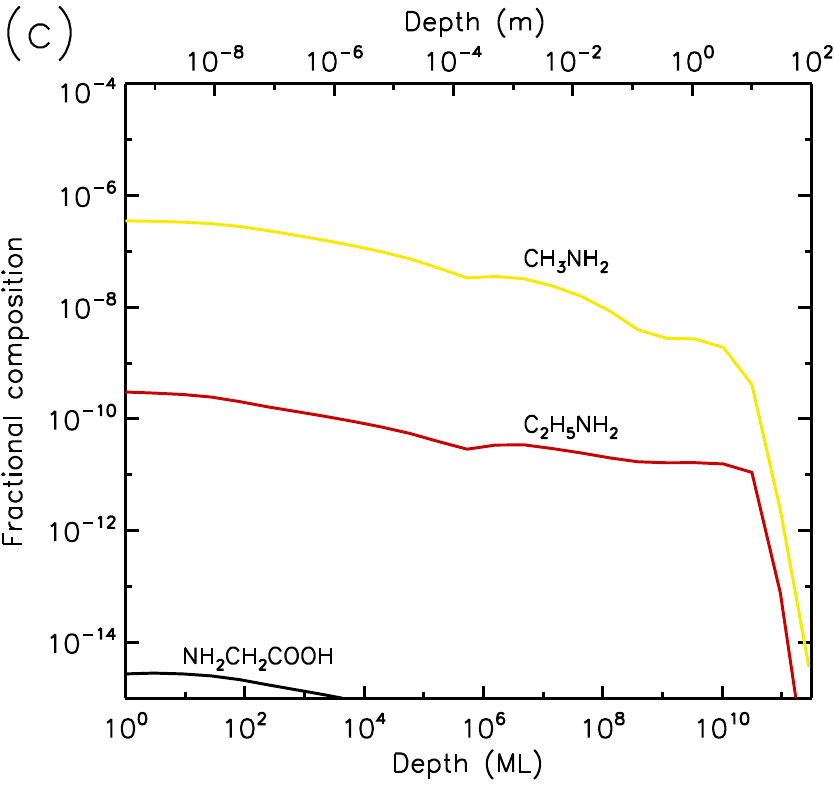}
\includegraphics[width=0.32\textwidth]{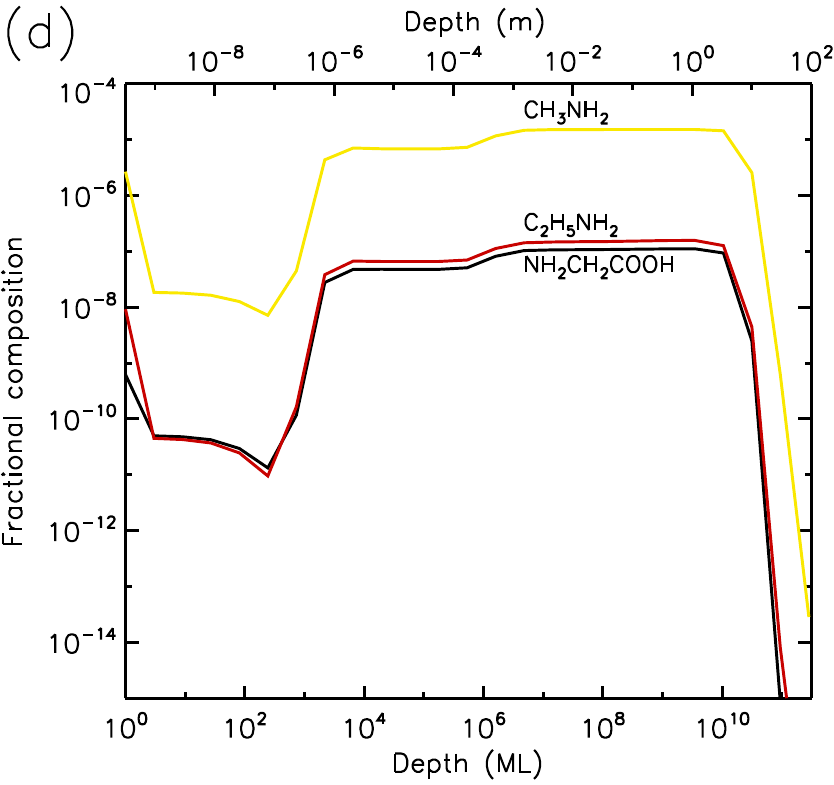}
\includegraphics[width=0.32\textwidth]{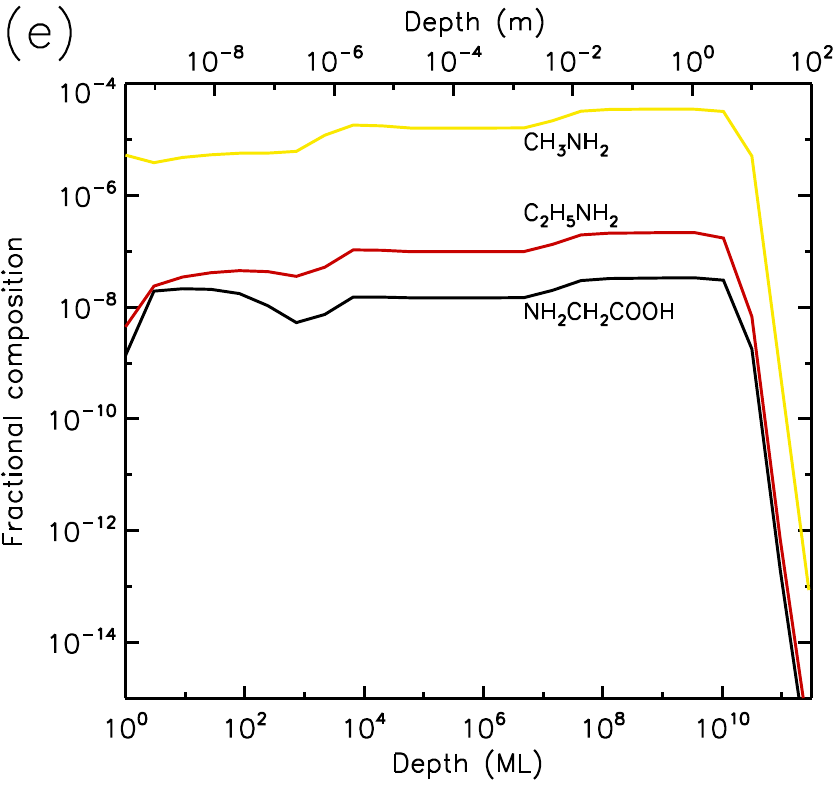}
\includegraphics[width=0.32\textwidth]{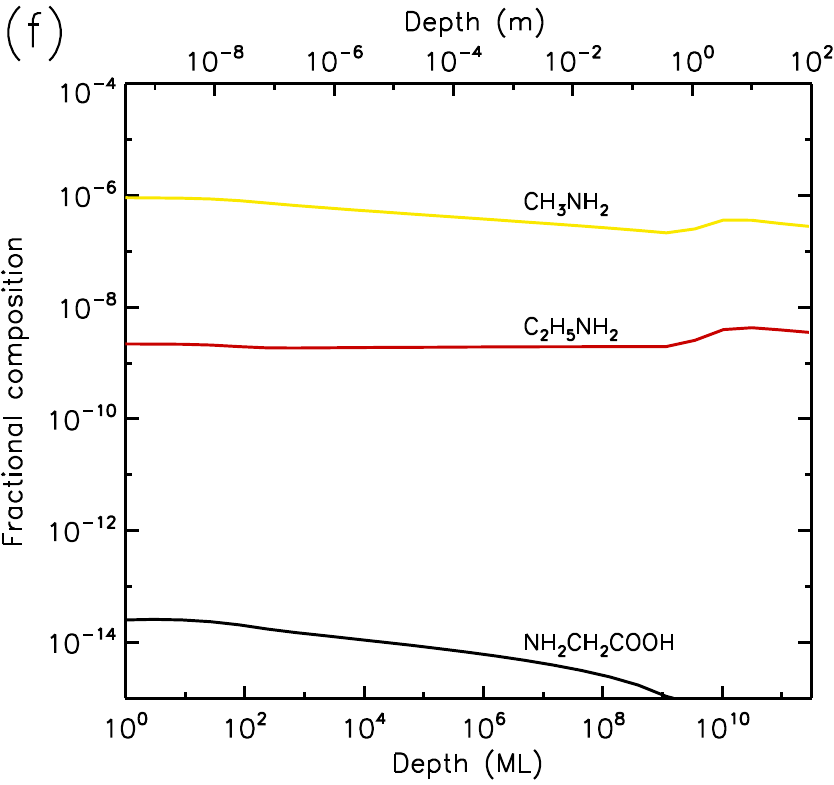}
\end{center}
\caption{\label{fig-glycine} Chemical abundances of glycine, methylamine and ethylamine, for cosmic-ray dissociation models at times $t=10^{6}$ yr (upper panels) and $t=5 \times 10^{9}$ yr (lower panels), assuming a temperature of 5~K (left panels), 10~K (middle panels), and 20~K (right panels).}
\end{figure*}

\subsection{Glycine}

Cosmic ray processing also produces COMs to large depths in these models. Glycine is not formed in great abundance (unlike other COMs); while this is technically consistent with the detected values for 67P, which range from 0 to 0.0025 with respect to water, the models do not reach anywhere close to the maximum detected value. Altwegg et al. (2016) suggest that the variation they detect is caused by changing conditions on the comet rather than uncertainty in the values. Furthermore, their results apparently suggest that the detected glycine is associated with its release from liberated dust, rather than ice, which is itself heated within the coma. Thus the ratio of glycine with respect to water may not be a meangful reflection of the overall ice composition. Altwegg et al. (2016) also compare the abundance of glycine with that of methylamine (CH$_3$NH$_2$) and of ethylamine (C$_2$H$_5$NH$_2$), finding ratios of 1.0$\pm$0.5 and 0.3$\pm$0.2, respectively. Each of these molecules are included in the present models, although ethylamine has only a sparse network in comparison with the other two. 

In the cosmic-ray processing models, glycine is at least an order of magnitude or so less abundant than methylamine at all times and depths. The production of glycine relies mostly on the addition reaction
NH$_2$ + CH$_2$COOH $\rightarrow$ NH$_2$CH$_2$COOH, as mentioned in Sec. 3.1,
rather than on the reaction NH$_2$CH$_2$ + COOH $\rightarrow$ NH$_2$CH$_2$COOH, which might be expected to follow more closely the abundance behavior of methylamine. However, the methylamine itself is formed by the addition of NH$_2$ and CH$_3$ radicals in the ice, so methylamine production should nevertheless be expected to show similar trends. This is indeed the case, as seen in Figure \ref{fig-glycine}, in which the abundances of glycine, methylamine and ethylamine are plotted for the cosmic-ray processing models. Although methylamine abundances are uniformly lower than those of glycine, the same general behavior is seen for both, especially after the full 5 Gyr timespan of the 5 and 10~K models. The models thus show some consistency with the detected behavior. Of course, if some or all of either molecule was formed in the ices prior to incorporation into the comet, then this analysis carries less weight. It may be seen that the abundance of ethylamine tracks very closely with that of glycine at 5 and 10~K, in good agreement with detected behavior. However, ethylamine in the present network does not have a formation mechanism directly involving NH$_2$, thus the precise agreement may simply be fortuitous, although, like methylamine, the general agreement in its abundance trend may be more reliable.

Based on the models presented here, if the true glycine abundance in the nucleus of 67P is close to the detected upper limit with respect to water, then its production is likely not the result of UV or cosmic-ray processing during cold storage.

\subsection{Observational consequences}

Setting aside that the present models do not consider a native composition at the start of the cold-storage phase that includes complex organic molecules, the broad observational implication of the UV and cosmic-ray processing investigated here is that new (and relatively dynamically young) comets in particular would show an enhanced complex molecular content versus older comets. Beyond a depth of perhaps several tens of meters, no effect is expected (although the influence of thermal diffusion is difficult to quantify, as discussed above). Molecules that are direct products of radicals formed from the most abundant ice components should be especially abundant in new comets, according to the models. These would include species already observed in comets (Hale-Bopp in particular) such as CH$_3$CHO, HCOOH, NH$_2$CHO, CH$_3$OCHO, and CH$_2$(OH)CHO, as well as related molecules including ethanol (C$_2$H$_5$OH) and dimethyl ether (CH$_3$OCH$_3$). Hydroxylamine (NH$_2$OH), although not yet detected in the interstellar medium, nor apparently in any comets, is predicted to have significant abundances also, as well as glyoxal (OHCCHO). 

The observed abundances of nitrile species (e.g. and HCN, CH$_3$CN) and the cyanopolyyne HC$_3$N are generally not well reproduced by the models. This may be a function of the simple starting ice composition. The inclusion of say, HCN or HNCO, in the initial ices may have a strong impact on the later products, although the more complex molecules may also have been present themselves in the comets without requiring further processing. Given the observed, substantial abundances of H$_2$S and some other sulfur-bearing species in comets, it might be expected that the large abundances of oxygen-bearing molecules like methyl formate could be accompanied by similar sulfuretted structures.

The various comparisons made here between the model abundances and the observed values must be tempered by the fact that the active phase itself may promote some degree of thermal chemistry within the ices, relying on the presence of unreacted radicals in the ice, that could become more mobile at elevated temperatures, and thus more reactive. Other thermally-activated reactions (including those between stable species) could be enhanced also. The production of yet more complex species than the typically observed ``hot-core'' type molecules, which includes the example of glycine as discussed above, may be more dependent on the elevated temperatures achieved during the active phase. Thus, the ideal comparison of models with observations would include the simulation of the time-dependent physical conditions associated with entry into the inner solar system.

\subsection{Other observational comparisons}

The chemical compositions of volatiles ejected from the two largest fragments of comet Schwassmann-Wachmann 3 (73P) were observed by Dello Russo et al. (2007). A Jupiter family comet with a likely origin in the Kuiper belt, 73P split into multiple fragments during its 1995 apparition, exposing material from deep within the comet nucleus. The observations by Dello Russo et al., taken during the comet's 2006 return, show that the two fragments are of very similar composition, indicating a lack of chemical differentiation within the nucleus. It is unclear whether the depth-scale of processing found in the chemical models presented here, of order 10~m, would be sufficient to differentiate material originating from the two fragments of 73P, which have sizes of hundreds of meters. The molecules observed by those authors are also fairly small compared with many of those investigated here. The modeled abundances of native ice species with respect to water, aggregated to a depth of 15~m as shown in tables \ref{tab-60K-hi} -- \ref{tab-20K-CR}, vary the most for methanol (CH$_3$OH), which is depleted by a factor of around 4 at the end of the 5~K and 10~K cosmic ray-processing models. Methanol varies by a factor around 1.3 between 73P-B and 73P-C. If any chemically processed material lay only in the upper 10\% of the nuclear material in those fragments, with the remaining material being otherwise identical, then the model results would seem compatible with those observations.

Gibb et al. (2007), on the other hand, found that the Oort cloud comet C/2001 A2 (LINEAR), which fragmented into two pieces, indeed showed evidence of nuclear heterogeneity, based on variability in the abundances of H$_2$CO and CH$_4$. These two molecules also are found to vary from their initial abundances in the 5~K cosmic-ray models, increasing their abundances by factors 1.7 and 2.2 respectively. Observations of the dynamically new comet C/2012 S1 (ISON) (Disanti et al. 2016; Dello Russo et al. 2016), which broke up upon its close approach to the Sun, showed strong variations in the abundances of some molecules (notably NH$_3$, H$_2$CO, HCN and C$_2$H$_2$) at heliocentric distances less than 0.5 au, versus those outside this radius. This may also suggest heterogeneity in the nuclear ices, although the precise origins of the detected molecules are uncertain.

Comets C/2001 A2 and C/2012 S1 would each appear to be a better dynamical match for the assumed physical and chemical conditions of the models presented here than 73P, but none of those cometary molecular observations would appear to rule out strongly the suggested scenario of chemical processing in the Oort cloud that is investigated in these simulations.

\section{Conclusions}

The model presented here is the first attempt to apply a chemical kinetics treatment of solid-phase chemistry, more typical in interstellar simulations, to the chemical, thermal and radiative processing of cometary nuclear material. Based on the results, the exposure of a cometary surface to the interstellar UV radiation field during ``cold storage'', nominally in the Oort cloud, may be sufficient to produce a range of quite complex organic molecules within the upper micron of material. The achievement of moderate temperatures as high as 60~K, which may be more typical of Kuiper Belt objects than of the Oort Cloud, could also allow such processing to extend to deeper layers, although this effect is strongly dependent on the process of bulk diffusion, which is poorly defined in both models and experiment, while diffusion within porous structures may also play a role that is not explicitly treated in these models.

The inclusion of cosmic-ray processing allows the production of both simple and complex molecules to depths on the order of 10~m, starting from only the basic ice components that are found in interstellar ice mantles. Processing at low temperatures in particular, over time periods on the order of 1 billion years, can produce complex organic molecules (COMs) to those depths, with abundances comparable to those obtained over much shorter periods in star-forming regions. The models show therefore that the complex molecular component of cometary ices may not be entirely inherited, but may be formed {\em in situ} to modest depths. 

Cosmic-ray processing may have the most important influence on the chemistry, versus UV, due to its penetration to large depths. In the models that treat cosmic-ray processing, the lowest temperatures (i.e. 5--10~K) appear to be better at reproducing observed abundances of both simple and complex molecules in active-phase comets. This may indicate that {\em diffusive} motions that could lead to molecular transport and/or reactions within the bulk ice are unimportant and/or strongly inhibited over macroscopic scales. Any complex organic material present in the deepest layers of the cometary ice nevertheless most likely reflects the native composition of the dust-grain ice mantles from which a comet is formed. Comets that have undergone many orbits entering the inner solar system would be unlikely to show any remaining evidence of the sort of processing investigated here during cold storage in the Oort cloud.

In contrast to other COMs, glycine is not so abundant in the models presented here, either as the result of cosmic-ray or UV processing, and does not approach the maximum abundance determined for 67P, either relative to water or to methylamine. An alternative scheme involving higher, or variable temperature, might allow a more optimal production rate to be found. However, based on the present models, high glycine abundances must either be inherited from the interstellar/protoplanetary stage or be formed through chemical pathways not included in the present network.

The good agreement of the aggregated abundances of COMs, in many cases, with those observed in the coma of Hale-Bopp may suggest that the prevalence of such molecules in that object was a result of long-term cosmic-ray induced processing in the upper 10+~m of material during cold storage. Dynamically new comets might therefore be expected, more generally, to exhibit higher COM abundances, if the production of those molecules {\em in situ} can be adequately distinguished from material inherited from earlier times.

The depth to which ice processing may extend is strongly dependent on both the porosity of the cometary nuclear material and on the rates of molecular dissociation caused by the high-energy end of the cosmic-ray flux distribution (here, up to 1~TeV). The treatment of cosmic-ray induced chemical processing used here is approximate, but is conservative in the assumed rates. The production of O$_2$ (and the chemically associated H$_2$O$_2$) at great depths in these models, caused by CR-induced dissociation of water, appears to be a plausible explanation for the anomalously-high abundance of O$_2$ detected in 67P, although this is dependent also on the rate of attrition of the comet; indeed, each of the parameters contributing to this possible explanation are subject to considerable errors. Nor does this conclusion preclude other contributions to the O$_2$ content, such as cosmic ray or UV-induced processing during the protoplanetary or interstellar stages of ice formation.

Just as O$_2$ may have plausible formation mechanisms in earlier stages, more complex molecules could also be incorporated into cometary material directly from icy dust grains. However, it is also possible to destroy these molecules, either by UV or cosmic-ray processing; future cometary models will include ice mantles composed of a range of organics, as well as sulfur-bearing species, to test their survival over long periods and not only their formation.

The present models consider chemistry occurring only during the cold storage phase. However, the heating and irradiation of active-phase comets could also be studied using these models. The elevated temperatures in particular, in the outer layers of an active-phase comet, could plausibly contribute to the production of simple and complex molecules from previously immobile radicals produced during cold storage, or embedded in the ice at the formation of the comet, with the products subsequently ejected into the coma. Barrier-mediated reactions between otherwise stable molecules could also become influential under such conditions (Theul{\'e} et al. 2013). The high solar flux of protons at 1~keV (vs. that originating from interstellar sources) may influence the chemical composition of the upper micron of a comet's surface during approach (e.g. Cooper et al. 2003) but it is unlikely that the abundances of the complex organics investigated here would be substantially affected at such depths, except either for (relatively) new comets that had not yet undergone sufficient surface processing to remove all icy material from that upper micron of material, or for comets that had lost a significant portion of their protective outer surface. In the models presented here, the photo-dissociation rate of water caused by interstellar UV at the surface of a comet, of order 10$^{-10}$ s$^{-1}$, would give a water molecule a lifetime against dissociation of a few hundred years. For a comet experiencing solar irradiation from a distance of 5 au, this would be cut to under a year (see also Sec. 1). However, as with the lower energy protons, the effect is unlikely to be significant beyond the upper micron of material at most, limiting the effect to new comets with relatively unprocessed surfaces when entering the inner solar system.

A selection of key conclusions from this preliminary study are summarized below:
\begin{enumerate}
\item UV and cosmic ray-induced processing of cometary ice may contribute to the production of complex organic molecules to depths on the order of 10~m or more over time periods on the order of 1 Gyr. The precise depths and production rates are, however, poorly determined.

\item The models suggest that O$_2$ and H$_2$O$_2$ are also produced to such depths. Under conditions where molecular diffusion is slow, abundances approaching detected values may be achieved.

\item Observable abundances of complex organic molecules may be enhanced, due to production during cold storage in the Oort cloud, for dynamically new (or young) comets.

\item Although glycine is produced, its abundance is modest and not sufficient to produce agreement with detected values in comet 67P.

\item The models may be improved by the adoption of time- and depth-dependent temperature profiles. The effects of nearby supernovae or the passage of hot stars, earlier in solar system history, also remain to be explored. 

\item Active-phase comets could also experience substantial chemical processing in their upper layers, due to solar approach. A similar chemical kinetic modeling study is required to determine the strength of such processing.

\item The direct, immediate reaction of radicals produced by photodissociation of mixed molecular ices, unmediated by thermal diffusion, is a mechanism missing from interstellar chemical kinetics models, and which could be easily incorporated into such treatments. This would provide a process by which complex molecules (as well as O$_2$) could be produced in ices at very low temperatures, in keeping with experimental evidence.

\end{enumerate}

\acknowledgements

This work was funded by the NASA Emerging Worlds Program, grant number NNX17AE23G. The author thanks C. Shingledecker and E. Herbst for helpful discussions, R. Johnson for a careful reading of the draft manuscript, and D. Atri for communication of the details of his simulations. The author also thanks the anonymous referee for very helpful comments and suggestions.

\appendix

\section{Appendix material}

\begin{figure*}
\begin{center}
\includegraphics[width=0.32\textwidth]{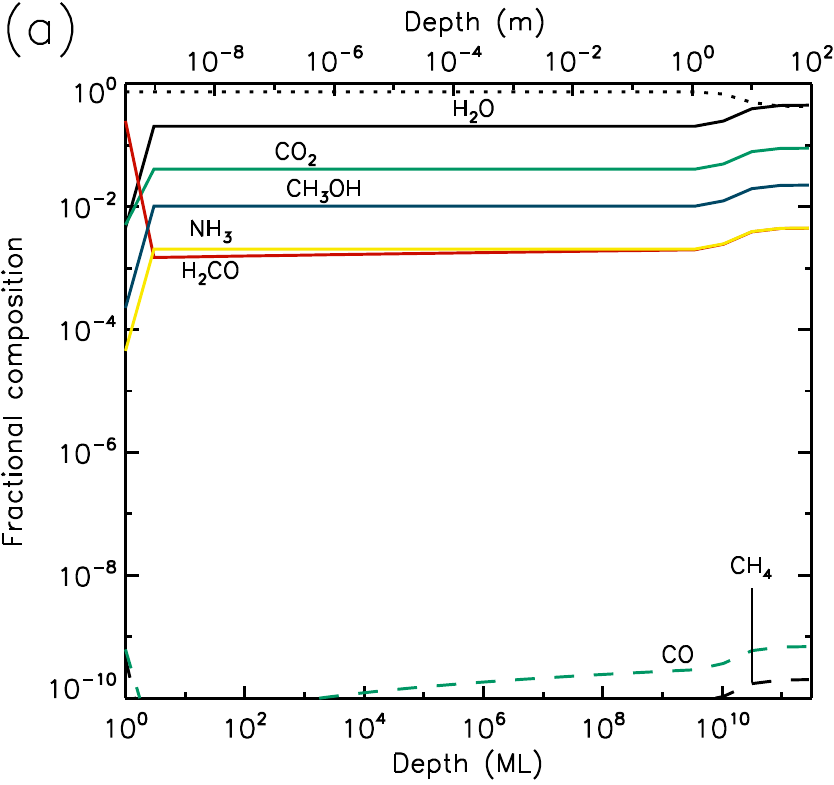}
\includegraphics[width=0.32\textwidth]{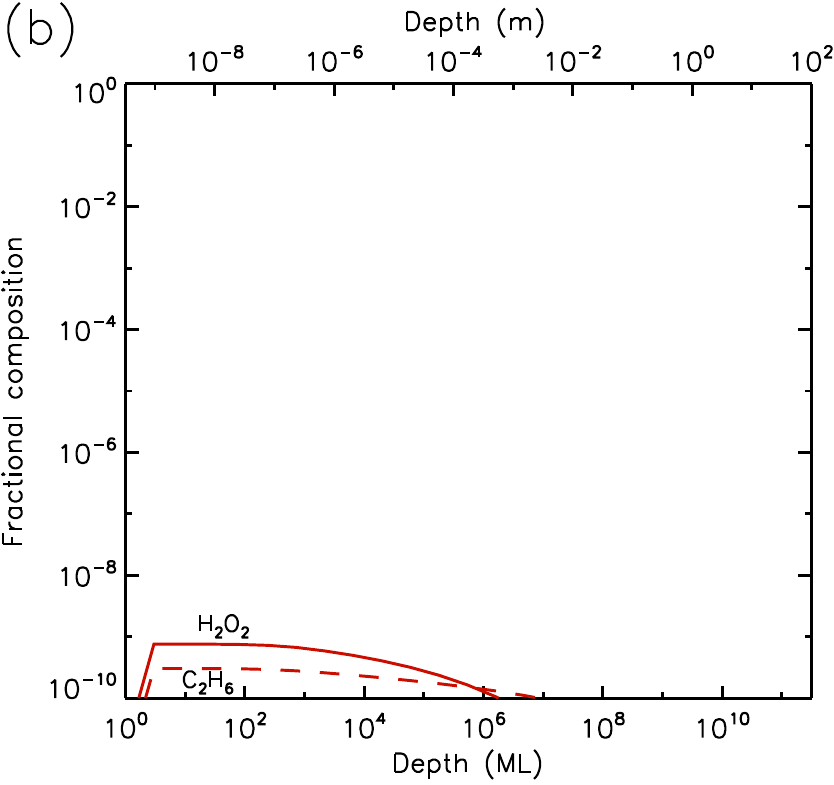}
\includegraphics[width=0.32\textwidth]{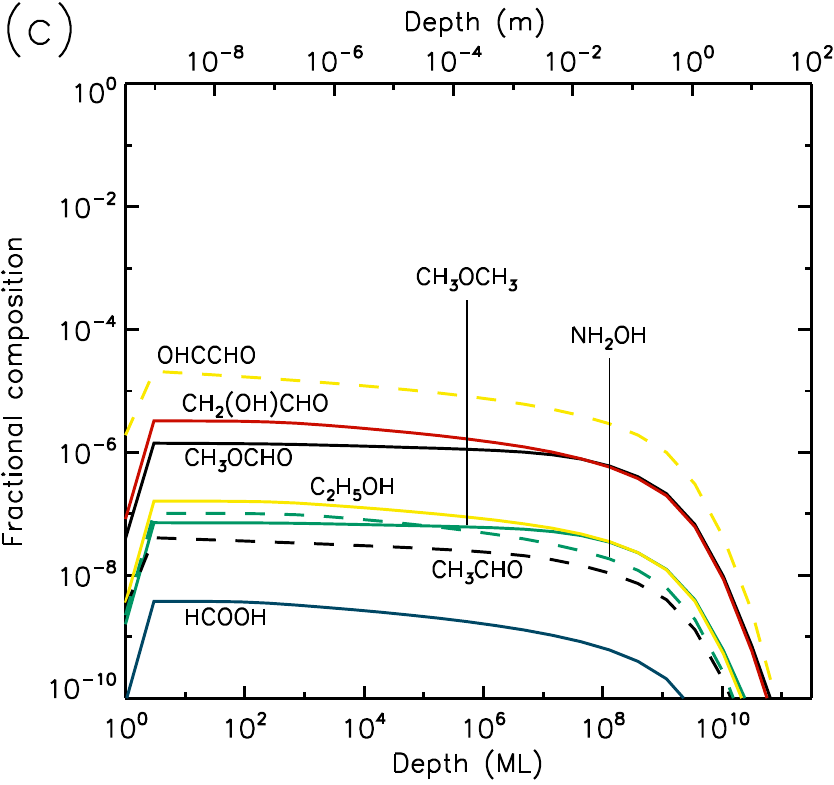}
\includegraphics[width=0.32\textwidth]{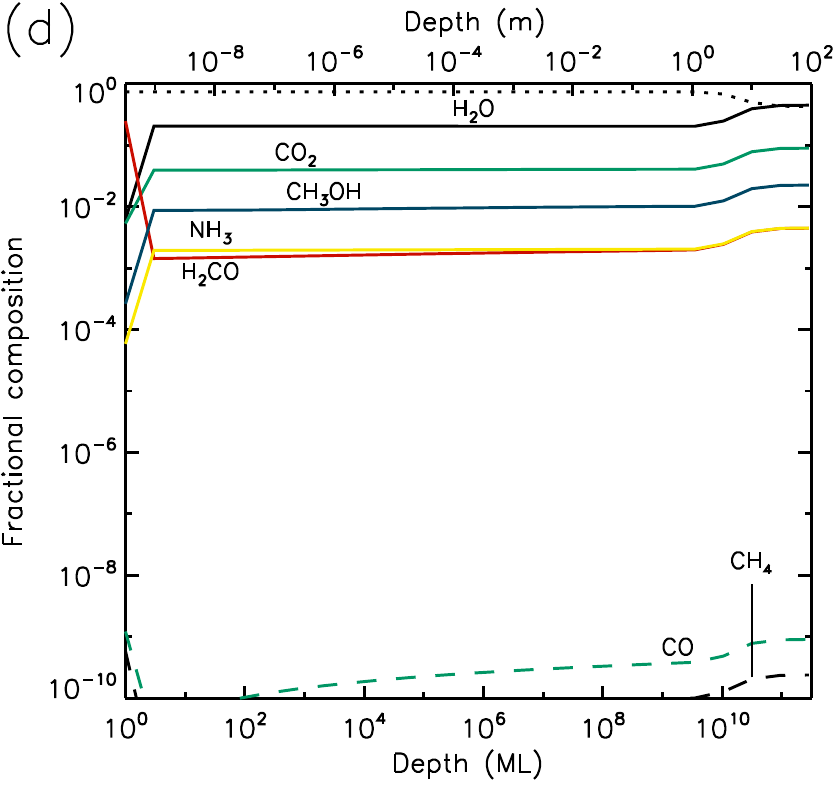}
\includegraphics[width=0.32\textwidth]{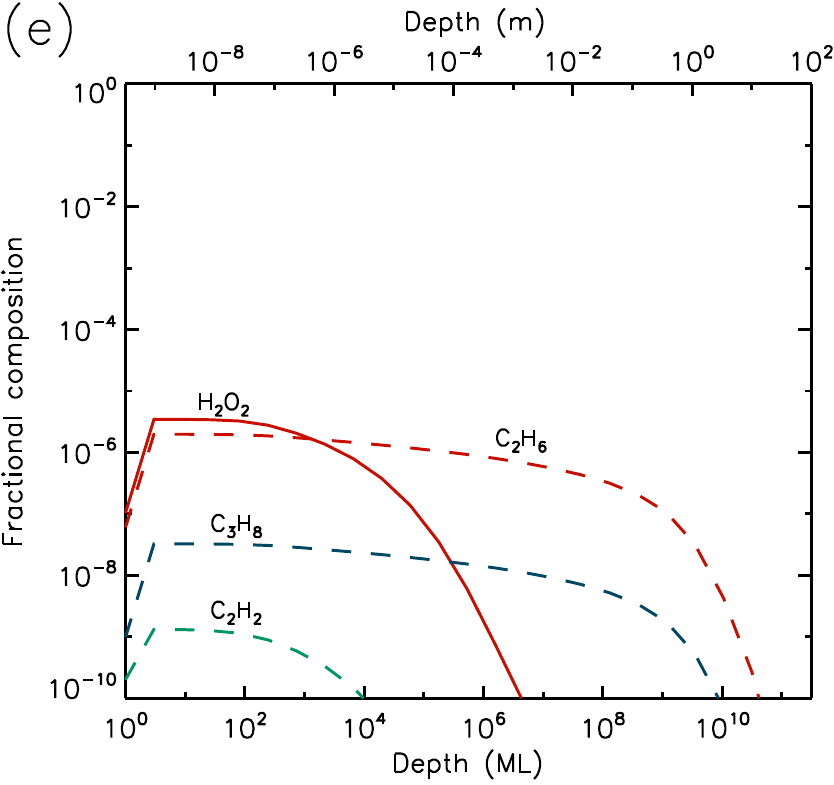}
\includegraphics[width=0.32\textwidth]{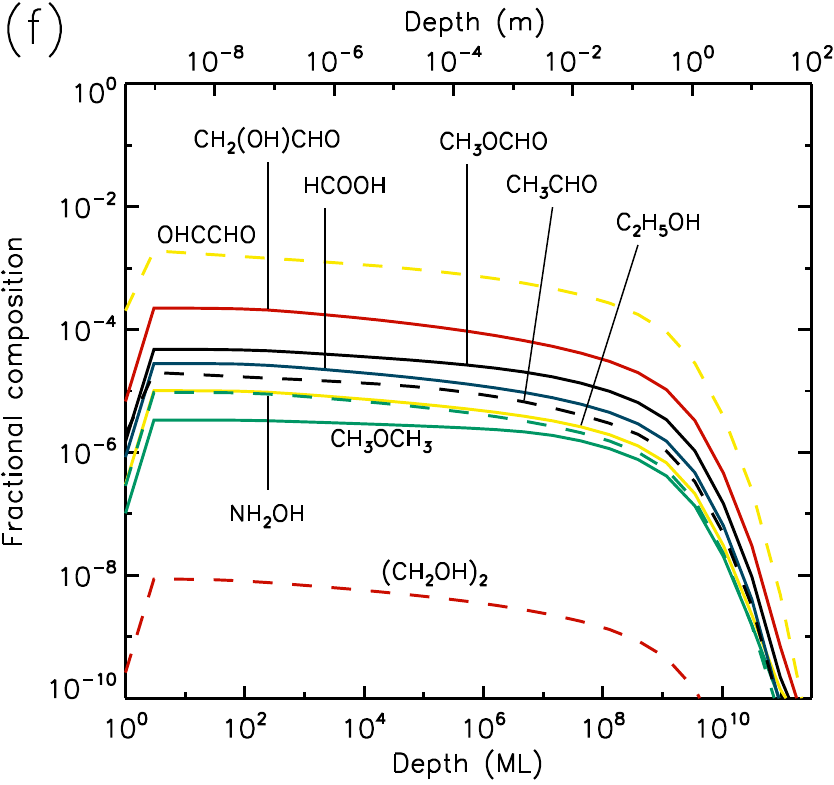}
\end{center}
\caption{\label{fig-50K-early} 50 K models at time $t=10^{6}$ yr. Abundances of the initial ice components are shown in the left panels, with dust shown as a dotted line. Results for two model setups are shown: low UV (upper panels) and high UV (lower panels).}
\end{figure*}

\begin{figure*}
\begin{center}
\includegraphics[width=0.32\textwidth]{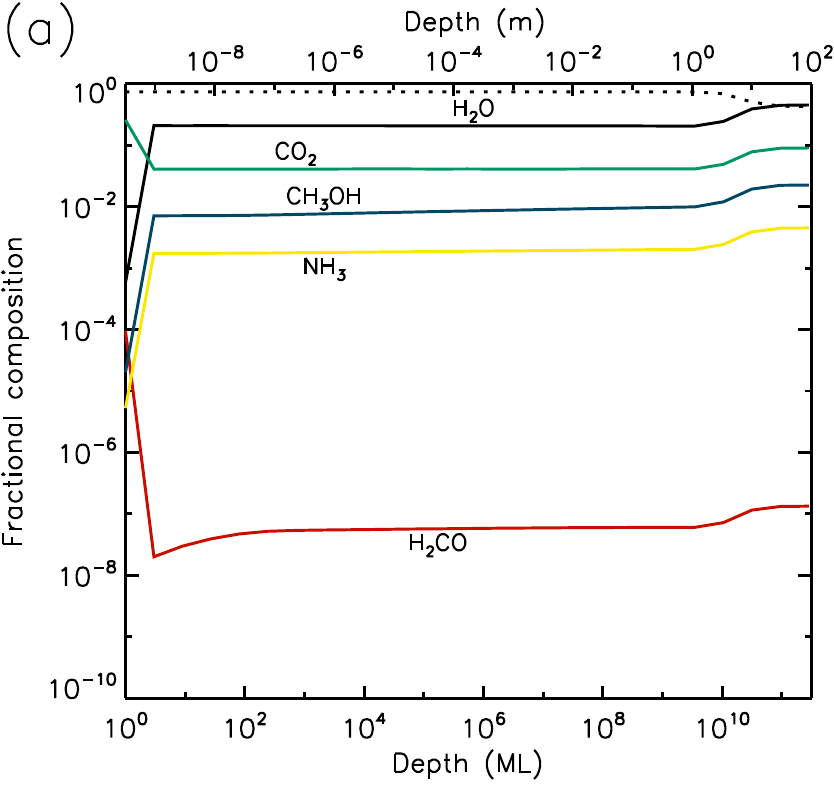}
\includegraphics[width=0.32\textwidth]{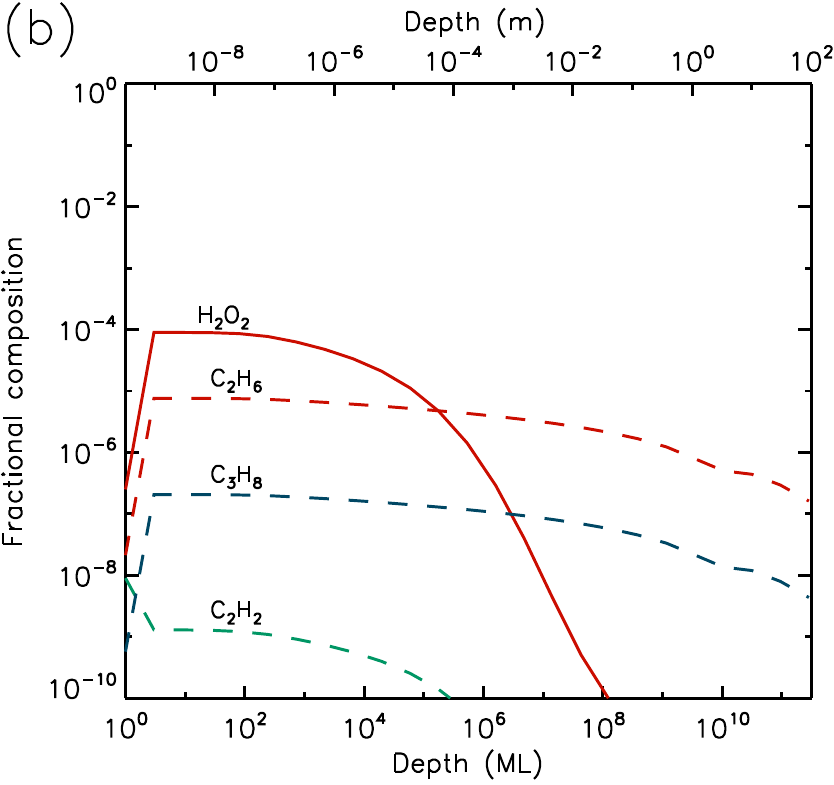}
\includegraphics[width=0.32\textwidth]{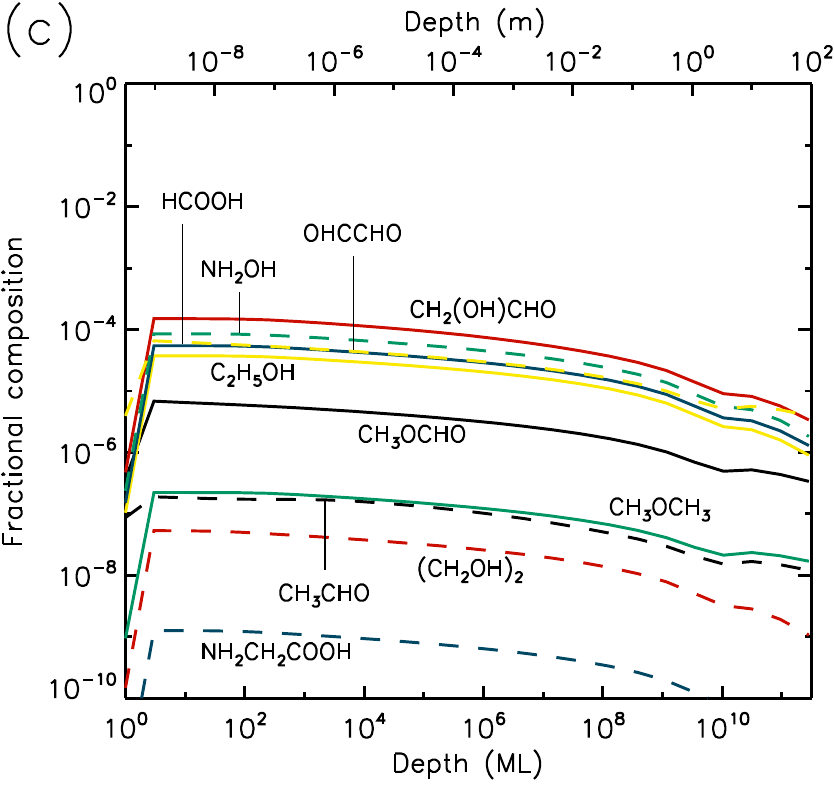}
\includegraphics[width=0.32\textwidth]{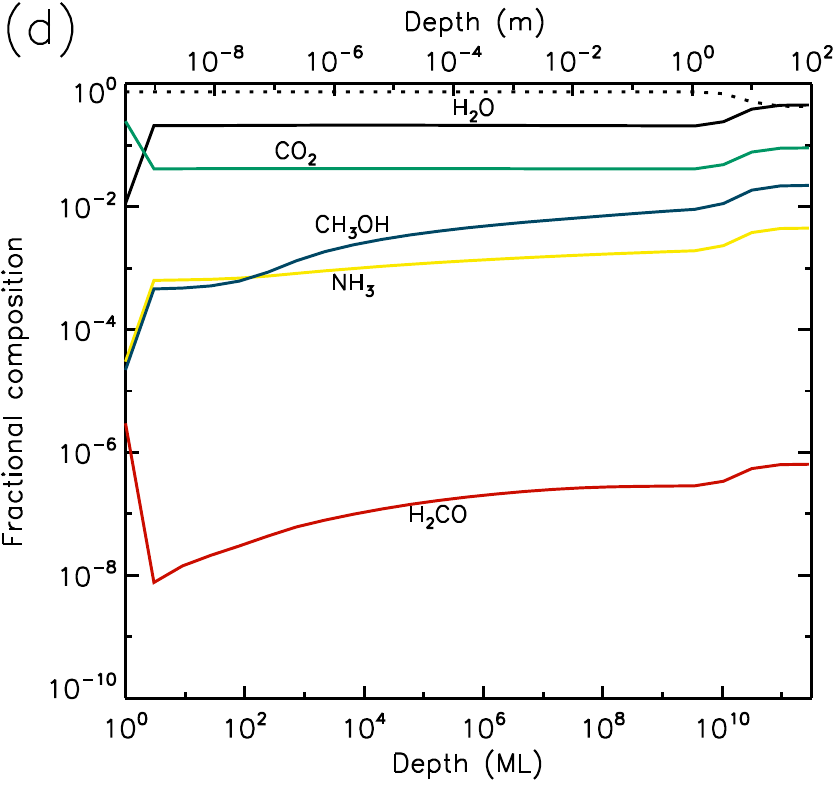}
\includegraphics[width=0.32\textwidth]{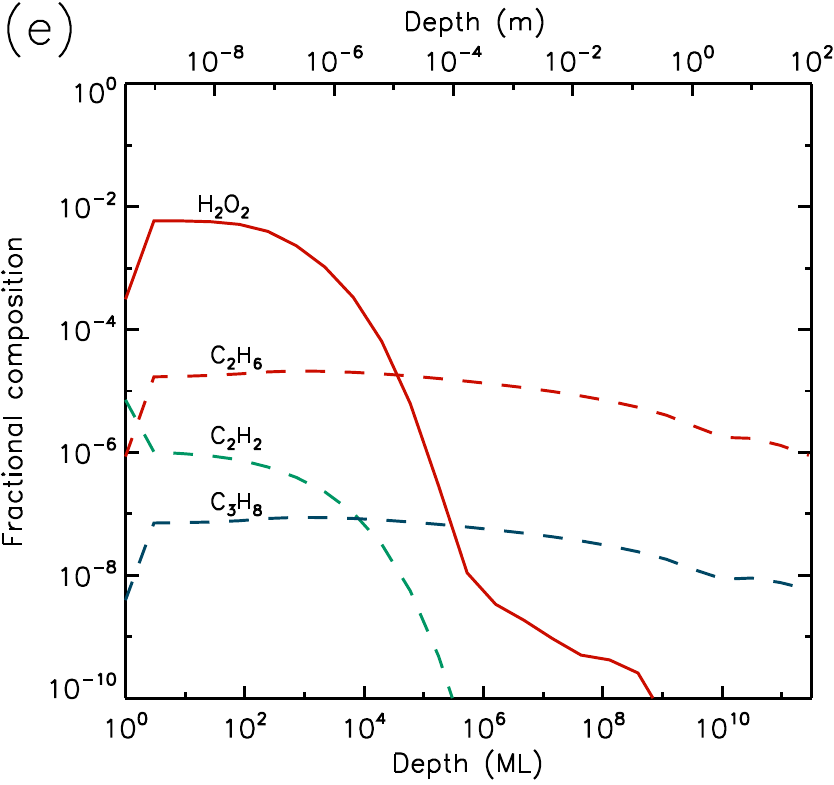}
\includegraphics[width=0.32\textwidth]{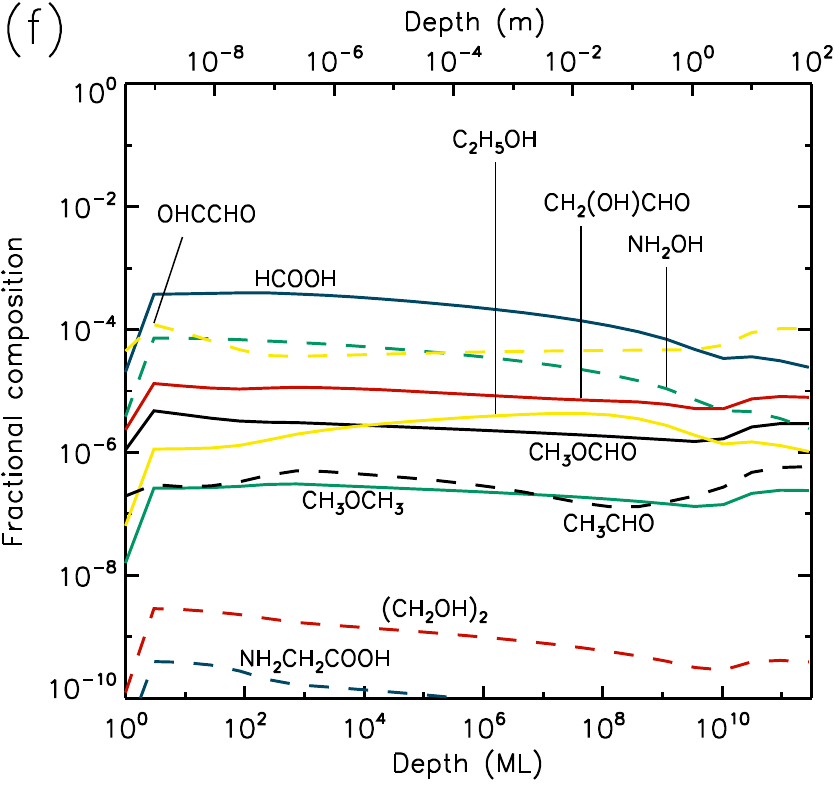}
\end{center}
\caption{\label{fig-50K-late} 50 K models at time $t=5 \times 10^{9}$ yr. Abundances of the initial ice components are shown in the left panels, with dust shown as a dotted line. Results for two model setups are shown: low UV (upper panels) and high UV (lower panels).}
\end{figure*}

\begin{figure*}
\begin{center}
\includegraphics[width=0.32\textwidth]{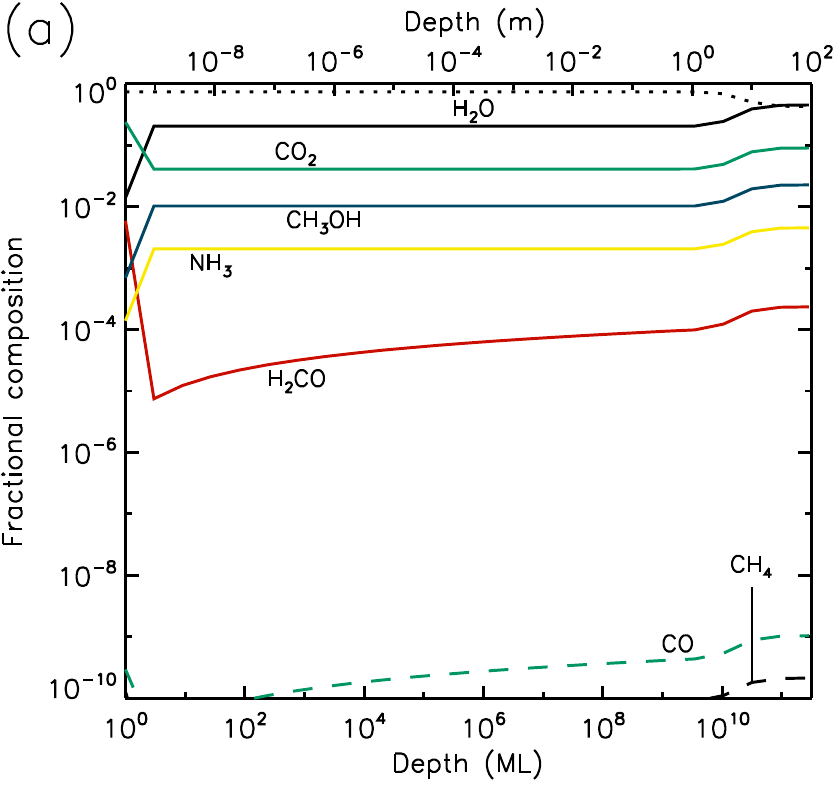}
\includegraphics[width=0.32\textwidth]{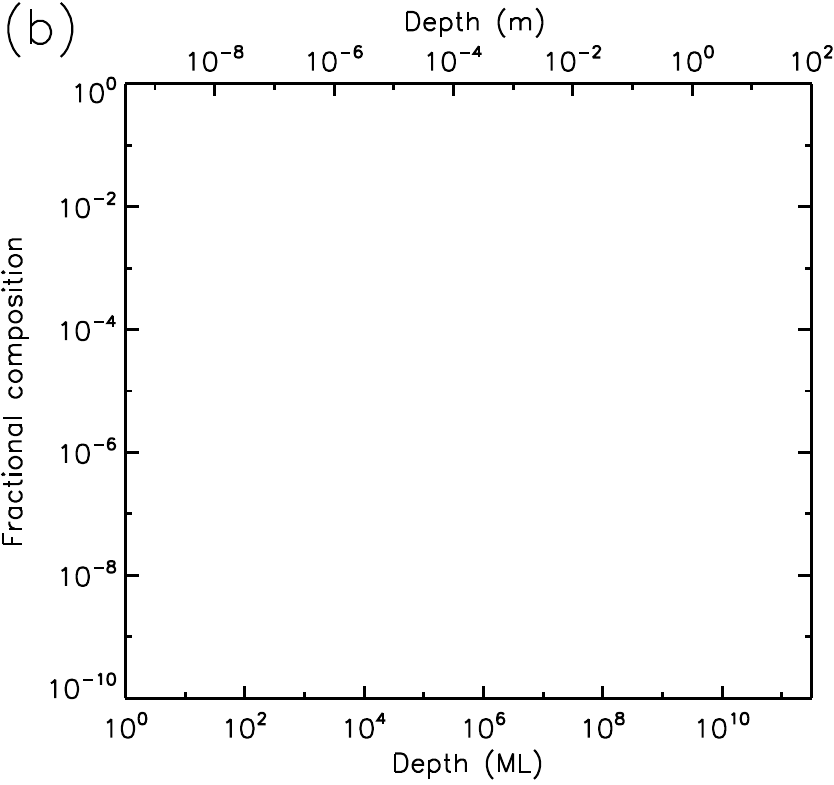}
\includegraphics[width=0.32\textwidth]{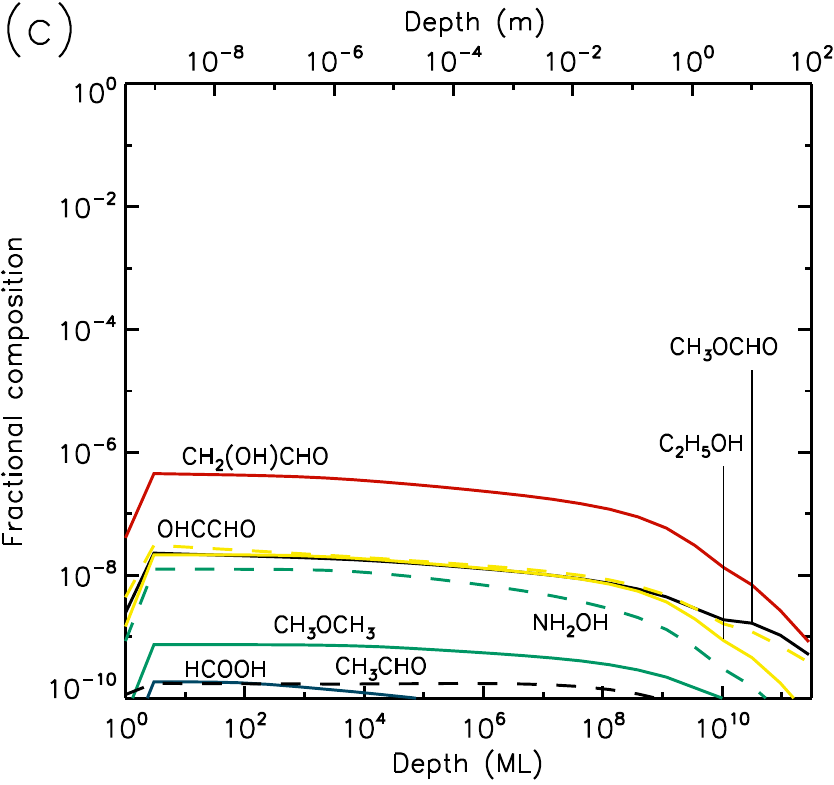}
\includegraphics[width=0.32\textwidth]{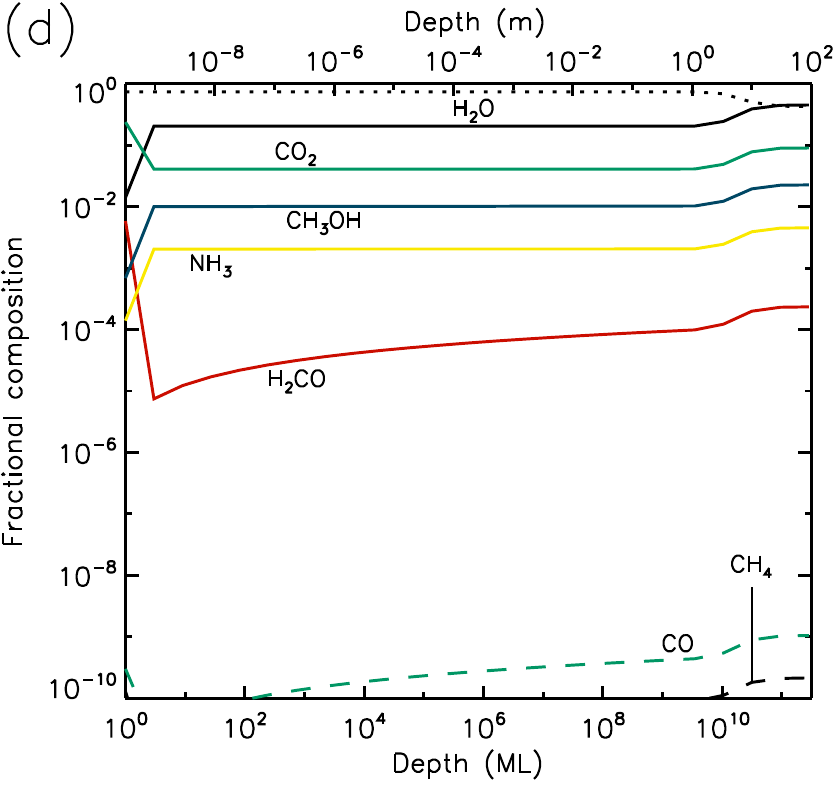}
\includegraphics[width=0.32\textwidth]{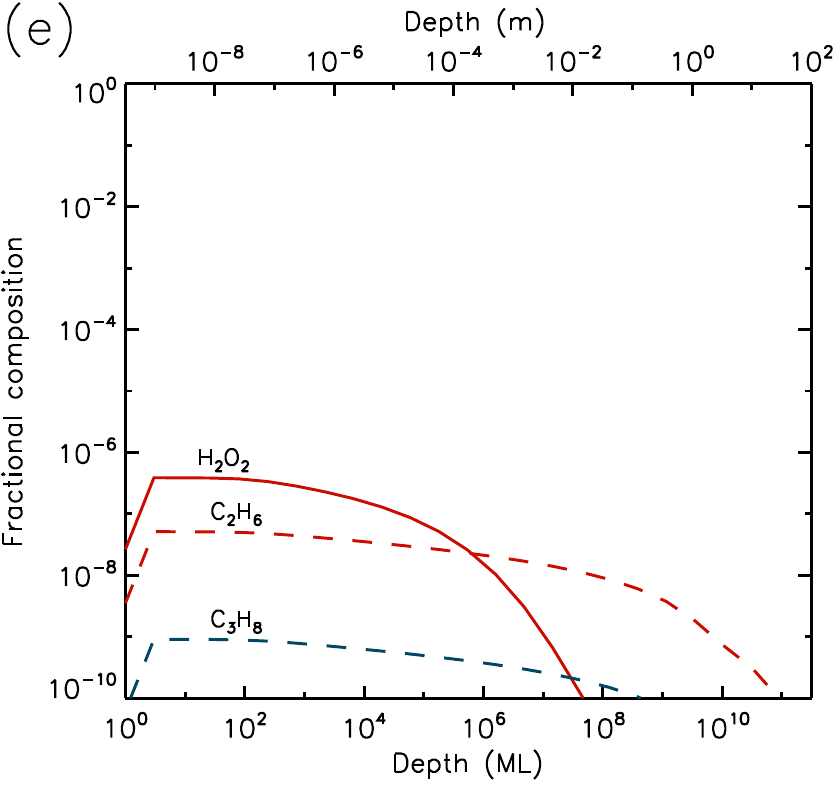}
\includegraphics[width=0.32\textwidth]{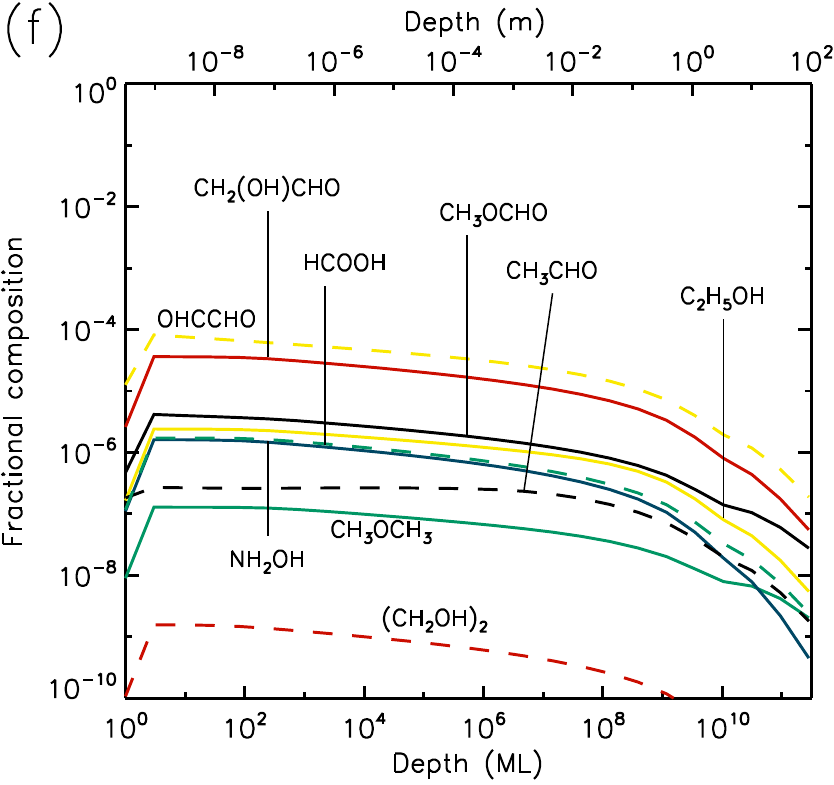}
\end{center}
\caption{\label{fig-60K-early} 60 K models at time $t=10^{6}$ yr. Abundances of the initial ice components are shown in the left panels, with dust shown as a dotted line. Results for two model setups are shown: low UV (upper panels) and high UV (lower panels).}
\end{figure*}

\begin{figure*}
\begin{center}
\includegraphics[width=0.32\textwidth]{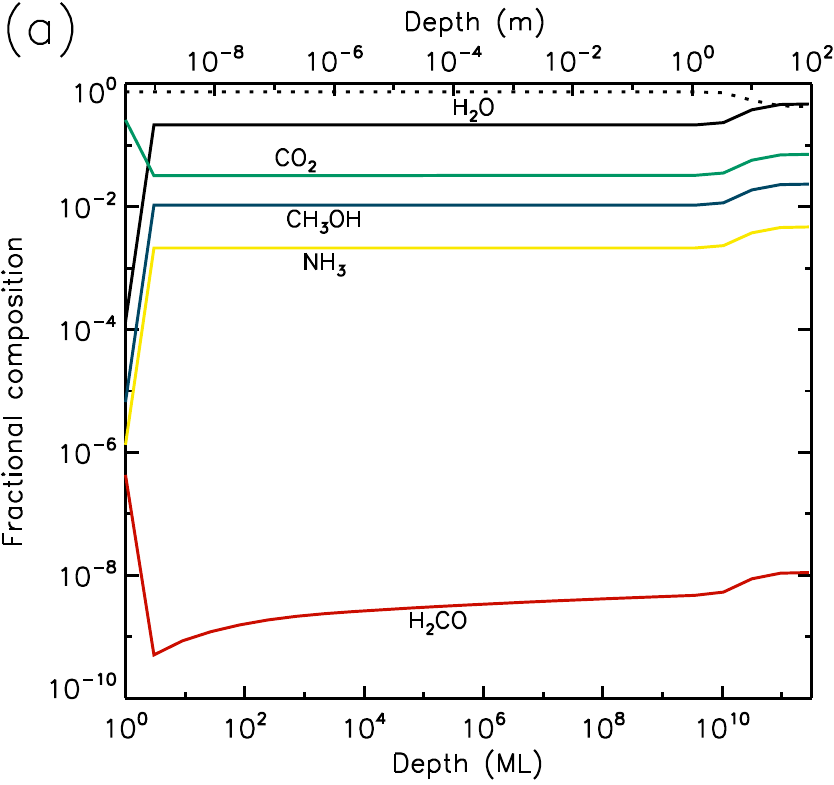}
\includegraphics[width=0.32\textwidth]{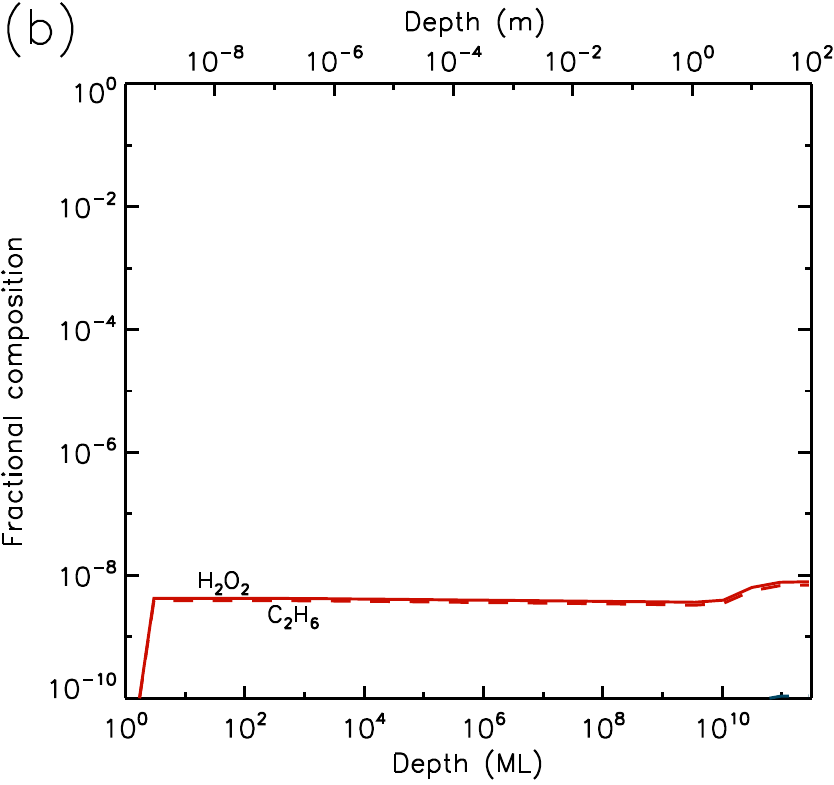}
\includegraphics[width=0.32\textwidth]{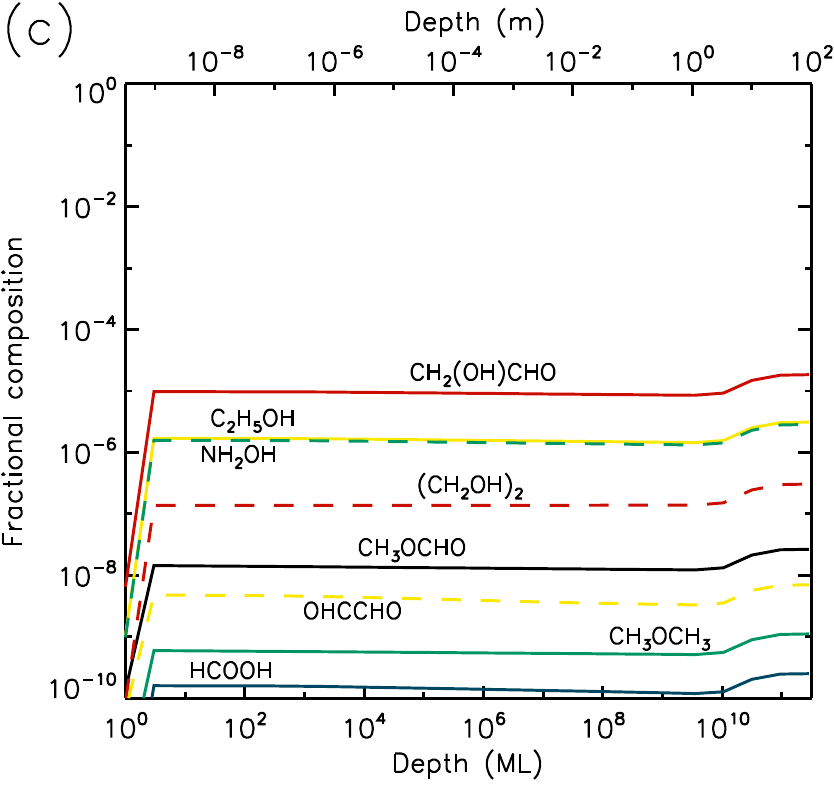}
\includegraphics[width=0.32\textwidth]{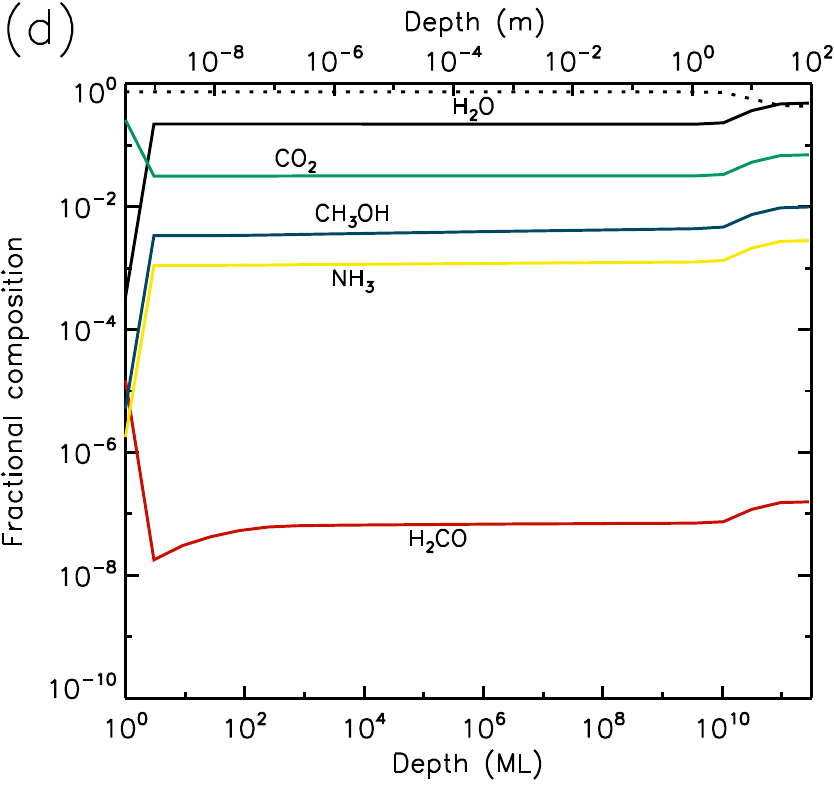}
\includegraphics[width=0.32\textwidth]{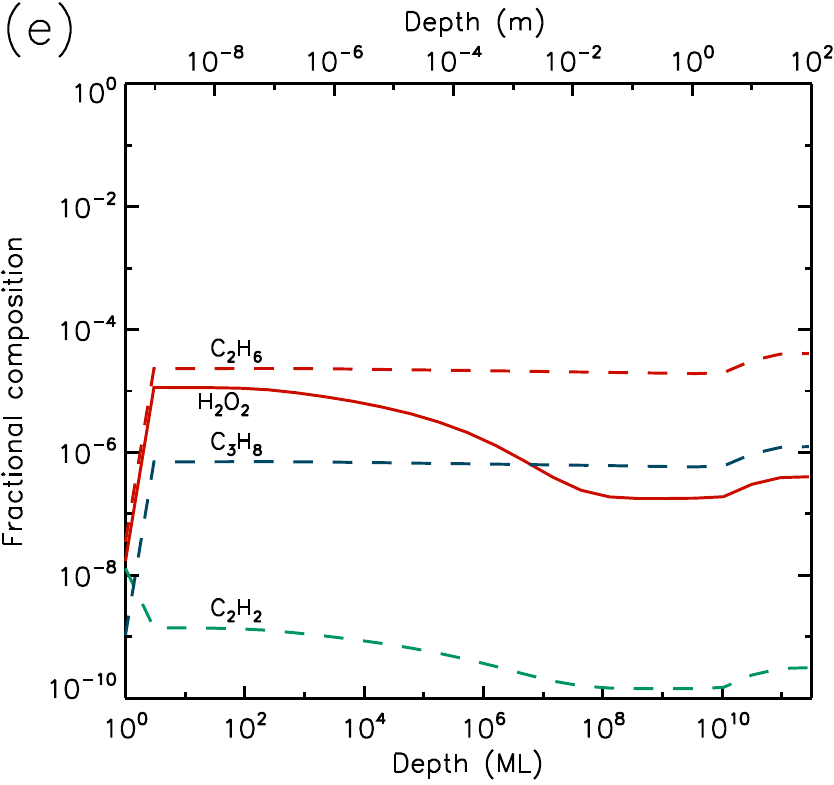}
\includegraphics[width=0.32\textwidth]{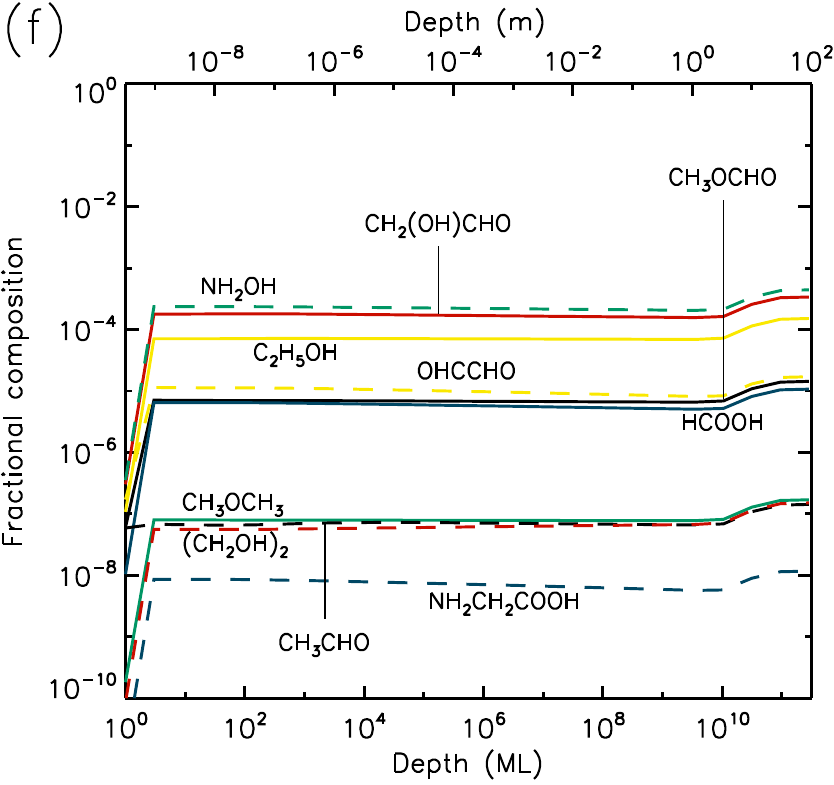}
\end{center}
\caption{\label{fig-60K-late} 60 K models at time $t=5 \times 10^{9}$ yr. Abundances of the initial ice components are shown in the left panels, with dust shown as a dotted line. Results for two model setups are shown: low UV (upper panels) and high UV (lower panels).}
\end{figure*}

\end{document}